\documentclass[11pt]{article}
\pdfoutput=1
\usepackage{jheppubme}
\usepackage[table]{xcolor}
\usepackage{graphicx, psfrag, dsfont}
\usepackage{float}
\usepackage{caption, comment}
\usepackage[mathscr]{euscript}
\usepackage{dynkin-diagrams}
\usepackage{appendix}
\usepackage{coolstr}
\usepackage{hyperref}
\usepackage[normalem]{ulem}
\hypersetup{pdfauthor={Name}}
\usepackage{color}
 \usepackage{tikz}
\usepackage[most]{tcolorbox}
\usepackage{amsmath}
\usepackage{mathtools}
\usepackage[bb=dsserif]{mathalpha}
\usepackage{bm}
\usetikzlibrary{arrows.meta,positioning}
\usetikzlibrary{shapes.geometric,decorations.pathmorphing,arrows.meta,calc}

\usetikzlibrary{
  calc,
  decorations.markings,
  decorations.pathreplacing,
  arrows.meta,
  patterns,
  fadings,
  shadows.blur,
  positioning
}

\definecolor{Cblue}{RGB}{42,100,210}
\definecolor{Cred}{RGB}{200,70,70}
\definecolor{EqGray}{RGB}{60,60,60}
\definecolor{Psi}{RGB}{130,0,180}
\definecolor{Gold}{RGB}{200,150,0}
\definecolor{Soft}{RGB}{230,230,230}

\tikzset{
  >={Latex[length=3.4mm,width=2.2mm]},
  sphere/.style={line width=0.8pt, draw=black},
  equator/.style={line width=2.0pt, draw=EqGray},
  equatorGlow/.style={line width=3.8pt, draw=EqGray!40},
  projector/.style={rounded corners=2pt, draw=black, fill=Soft, line width=0.6pt},
  labelbox/.style={rounded corners=2pt, draw=none, fill=white, inner sep=1.6pt},
  tinybullet/.style={circle, fill=black, inner sep=0.8pt},
  psiRing/.style={line width=2.2pt, draw=Psi, dash pattern=on 6pt off 3.3pt},
  psiGlow/.style={line width=4.8pt, draw=Psi!30},
  formula/.style={align=center, inner sep=2pt, fill=white, draw=none},
  caption/.style={align=center, font=\footnotesize},
}

\pgfdeclarehorizontalshading{hemifadeBlue}{100bp}{
  color(0bp)=(Cblue!15); color(100bp)=(Cblue!65)
}
\pgfdeclarehorizontalshading{hemifadeRed}{100bp}{
  color(0bp)=(Cred!15); color(100bp)=(Cred!65)
}

\usepackage{chngcntr}
\counterwithin{table}{subsection}
\usepackage{booktabs,caption}
\usepackage{lscape}
\definecolor{Mygrey}{gray}{0.8}
\definecolor{Mywhite}{gray}{1.0}

\newcommand{\be}{\begin{equation}}
\newcommand{\ee}{\end{equation}}
\newcommand{\bea}{\begin{eqnarray}}
\newcommand{\eea}{\end{eqnarray}}
\newcommand*{\matminus}{%
  \leavevmode
  \hphantom{0}%
  \llap{%
    \settowidth{\dimen0 }{$0$}%
    \resizebox{1.1\dimen0 }{\height}{$-$}%
  }%
}
\usepackage{amsmath,amssymb,epsfig,jheppubme}
\usepackage{multirow,longtable, enumerate, bm}
\usepackage[flushleft]{threeparttable}

\newcommand\bi{\begin{itemize}}
\newcommand\ei{\end{itemize}}

\newcommand\cC{{\cal C}}

\newcommand{\cH}{{\cal H}}

\newcommand\ZZ{\hbox{Z\kern-.4emZ}}
\newcommand\sZZ{\hbox{\sevenfont Z\kern-.4emZ}}

\newcommand{\cM}{{\cal M}}

\newcommand{\cE}{{\cal E}}

\newcommand{\Comment}[1]{{}}

\definecolor{Mygrey}{gray}{0.8}
\definecolor{Mywhite}{gray}{1.0}

\colorlet{dred}{red!70!black!100!}

\def\IB{\relax{\rm I\kern-.18em B}}

\def\ID{\relax{\rm I\kern-.18em D}}
\def\IE{\relax{\rm I\kern-.18em E}}
\def\IF{\relax{\rm I\kern-.18em F}}
\def\II{\relax{\rm I\kern-.18em I}}

\def\Id{\relax{1\kern-.32em 1}}
\def\IG{\relax\hbox{$\inbar\kern-.3em{\rm G}$}}
\def\IR{\relax{\rm I\kern-.18em R}}

\newcommand{\AD}[1]{\textcolor{blue}{\textsf{[AD: #1]}}}
\newcommand{\NBU}[1]{\textcolor{teal}{\textsf{[NBU: #1]}}}

\title{Verlinde lines, anyon permutations and commutant pairs inside $E_{8,1}$ CFT} 
\author[a]{Naveen Balaji Umasankar}
\author[bc]{and Arpit Das}

\affiliation[a]{Department of Physics, Yale University, 217 Prospect St, New Haven, CT 06511}
\affiliation[b]{School of Mathematics, University of Edinburgh, EH9 3FD, U.K.}
\affiliation[c]{Higgs Centre for Theoretical Physics, University of Edinburgh, Edinburgh, EH8 9YL, U.K.}

\emailAdd{naveen.umasankar@yale.edu}
\emailAdd{arpit.das@ed.ac.uk}

\abstract{
We develop a defect-theoretic refinement of meromorphic 2d CFTs in which the ordinary torus partition function---often just the vacuum character---does not reveal how states organize under symmetry lines. Our central proposal is an \emph{equatorial projection} framework: from a commutant decomposition into commuting rational chiral algebras with categories $\mathcal{C}$ and $\widetilde{\mathcal{C}}$, we encode genus-one couplings by a non-negative integer matrix $M$ pairing characters and satisfying modular intertwiner relations. Invertible topological defect lines act directly on this gluing data (Verlinde lines diagonally via $S$-matrix eigenvalues, and anyon-permuting lines by braided-autoequivalence permutations), making modular covariance of defect amplitudes automatic and sharply distinguishing insertions that yield genuine modular invariants from those defining consistent non-holomorphic interfaces. We further show that the \emph{replacement rules} of \cite{Hegde:2021sdm, Lin:2019hks} arise as equatorial projections of defect actions, and we extend these constructions beyond two-character examples by systematically treating three-character commutant pairs in the $E_{8,1}$ theory. The unique $c=8$ meromorphic CFT $E_{8,1}$ serves as a universal testbed, producing new defect partition functions and clarifying the roles of $\mathrm{Pic}(\mathcal{C})$ and $\mathrm{Aut}^{\mathrm{br}}(\mathcal{C})$. Finally, we outline extensions to higher central charges (e.g.\ $c=32,40$), yielding modular-invariant non-meromorphic theories beyond the $c=24$ Schellekens landscape \cite{Schellekens:1992db} as defect/interface descendants of meromorphic parents.
}

\keywords{Topological Defect Lines, Defect Partition Functions, Conformal Field Theory}

\begin{document}

\maketitle

\section{Introduction}
Conformal field theories (CFTs) in two dimensions have long served as a crucial testing ground for ideas in string theory, algebraic geometry, representation theory, and classification of rational CFTs \cite{Polyakov:1981rd, Moore:1988qv, Mathur:1988na, Bantay:2005vk}. Of particular interest are self-dual, under charge conjugation, meromorphic theories, often called \textit{holomorphic} CFTs, which are characterized by their chiral algebra having only a single irreducible module, so the genus-one amplitude is determined entirely by the vacuum character (equivalently, the anti-holomorphic chiral algebra is trivial). This extreme chirality makes their modular properties exceptionally rigid, and places strong constraints on the admissible symmetry algebras and spectra. A canonical benchmark is the classification of meromorphic theories up to central charge $c = 24$ by Schellekens \cite{Schellekens:1992db}, who identified $71$ distinct theories, each corresponding to a unique combination and/or extension of Kac-Moody algebras consistent with modular invariance and other physical requirements.\\
\newline
\noindent
Despite this comprehensive result, many structural aspects of meromorphic CFTs remain poorly understood. In physics literature, Schellekens' $c = 24$ benchmark is often treated as a rare case when the meromorphic landscape can be sharply constrained, yet it has also served to highlight how limited our general structural control remains once one moves beyond a small set of special constructions. In \cite{Chandra:2018ezv}, the authors emphasize that modular considerations can produce infinitely many admissible character candidates above $c = 24$, while the existence and systematic organization of corresponding meromorphic CFTs is far from automatic, so the space of modularly allowed data is much larger than the space of known consistent theories. A closely related theme appears in AdS$_{3}$/CFT$_{2}$ discussion of holomorphically factorized candidates, where extremal meromorphic theories at $c = 24k$ were framed as natural targets \cite{Witten:2007kt}, but identifying the actual dual CFTs is highly non-trivial. Even the most direct gravitational attempt to compute the partition function runs into basic consistency problems, which again signals that modular covariance and semiclassical reasoning do not straightforwardly determine a consistent meromorphic CFT \cite{Maloney:2007ud}. On the CFT side, constraints on putative extremal self-dual theories from modular differential equations were derived, giving evidence that entire infinite families may fail to exist beyond some range, and hence, existence itself becomes a structural question rather than a corollary of modular invariance \cite{Gaberdiel:2007ve}. In particular, although the ordinary torus partition function is a powerful invariant, it can be too coarse. In a holomorphic theory, it reduces to a vacuum character and therefore does not resolve how states organize under nonlocal symmetries. A natural refinement is obtained by inserting topological defect lines (TDLs). Threading an invertible TDL along the temporal cycle produces a twinned genus-one amplitude, which refines the vacuum character by weighting states by their eigenvalues. Modular transformations exchange temporal and spatial cycles, so the family of defect amplitudes close under the modular group by conjugation of the inserted line; the central issue is then to determine when such modular-covariant data can be completed into a genuine modular invariant with non-negative multiplicities (as in orbifold-like constructions), versus when it should be interpreted as a consistent but non-holomorphic interface amplitude rather than a partition function of a new full CFT. These defect insertions don't just decorate the partition functions, but also study the internal organization of the theory, revealing how chiral sectors reorganize under symmetry lines encoded by Verlinde lines and anyon-permuting defects encoded by braided autoequivalences. Hence, by analyzing how the theory responds to these insertions, one can gain insights into its hidden structures and symmetries. In a holomorphic theory, this is particularly powerful because, although the defect-free partition function collapses to a single vacuum character, defect insertions can reveal non-trivial decompositions and subsector dynamics that would otherwise remain invisible.\\
\newline
\noindent
The central idea of this paper is that the defect partition functions of meromorphic theories can be computed and organized efficiently by adopting an \textit{equatorial projection} perspective. We start from a commutant decomposition inside the meromorphic theory, identifying a pair of commuting rational chiral algebras with representation categories $\mathcal{C}$ and $\widetilde{\mathcal{C}}$. The associated genus-one amplitude is encoded by a non-negative integer matrix $M$ via the sequilinear pairing
\begin{align}
    Z_{M}(\tau, \overline{\tau}) = \sum\limits_{i\in I_{\mathcal{C}}}\sum\limits_{j\in I_{\widetilde{\mathcal{C}}}}M_{ij}\chi_{i}(\tau)\overline{\widetilde{\chi}_{j}(\tau)},
\end{align}
where consistency requires that $M$ intertwines the modular actions, schematically $S_{\mathcal{C}}M = MS_{\widetilde{\mathcal{C}}}$, and  $T_{\mathcal{C}}M = MT_{\widetilde{\mathcal{C}}}$, with the precise conjugation fixed by convention. The equatorial projection principle asserts that threading an invertible TDL along the temporal cycle acts on the coupling data by explicit operations: Verlinde lines act diagonally through the modular $S$-matrix eigenvalues, while anyon-permuting defects act by permutation matrices associated with braided autoequivalences. In this language, the modular covariance of defect amplitudes becomes automatic as modular transformations simply conjugate the inserted line operator, keeping the amplitude within a closed family. The central question is therefore when an insertion produces a genuine modular invariant, i.e., a matrix $M'$ with non-negative integer entries that satisfies the modular intertwiner equations, and when it instead yields a non-holomorphic interface amplitude, which is a consistent genus-one observable but not the partition function of a new full CFT.\\

\noindent This framework is specifically well-suited to studying and extending the “replacement rule” philosophy of \cite{Hegde:2021sdm}. Roughly, replacement rules allow one to generate new modular data and new couplings by systematically modifying one side of a commutant pairing while preserving modular constraints. We show that such replacement operations can be realized as equatorial projections of defect insertions acting on commutant data, thereby unifying defect actions and replacement rules within a single algebraic mechanism. We implemented this beyond the two-character setting studied in \cite{Hegde:2021sdm, Burbano:2021loy} by treating three-character commutant pairs, including tensor-product and IVOA-type cases that were omitted in earlier treatments. This resolves gaps in the defect landscape of the unique $c = 8$ meromorphic theory, which produces a systematic supply of new defect partition functions at higher central charge. Concretely, the present work focuses on three interconnected themes:
\begin{enumerate}
    \item \textbf{Defect partition functions as refined probes}\\
    We compute torus amplitudes in the presence of invertible TDLs and use them to isolate subsectors that do not appear in the defect-free torus partition function. These subsectors can be modular covariant in general and, in favorable cases, can reorganize into modular invariant, non-holomorphic couplings that merit interpretation as emergent RCFT data or as consistent interface observables.
    \item \textbf{Commutant pairs and equatorial projections}\\
    We construct and analyze commutant pairs of rational chiral algebras inside meromorphic theories and encode their equatorial gluing by modular intertwiners $M$. This provides a concrete bookkeeping device for admissible genus-one couplings and a natural arena in which defect insertions act by explicit transformations of $M$. 
    \item \textbf{Replacement rules via equatorial projections}\\
    We reinterpret replacement rules as controlled deformations of $M$ induced by invertible defects, providing a clearer understanding of both the algebraic mechanism and the modular structure. This makes transparent which operations are “safe” that produce genuine modular invariants and which generically produce defect interfaces.
    %\item \textbf{Beyond $c = 24$: defect structures at higher central charges}\\
    %While much of the foundational work has focused on $c=24$, we aim to extend these analyses to meromorphic CFTs with higher central charges, such as $c=32,40$, and beyond. This involves studying the associated Kac-Moody algebras and their defect structures. We then identify novel non-meromorphic but modular-invariant CFTs that do not appear in Schellekens' classification, thereby extending the landscape of consistent theories. Finally, we introduce a framework that interprets these theories arising from defect lines or interfaces, in analogy with the ideas from boundary and defect CFT.
\end{enumerate}
\noindent
These ideas build on and generalize earlier studies in several directions. The classification of meromorphic theories t $c = 24$ by Schellekens \cite{Schellekens:1992db}, the study of holomorphic factorization in RCFTs \cite{Stigner:2010nz}, and recent work on modular differential linear equations and character-based classification \cite{Bantay:2001ni, Das:2022bxm, Umasankar:2022kzs}, all provide backdrop to this work. In particular, the perspective on modular invariant but non-holomorphic partition functions resonates with recent findings in \cite{Barkeshli:2014cna}, while the emphasis on defects and topological lines connects to the broader literature on higher-form symmetries and algebraic QFT \cite{Gaiotto:2014kfa, Benedetti:2024dku, Shao:2023gho, Iqbal:2024pee}.\\

\noindent 
A central contribution of this work is to develop a conceptually transparent and computationally effective framework for computing defect partition functions of meromorphic CFTs using equatorial projections of TDLs. This perspective clarifies how Verlinde lines and automorphism defects act on commutant pairs, and distinguishes insertions that produce bona fide modular invariant partition functions from those that instead define non-holomorphic defect interfaces. The case $c = 8$ plays a key role in this narrative. First, $E_{8,1}$ is the unique meromorphic CFT at central charge eight, making it an ideal setting in which to isolate the core mechanism without the additional ambiguities present at higher central charges. Second, every three-character commutant pair in the classification of \cite{Das:2022slz} can be matched to a bilinear pairing whose equatorial projection yields the $E_{8,1}$ vacuum character, so $c = 8$ serves as a universal testbed for extending replacement rules beyond two-character theories. Third, several commutant pairs relevant to $c = 8$ involve tensor-product RCFTs or IVOA-type constructions that were left unexplored in prior work; resolving these cases yields new defect partition functions and clarifies the role of the braided autoequivalence group $\text{Aut}^{\text{br}}(\mathcal{C})$ and the Picard group $\text{Pic}(\mathcal{C})$ in governing invertible TDLs in meromorphic theories. For these reasons, $c = 8$ not only provides the cleanest setting in which to articulate the equatorial projection principle but also serves as a foundational example from which higher central charge cases naturally generalize.\\

\noindent Recently, a complementary prescription for constructing two-dimensional CFTs from chiral data was proposed in \cite{Rastelli:2025nyn}, where the authors provide a physically motivated prescription of canonical gluing arising from four-dimensional $\mathcal{N} = 2$ supersymmetric field theories placed on a curved background containing a cigar or a hemisphere geometry (see also \cite{Chandra:2025qpv, Deb:2025ddc}). Reduction on a single cigar produces a chiral algebra sector described by a VOA localized at the tip of the cigar. A full non-chiral two-dimensional CFT is then obtained by gluing two such cigars into a sphere, where the gluing amounts to summing the modulus squares of conformal blocks, and hence, is fixed by an underlying four-dimensional path integral that introduces a sesquilinear pairing on the space of conformal blocks. By contrast, the equatorial projection framework developed in this paper is intrinsically two-dimensional and categorical. We start from two (not necessarily identical) vertex operator algebras (VOAs) $V$ and $\widetilde{V}$ with representation categories $\mathcal{C} = \text{Rep}(V)$ and $\widetilde{\mathcal{C}} = \text{Rep}(\widetilde{V})$, and regard them as living on two hemispheres of a sphere. Gluing along the equator is encoded by a multiplicity matrix $M\in \text{Mat}_{I_{\mathcal{C}}\times I_{\widetilde{\mathcal{C}}}}(\mathbb{Z}_{\geq0})$, where $I_{\mathcal{C}}$ and $I_{\widetilde{\mathcal{C}}}$ label the simple objects (primaries) of $\mathcal{C}$ and $\widetilde{\mathcal{C}}$. This matrix specifies how left- and right-moving primaries are paired in the full Hilbert space, and yields a genus-one amplitude $Z_{M}(\tau, \overline{\tau})$. The modular consistency of the gluing is imposed through intertwiner relations between the modular data of $\mathcal{C}$ and $\widetilde{\mathcal{C}}$; in particular, our framework parameterizes the space of admissible gluings rather than selecting a unique one. From this viewpoint, the hemisphere gluing of \cite{Rastelli:2025nyn} can be regarded as specifying a preferred sesquilinear pairing within this larger space of equatorial gluings. Our formalism is deliberately broader since it allows for diagonal, simple currents, and exceptional modular invariants, as well as pairings between different chiral theories $(\mathcal{C}, \widetilde{\mathcal{C}})$.\\

\noindent Another important distinction concerns the role of defects. In our formalism, both invertible and non-invertible TDLs act naturally on the gluing data. Concretely, a defect insertion represented on the left and right chiral halves by matrices $R_{L}$ and $R_{R}$ acts by
\begin{align}
    M\mapsto M^{(R_{L}, R_{R})} = R_{L}^{T}M\overline{R_{R}},    
\end{align}
with the transpose/conjugation dictated by the left-right conventions. This provides a systematic way to generate modular-covariant defect amplitudes, to organize genus-one observables into orbits under defect actions, and to identify when such deformations correspond to genuine modular invariant versus non-holomorphic defects. Such defect-theoretic deformations of the gluing do not play a central role in \cite{Rastelli:2025nyn}, where the emphasis is instead on the canonical gluing dictated by the four-dimensional origin of the theory.

\subsection*{Outline}
This paper is organized as follows. In \S2, we collect the necessary background---to keep the presentation self-contained---on vertex operator algebras (VOAs), unitary modular tensor categories (MTCs), modular data and topological defect lines (TDLs), torus amplitudes and orbifold sectors, and the basics of vector-valued modular forms. In \S3, we introduce the equatorial projection framework and illustrate it through a number of representative examples. \S4 is devoted to orbifold constructions: we describe the relevant procedures and compute the corresponding orbifold partition functions in a variety of explicit cases. In \S5, we apply our methods to the meromorphic theory $E_{8,1}$ by computing defect partition functions generated by the defect data of the commutant/coset pairs that realize $E_{8,1}$. We conclude in \S6 with a discussion and outlook, with additional technical details collected in the appendices.

\begin{comment}
\section{Literature overview}
We have to read thoroughly \cite{Burbano:2021loy, Rayhaun:2023pgc} to check overlap of results with us (after we do our calculations, obviosuly!).
\begin{itemize}
    \item TDLs in RCFT \cite{Chang:2018iay}
    \item Defect partition functions
    \item Modular invariants \cite{Gannon:1999cp}
    \item Commutant pairs \cite{Das:2022slz}
\end{itemize}
\end{comment}

\section{Background}
%\AD{Write this starting from TDLs, Verlinde formula, $S$ and $T$ matrices, modular transformations, vvmfs, modular data, etc...\cite{Chang:2018iay, Haghighat:2023sax, Gu:2023yhm} have nice intro to TDLs..Add section on S-matrix computation and motivate this since we are using the MMS construction (to tie in with the explicit computations presented in the appendix)}
%\NBU{Update: done! Feel free to add/edit with references.}

\noindent 
Two-dimensional RCFTs admit a particularly rigid set of non-local observables called Topological Defect Lines (TDLs) \cite{Verlinde:1988sn, Drukker:2010jp, Gaiotto:2014lma}. These defects are termed “topological” because they are tensionless and admit smooth deformations on the world sheet that leave correlation functions unchanged, as long as no local fields or other defects are crossed \cite{Frohlich:2009gb}. In this sense, TDLs are freely deformable one-dimensional operator characterized by isotopy invariance (see \cite{Chang:2018iay} for more details). When placed on a constant-time slice, a TDL defines a linear operator $\mathcal{L}$ on a Hilbert space $\mathcal{H}$ of the CFT as $\mathcal{L}:\mathcal{H}\to \mathcal{H}$. Because it is topological, it commutes with the action of left and right chiral algebras, particularly the Virasoro modes as follows
\begin{align}
    \left[\mathcal{L}, L_{n}\right] = \left[\mathcal{L}, \overline{L}_{n}\right] = 0,\qquad \forall n\in\mathbb{Z}.
\end{align}
Fusion of defect lines is realized as operator composition. If two TDLs $\mathcal{L}_{1}$ and $\mathcal{L}_{2}$ are placed on the same constant-time slice, their fusion corresponds to multiplying the corresponding operators,
\begin{align}
    \mathcal{L}_{1}\times\mathcal{L}_{2}\longleftrightarrow \mathcal{L}_{1}\mathcal{L}_{2}: \mathcal{H}\to \mathcal{H}.
\end{align}
This operator algebra is reflected in the fusion category structure of TDLs in the RCFT, i.e., 
\begin{align}
    \mathcal{L}_{i}\mathcal{L}_{j} = \sum\limits_{k}N_{ij}^{k}\mathcal{L}_{k},
\end{align}
where $\mathcal{N}_{ij}^{k}\in\mathbb{Z}_{\geq0}$ are the fusion coefficients. Viewing TDLs as symmetry-like operators that generalize ordinary global symmetries and encode both invertible and non-invertible “generalized symmetries” has been emphasized in modern literature, including applications to defect RG flows, fermionic CFTs, and parafermionic/bosonized models \cite{Chen:2023jht, Chang:2018iay}.\\

\subsection*{VOAs and UMTCs}
A VOA \cite{Frenkel2000VertexAA, Frenkel:1988xz, Kac1997VertexAF} is the algebraic packaging of the chiral operator product expansion in a two-dimensional CFT. Concretely, it consists of a vector space $V$ of chiral states, a vacuum $\vert\Omega\rangle\in V$, a translation operator $T\in \text{End}(V)$, and a state field correspondence
\begin{align}
    a\longmapsto Y(a,z)\in \text{End}(V),\qquad Y(a,z) = \sum\limits_{n\in\mathbb{Z}}a_{(n)}z^{-n-1},
\end{align}
which assigns to each state a holomorphic field expanded in Laurent modes. The truncation condition $a_{(n)}b = 0$ for all sufficiently large $n$ is the statement that OPEs have only finitely many singular terms. Here, $b\in V$ is arbitrary, and $a_{(n)}\in \text{End}(V)$ denotes the $n^{\text{th}}$ mode   of the field $Y(a,z)$, so that
\begin{align}
    Y(a,z)b = \sum\limits_{n\in\mathbb{Z}}\left(a_{(n)}b\right)z^{-n-1}.
\end{align}
The vacuum axioms say that the vacuum field is the identity operator, and that inserting the vacuum recovers the original state, so 
\begin{align}
    Y(\vert\Omega\rangle, z) = \mathbb{1}_{V},    
\end{align}
 and $Y(a,z)\vert\Omega\rangle$ is regular at $z = 0$ with leading term $a$. Translation covariance is encoded by 
 \begin{align}
     \left[T, Y(a,z)\right] = \partial_{z}Y(a,z),
 \end{align}
which is the chiral version of momentum-generating infinitesimal translations. The final axiom is locality in the CFT sense, namely that for any $a,b\in V$, the products $Y(a,z)Y(b,w)$ and $Y(b,w)Y(a,z) $ are boundary expansions of the same underlying rational function with possible singularities only at $z = 0$, $w = 0$, and $z = w$. This is the precise algebraic version of mutual locality and OPE associativity. In this paper, we restrict to the rational (and in examples, unitary) setting, so that $V$ has finitely many simple modules and semisimple module categories. In particular, we may regard $\mathcal{C} = \text{Rep}(V)$ as a unitary modular tensor category capturing the chiral topological lines (Verlinde lines) of the theory.\\

\noindent 
The representation theory for RCFTs is expected to form a semisimple tensor category, whose simple objects, or primaries, label superselection sectors, or equivalently, irreducible topological lines, and whose fusion encodes the operator product of these lines. A unitary modular tensor category is the abstract structure capturing precisely this data. It is a finite semisimple rigid braided tensor category with a compatible twist, together with a unitarity structure so that the associativity and braiding isomorphisms can be chosen unitary, and with the additional non-degeneracy property that the braiding is “maximally non-trivial,” equivalently that the modular $S$-matrix is invertible. In this paper, the unitary modular tensor category is not introduced abstractly but arises as the representation category of a rational chiral algebra. Concretely, given a vertex operator algebra $V$ satisfying the standard finiteness and semisimplicity conditions used in RCFTs, we consider the category $\text{Rep}(V)$ whose objects are $V$-modules and whose morphisms are $V$-intertwiners. The fusion product in $\text{Rep}(V)$ is defined by the VOA tensor product theory, meaning it is characterized by the appropriate universal property for intertwining operators, and physically it encodes the chiral OPE of primary fields. This tensor product theory was developed to construct a vertex tensor category structure on $V$-modules, which then produces a braided tensor category structure, so braiding is obtained from the analytic continuation and exchange of chiral conformal blocks, and rigidity expresses the existence of conjugate sectors \cite{Huang1993ATO, Huang2014BraidedTC}.

\subsection*{Modular data and the Verlinde Formula}
Let $\mathcal{C} = \text{Rep}(V)$ be the modular tensor category (MTC) of a rational VOA $V$. Denoting the primaries or the simples by $I_{\mathcal{C}}$, with corresponding characters $\chi_{i}(\tau)$. Since $V$ is rational, the space spanned by the characters $\mathcal{H}_{\mathcal{C}}\equiv \text{Span}_{\mathbb{C}}\{\chi_{i}(\tau)\vert i\in I_{\mathcal{C}}\}$, is a finite-dimensional complex vector space with $\text{dim}\ \mathcal{H}_{\mathcal{C}}\leq \vert I_{\mathcal{C}}\vert$, and is $\text{SL}(2, \mathbb{Z})$-stable \cite{Zhu:1996gaq}. Assuming the irreducible characters are linearly independent, we have $\text{dim}\ \mathcal{H}_{\mathcal{C}} = \vert I_{\mathcal{C}}\vert$. Choosing the basis given by simple modules, the modular group acts by linear transformations on this space. The matrices $S_{\mathcal{C}}$ and $T_{\mathcal{C}}$ may be taken unitary with respect to the natural Hermitian form coming from the MTC. Hence, the image of the modular representation lies in the unitary group
\begin{align}
    U(\vert I_{\mathcal{C}}\vert) = \{U\in \text{GL}(\vert I_{\mathcal{C}}\vert, \mathbb{C})\vert U^{\dagger}U = \mathbb{1}\},
\end{align}
the group of unitary transformations of a $\vert I_{\mathcal{C}}\vert$-dimensional complex vector space. Modular covariance of torus one-point functions implies that the character vector $\chi(\tau) = (\chi_{i}(\tau))_{i\in I_{\mathcal{C}}}$ furnishes a projective representation
\begin{align}
    \rho_{\mathcal{C}}: \text{SL}(2, \mathbb{Z})\to U(\vert I_{\mathcal{C}}\vert),\qquad \rho_{\mathcal{C}}(S) = S_{\mathcal{C}},\ \ \rho_{\mathcal{C}}(T) = T_{\mathcal{C}},
\end{align}
with $T_{\mathcal{C}}$ diagonal and entries $T_{ii} = e^{2\pi i\left(h_{i} - \tfrac{c}{24}\right)}$. The pair $(S_{\mathcal{C}}, T_{\mathcal{C}})$ encodes essentially all the genus-one data of the chiral theory. A distinguished family of purely chiral TDLs are the Verlinde lines associated with objects $a\in\mathcal{C}$. On the character vector, they act diagonally by
\begin{align}
    D_{a} = \text{Diag}\left(\frac{\left(S_{\mathcal{C}}\right)_{ai}}{\left(S_{\mathcal{C}}\right)_{0i}}\right)_{i\in I_{\mathcal{C}}}.
\end{align}
These lines are invertible precisely when $a$ is a simple current, i.e., $a\in \text{Pic}(\mathcal{C})$. The Verlinde formula then recovers the fusion coefficients from the $S$-matrix as follows \cite{Verlinde:1988sn}
\begin{align}
    N_{ab}^{c}  = \sum\limits_{i\in I_{\mathcal{C}}}\frac{S_{ai}S_{bi}S^{*}_{ci}}{S_{0i}}.
\end{align}
This is the bridge between genus-one modular information and operator-algebraic fusion information, and this is the basic reason why defect insertions are so tightly controlled once we know the data $(S_{\mathcal{C}}, T_{\mathcal{C}})$.

\subsection*{Invertible TDLs}
Among all defects, the invertible ones form a group under fusion, commonly organized as
\begin{align}
    \text{Pic}(\mathcal{C})\rtimes\text{Aut}^{\text{br}}(\mathcal{C}),
\end{align}
where $\text{Pic}(\mathcal{C})$ is generated by simple currents that are invertible objects, and $\text{Aut}^{\text{br}}$ is the group of braided autoequivalences acting as permutations of primaries. These are also called anyon-permuting invertible TDLs, and on characters, these act by permutation matrices $P_{g}$ with $g\in \text{Aut}^{\text{br}}$ commuting with the modular representation 
\begin{align}
    P_{g}\rho(\gamma) = \rho(\gamma)P_{g},\qquad \gamma\in\text{SL}(2, \mathbb{Z}),
\end{align}
when $g$ is realized as an actual symmetry of the modular data. The combined action of an invertible defect labeled by $(a,g)$ is packaged as the endomorphism
\begin{align}
    \rho(a,g) = D_{a}P_{g}\in\text{End}(\mathbb{C}^{I_{\mathcal{C}}}).
\end{align}
This realization of the diagonal Verlinde operator times the permutation operator is exactly what makes defect computations tractable at genus-one since the action of invertible defects on characters is explicit as soon as we know the data $(S_{\mathcal{C}}, T_{\mathcal{C}})$.

\subsection*{Defect partition functions, twisted traces, and orbifold sectors}
Placing a TDL along the temporal cycle of the torus produces a defect partition function, or twisted torus amplitude, defined as follows
\begin{align}
    \mathcal{Z}^{\mathcal{L}}(\tau, \overline{\tau}) = \text{Tr}_{\mathcal{H}}\left(\mathcal{L}q^{L_{0} - \frac{c}{24}}\overline{q}^{\overline{L}_{0} - \frac{\overline{c}}{24}}\right).
\end{align}
More generally, for a finite symmetry group $G$ of invertible lines, we can define the standard family $Z_{h,g}$, with an $h$-twist along the spatial cycle and a $g$-twist along the temporal cycle. These partial traces transform among themselves under $\text{SL}(2, \mathbb{Z})$ by the $S$ and $T$ reshufflings, and modular invariance is restored only after summing over all sectors in the orbifold partition function
\begin{align}
    Z_{\text{orb}} = \frac{1}{\vert G\vert}\sum\limits_{g,h\in G}Z_{h,g}.
\end{align}
On the torus, this illustrates the familiar principle that gauging a symmetry entails including twisted sectors. We will repeatedly contrast genuine modular invariant CFT partition functions with modular covariant defect torus amplitudes in later sections.

\section*{Vector-valued modular forms}
A powerful way to organize character data is to view it as a vector-valued modular form or vvmf for the representation $\rho_{\mathcal{C}}$ \cite{Mason:2007, Bantay:2007zz, Bantay:2005vk}. In many classification programs, and most notably in the Mathur-Mukhi-Sen (MMS) program \cite{Mathur:1988na, Mathur:1988rx, Mathur:1988gt} and its modern extensions \cite{Hampapura:2015cea, Das:2022uoe, Bantay:2005vk, Mason:2007, Gannon:2013jua, Das:2020wsi, Umasankar:2022kzs, Das:2022bxm, Naculich:1988xv, Pan:2023jjw, Das:2023qns, Govindarajan:2025jlq}, the characters are characterized as solutions to a modular linear differential equation (MLDE) whose monodromy furnishes $\rho_{\mathcal{C}}$. This approach explains why we can reconstruct the modular representation, and hence, fusion and defect data from a small amount of spectral input for low theories with low character count. The original MMS viewpoint and subsequent developments form part of the standard toolkit for character-based RCFT classification and will be used implicitly throughout our explicit examples.

\subsection*{Computing $S$ and $T$}
In applications of this paper, we will frequently encounter commutant pair constituents that are primarily presented through central charge $c$, conformal weights $\{h_{i}\}$, and $q$-series expansions. In order to feed these theories into the equatorial framework, we require modular data. The diagonal $T$-matrix is immediate from $(c,h_{i})$, while the non-trivial input is the $S$-matrix. There are two complementary motivations for the MMS-style extraction of $S$:
\begin{enumerate}
    \item Genus-one consistency:\\
    \noindent The pair $(S, T)$ must satisfy the $\text{SL}(2, \mathbb{Z})$ relations $S^{2} = (ST)^{3} = C$, up to phases, and unitarity. This must then be compatible with the positivity or integrality of fusion via the Verlinde formula. This makes $S$ highly constrained once $T$ and a small amount of spectral data are fixed.
    \item Monodromy MLDE construction:\\
    \noindent When characters satisfy an MLDE, the modular group action is realized as monodromy of the differential equation, which directly yields $\rho_{\mathcal{C}}(S)$ and $\rho_{\mathcal{C}}(T)$. This route outlined in \cite{Mathur:1988gt} is what we adopt in appendix \ref{sec:appendix_S}. This provides precisely the modular input needed to construct Verlinde line eigenvalues, determine defect actions on character vectors, and implement the equatorial projection computations.
\end{enumerate}

\section{Equatorial projection framework}

\begin{figure}[htb!]
\centering
\begin{tikzpicture}[
    wiggly/.style={decorate, decoration={snake, amplitude=0.7mm, segment length=3mm}},
    >=Stealth
]

% === LEFT SIDE: Two separated hemispheres with stitching ===

% Northern hemisphere (top) - tilted
\begin{scope}[shift={(0,2.2)}, rotate=-15]
    % Soft shadow under rim
    \fill[black!15, opacity=0.15] (0,-0.15) ellipse (2.1 and 0.65);
    
    % Hemisphere shading - blue
    \shade[ball color=blue!35!white, opacity=0.75]
      (-2,0) arc (180:360:2 and 0.6) arc (0:180:2 and 2);
    
    % Specular highlight (glossy effect)
    \fill[white, opacity=0.3] (-0.6,1.3) ellipse (0.5 and 0.3);

    % Boundary ellipse (equator edge) - thicker outline
    \draw[line width=1.2pt, blue!65!black] (-2,0) arc (180:360:2 and 0.6);
    \draw[line width=1.2pt, blue!65!black, dashed] (-2,0) arc (180:0:2 and 0.6);

    % Dome outline - thicker
    \draw[line width=1.2pt, blue!65!black] (-2,0) arc (180:0:2 and 2);

    % North pole marker
    \fill[red!75!black] (0,2) circle (2pt);
    \node[above right, red!75!black, font=\small] at (0,2) {$N$};

    % Labels (muted colors)
    \node[blue!70!black, font=\large] at (0,1.5) {$V$};
    \node[blue!55!black, font=\small] at (0,1.1) {$\mathrm{Rep}(\mathcal{C})$};
    \node[blue!55!black, font=\footnotesize] at (1.9,1.5) {$(S,T)$};

    % Index label (muted)
    \node[blue!60!black, font=\small] at (-2.8,1.2) {$i \in I_{\mathcal{C}}$};
    \node[blue!60!black, font=\small] at (-2.8,0.6) {$\chi_i(\tau)$};

    % --- Seam points ON THE RIM with punctures
    \foreach \k/\ang in {1/205,2/235,3/265,4/295,5/325}{
      \coordinate (u\k) at ({2*cos(\ang)}, {0.6*sin(\ang)});
      \fill[purple!70!black] (u\k) circle (1.5pt);
    }
    
    % Label representative indices
    \node[purple!70!black, font=\tiny] at ({2*cos(235)+0.3}, {0.6*sin(235)+0.25}) {$i_2$};
    \node[purple!70!black, font=\tiny] at ({2*cos(295)-0.3}, {0.6*sin(295)+0.25}) {$i_4$};
\end{scope}

% Southern hemisphere (bottom) - tilted
\begin{scope}[shift={(0,-2.2)}, rotate=-15]
    % Soft shadow under rim
    \fill[black!15, opacity=0.15] (0,0.15) ellipse (2.1 and 0.65);
    
    % Hemisphere shading - orange/red
    \shade[ball color=orange!45!red!35!white, opacity=0.75]
      (-2,0) arc (180:0:2 and 0.6) arc (0:-180:2 and 2);
    
    % Specular highlight
    \fill[white, opacity=0.3] (-0.6,-1.3) ellipse (0.5 and 0.3);

    % Boundary ellipse - thicker (SWAPPED: front solid, back dashed)
    \draw[line width=1.2pt, orange!65!red] (-2,0) arc (180:360:2 and 0.6);
    \draw[line width=1.2pt, orange!65!red, dashed] (-2,0) arc (180:0:2 and 0.6);

    % Dome outline - thicker
    \draw[line width=1.2pt, orange!65!red] (-2,0) arc (180:360:2 and 2);

    % South pole marker
    \fill[red!75!black] (0,-2) circle (2pt);
    \node[below right, red!75!black, font=\small] at (0,-2) {$S$};

    % Labels (muted)
    \node[orange!70!red, font=\large] at (0,-1.5) {$\widetilde{V}$};
    \node[orange!55!red, font=\small] at (0,-1.1) {$\mathrm{Rep}(\widetilde{\mathcal{C}})$};
    \node[orange!55!red, font=\footnotesize] at (1.9,-1.5) {$(\widetilde{S},\widetilde{T})$};

    % Index label (muted)
    \node[orange!60!red, font=\small] at (-2.8,-1.2) {$j \in I_{\widetilde{\mathcal{C}}}$};
    \node[orange!60!red, font=\small] at (-2.8,-0.6) {$\widetilde{\chi}_j(\bar{\tau})$};

    % --- Seam points with punctures
    \foreach \k/\ang in {1/205,2/235,3/265,4/295,5/325}{
      \coordinate (d\k) at ({2*cos(\ang)}, {0.6*sin(\ang)});
      \fill[purple!70!black] (d\k) circle (1.5pt);
    }
    
    % Label representative indices
    \node[purple!70!black, font=\tiny] at ({2*cos(235)+0.3}, {0.6*sin(235)-0.25}) {$j_2$};
    \node[purple!70!black, font=\tiny] at ({2*cos(295)-0.3}, {0.6*sin(295)-0.25}) {$j_4$};
\end{scope}

% === Wiggly gluing threads with Bézier curves and varying opacity ===
\foreach \k/\opa in {1/0.5,2/0.7,3/0.9,4/0.7,5/0.5}{
  \draw[line width=0.9pt, purple!70!black, wiggly, opacity=\opa] 
    (u\k) .. controls ($(u\k)!0.5!(d\k) + (-0.3,0)$) .. (d\k);
}

% M_ij label on middle thread
\node[purple!75!black, font=\footnotesize, fill=white, fill opacity=0.7, text opacity=1, inner sep=1pt] 
  at (-0.5,0) {$M_{ij}$};

% Gluing matrix label - prominent
\node[purple!80!black, font=\Large] at (1.7,0) {$M$};

% === CLEANER ARROW with better alignment ===
\draw[line width=2.5pt, -{Stealth[length=5mm, width=4mm]}, purple!65!black] (4.5,0) -- (6.5,0);
\node[above, purple!65!black, font=\small] at (5.5,0.35) {glue along $S^1$};

% === RIGHT SIDE: Resulting sphere (ULTRA-ENHANCED) ===
\begin{scope}[shift={(9.5,0)}, rotate=-15]
    
    % Soft vignette behind sphere (radial gradient halo)
    \shade[inner color=purple!8!white, outer color=white, opacity=0.6] (0,0) circle (2.8);
    
    % CLIP EVERYTHING TO SPHERE
    \begin{scope}
        \clip (0,0) circle (2);
        
        % Base sphere shading
        \shade[ball color=blue!25!orange!25!white, opacity=0.7] (0,0) circle (2);
        
        % Northern hemisphere - cooler, glossier
        \shade[top color=blue!40!white, bottom color=blue!18!orange!18!white, opacity=0.7]
            (-2,0) arc (180:360:2 and 0.5) arc (0:180:2 and 2) -- cycle;
        
        % Northern hemisphere - BRIGHTER glossy highlight (more reflective)
        \fill[white, opacity=0.4] (-0.7,0.9) ellipse (0.6 and 0.35);
        
        % Southern hemisphere - warmer, more matte
        \shade[bottom color=orange!50!red!40!white, top color=orange!18!blue!18!white, opacity=0.7]
            (-2,0) arc (180:0:2 and 0.5) arc (0:-180:2 and 2) -- cycle;
        
        % Southern hemisphere - BROADER, dimmer highlight (matte finish)
        \fill[white, opacity=0.18] (-0.5,-1.0) ellipse (0.9 and 0.5);
        
        % Terminator curve (day/night boundary)
        \draw[black, opacity=0.12, line width=5pt] (0.4,0) circle (1.7);
        
        % Soft shadow just below equator
        \fill[black, opacity=0.08] 
            (-2,-0.05) arc (180:360:2 and 0.52) arc (360:180:2.05 and 0.58) -- cycle;
        
        % Ghosted cutaway hint - northern cap boundary
        \draw[blue!40!black, opacity=0.06, line width=0.8pt] 
            (-2,0) arc (180:0:2 and 2);
        
        % Ghosted cutaway hint - southern cap boundary  
        \draw[orange!50!red, opacity=0.06, line width=0.8pt] 
            (-2,0) arc (180:360:2 and 2);
        
        % Meridian curve (vertical ellipse) - partial, mostly dashed
        \draw[gray!50, opacity=0.08, line width=0.6pt] 
            (0,-2) arc (270:360:0.45 and 2);
        \draw[gray!50, opacity=0.08, line width=0.6pt] 
            (0,2) arc (90:180:0.45 and 2);
        \draw[gray!50, opacity=0.05, line width=0.6pt, dashed] 
            (0,-2) arc (270:180:0.45 and 2);
        \draw[gray!50, opacity=0.05, line width=0.6pt, dashed] 
            (0,2) arc (90:0:0.45 and 2);
        
        % Latitude curve above equator (at height +0.7)
        \draw[gray!50, opacity=0.06, line width=0.5pt] 
            (-1.43,0.7) arc (180:360:1.43 and 0.36);
        \draw[gray!50, opacity=0.03, line width=0.5pt, dashed] 
            (-1.43,0.7) arc (180:0:1.43 and 0.36);
        
        % Inner concentric equator ring
        \draw[green!45!black, opacity=0.25, line width=0.8pt] 
            (-1.85,0) arc (180:360:1.85 and 0.46);
        \draw[green!45!black, opacity=0.15, line width=0.8pt, dashed] 
            (-1.85,0) arc (180:0:1.85 and 0.46);
    \end{scope}
    
    % Sphere outline with micro-contrast (double outline effect)
    \draw[line width=1.2pt, gray!55!black] (0,0) circle (2);
    \draw[line width=0.4pt, gray!70!black, opacity=0.5] (0,0) circle (1.97);
    
    % Rim light on far edge (studio lighting)
    \draw[white, opacity=0.20, line width=1.8pt] 
        (0,0) ++(-155:2) arc (-155:-25:2);
    
    % LUMINOUS EQUATOR BAND (asymmetric halo + crisp core)
    % Layer 1: Asymmetric soft halo (brighter front-left, dimmer back-right)
    \draw[line width=6pt, green!50!yellow!30, opacity=0.20] 
        (-2,0) arc (180:270:2 and 0.5);
    \draw[line width=6pt, green!50!yellow!30, opacity=0.12] 
        (-2,0) arc (180:360:2 and 0.5) arc (0:90:2 and 0.5);
    \draw[line width=6pt, green!50!yellow!30, opacity=0.06, dashed] 
        (-2,0) arc (180:0:2 and 0.5);
    
    % Layer 2: Medium band with seam texture (micro-dashed)
    \draw[line width=3pt, green!55!black, opacity=0.45, 
          dash pattern=on 1.5pt off 0.8pt] 
        (-2,0) arc (180:360:2 and 0.5);
    \draw[line width=3pt, green!55!black, opacity=0.22, 
          dash pattern=on 1.5pt off 0.8pt, dashed] 
        (-2,0) arc (180:0:2 and 0.5);
    
    % Layer 3: Crisp core seam
    \draw[line width=2.2pt, green!60!black] 
        (-2,0) arc (180:360:2 and 0.5);
    \draw[line width=1.5pt, green!60!black, opacity=0.6, dashed] 
        (-2,0) arc (180:0:2 and 0.5);
    
    % Tiny tick marks / punctures on equator (structural)
    \foreach \ang in {200, 240, 280, 320, 340}{
        \fill[green!60!black, opacity=0.7] ({2*cos(\ang)}, {0.5*sin(\ang)}) circle (1pt);
    }
    
    % Echo LHS sewing: tiny curved segments crossing the seam
    \draw[purple!70!black, line width=0.7pt, opacity=0.6]
        ({2*cos(240)-0.1}, {0.5*sin(240)+0.15}) 
        .. controls ({2*cos(240)}, {0.5*sin(240)}) .. 
        ({2*cos(240)+0.1}, {0.5*sin(240)-0.15});
    \draw[purple!70!black, line width=0.7pt, opacity=0.6]
        ({2*cos(320)-0.1}, {0.5*sin(320)+0.15}) 
        .. controls ({2*cos(320)}, {0.5*sin(320)}) .. 
        ({2*cos(320)+0.1}, {0.5*sin(320)-0.15});
    
    % Poles
    \fill[red!75!black] (0,2) circle (2.5pt);
    \node[above, red!75!black, font=\small] at (0,2.15) {$N$};
    
    \fill[red!75!black] (0,-2) circle (2.5pt);
    \node[below, red!75!black, font=\small] at (0,-2.15) {$S$};
    
    % Equator label with leader line
    \coordinate (eqpoint) at (2,0);
    \draw[green!55!black, line width=0.6pt] (eqpoint) -- ++(0.3,0);
    \node[right, green!55!black, font=\Large] at (2.3,0) {$S^1_{\mathrm{eq}}$};
    
    % Hemisphere labels on sphere (muted)
    \node[blue!65!black, font=\large] at (0,1.1) {$V$};
    \node[orange!65!red, font=\large] at (0,-1.1) {$\widetilde{V}$};
    
\end{scope}

\end{tikzpicture}
\caption{Gluing chiral hemispheres: a VOA $V$ with $\text{Rep}(\mathcal{C})$ on the northern hemisphere and $\widetilde{V}$ with $\text{Rep}(\widetilde{\mathcal{C}})$ are sewn along the boundaries at points $i\in I_{\mathcal{C}}$ and $j\in I_{\widetilde{\mathcal{C}}}$ by gluing matrix $M$, yielding the glued sphere with equator $S_{\text{eq}}^{1}$.}
\label{fig:equatorial_gluing}
\end{figure}
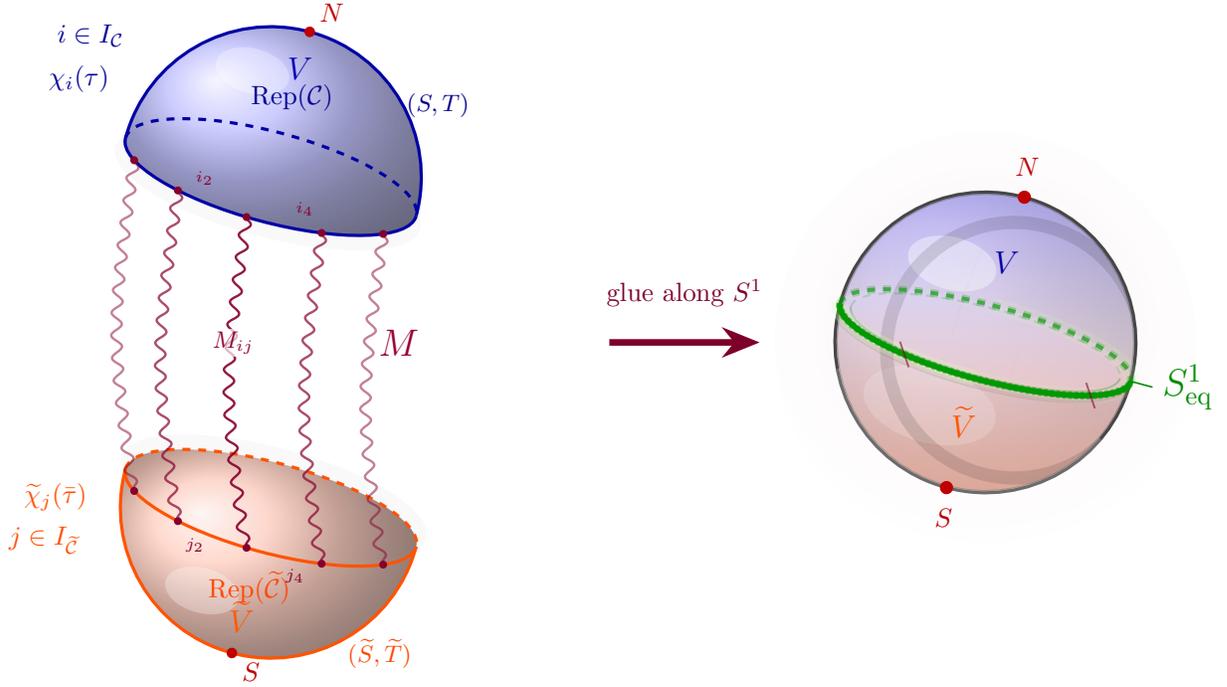

\noindent A genus-one amplitude of a (not necessarily meromorphic) RCFT may be viewed as arising from gluing a left-moving and a right-moving chiral block along an equator. A convenient way to formalize this is to work on the two–sphere
\begin{align}
  S^{2} = H_{N} \cup H_{S}, \qquad H_{N} \cap H_{S} = S^{1}_{\mathrm{eq}},
\end{align}
where $H_{N}$ and $H_{S}$ are the northern and southern closed hemispheres, and $S^{1}_{\mathrm{eq}}$ is their common boundary circle (the equator). We place a chiral VOA $V$ with representation category $\mathcal{C} = \mathrm{Rep}(V)$ on the northern hemisphere $H_{N}$, and a (possibly different) chiral VOA $\widetilde{V}$ with representation category $\widetilde{\mathcal{C}} = \mathrm{Rep}(\widetilde{V})$ on the southern hemisphere $H_{S}$. The pair $(V,\widetilde{V})$ are the chiral algebras underlying our RCFTs with central charges $c$ and $\widetilde{c}$, and their representation categories $\mathcal{C}, \widetilde{\mathcal{C}}$ are the corresponding unitary modular tensor categories (UMTCs) \cite{Moore:1988qv, Turaev:1994xb, Bakalov2000LecturesOT, Kitaev:2005hzj}. While a complete axiomatic classification of full RCFTs is still under development, it is widely conjectured that consistent genus-one couplings are governed by the modular data of these chiral categories (e.g., via the TFT/RCFT correspondence) \cite{Fuchs:2002cm, Fuchs:2003id, Felder:1999mq}. We denote by $I_{\mathcal{C}}$ and $I_{\widetilde{\mathcal{C}}}$ the finite index sets of isomorphism classes of irreducible $V$- and $\widetilde{V}$-modules (the left- and right-moving primaries). For each $i\in I_{\mathcal{C}}$ and $j\in I_{\widetilde{\mathcal{C}}}$, let $\chi_{i}(\tau)$ and $\widetilde{\chi}_{j}(\overline{\tau})$ be the corresponding characters, and assemble them into vectors
\begin{align}
  \begin{split}
    \chi(\tau) &= \bigl(\chi_{i}(\tau)\bigr)_{i\in I_{\mathcal{C}}},\\
    \widetilde{\chi}(\overline{\tau}) &= \bigl(\widetilde{\chi}_{j}(\overline{\tau})\bigr)_{j\in I_{\widetilde{\mathcal{C}}}}.
  \end{split}
\end{align}
The full genus-one amplitude is then sesquilinear in these vectors, and the anti-holomorphic dependence is implemented by complex conjugation $\overline{\widetilde{\chi}_{i}(\tau)} = \widetilde{\chi}_{i}(\overline{\tau})$. The modular data of $\mathcal{C}$ and $\widetilde{\mathcal{C}}$ are the $S$ and $T$ matrices acting on these characters, denoted by $S_{\mathcal{C}}, T_{\mathcal{C}}$ and $S_{\widetilde{\mathcal{C}}}, T_{\widetilde{\mathcal{C}}}$, respectively. Gluing the two hemispheres along the equator $S^{1}_{\mathrm{eq}}$ amounts to specifying how left- and right-moving chiral boundary states are paired. This pairing is encoded by a matrix of non-negative integers
\begin{align}
  M \in \mathrm{Mat}_{I_{\mathcal{C}}\times I_{\widetilde{\mathcal{C}}}}\bigl(\mathbb{Z}_{\geq 0}\bigr).
\end{align}
This gluing is shown in figure \ref{fig:equatorial_gluing}.
\\

\noindent The rows of $M$ are labeled by the left primaries $i\in I_{\mathcal{C}}$, and the columns by the right primaries $j\in I_{\widetilde{\mathcal{C}}}$. The entry $M_{ij}$ records the multiplicity with which the left sector $i$ is paired with the right sector $j$ along the equator. Passing to the torus by taking $S^{1}_{\mathrm{eq}}$ as the spatial circle and evolving in Euclidean time, the bilinear pairing encoded by $M$ produces the torus partition function 
\begin{align}\label{part}
  Z_{M}(\tau,\overline{\tau}) = \text{Tr}_{\mathcal{H}}\left(q^{L_{0} - \frac{c}{24}}\overline{q}^{\overline{L}_{0} - \frac{\overline{c}}{24}}\right) = \sum_{i\in I_{\mathcal{C}}}\sum_{j\in I_{\widetilde{\mathcal{C}}}}
    M_{ij}\,\chi_{i}(\tau)\,\widetilde{\chi}_{j}(\overline{\tau})
  = \chi(\tau)^{T} M\,\widetilde{\chi}(\overline{\tau}).
\end{align}
This ``equatorial pairing'' viewpoint is standard in the 3D TFT interpretation of RCFT \cite{Fuchs:2002cm}. In the TFT construction of correlators, one associates to the circle $S^{1}_{\mathrm{eq}}$ a state space spanned by simple objects of a modular tensor category, and the full partition function is obtained by gluing two such chiral boundary conditions along the common circle via a suitable Frobenius algebra object. In our language, the gluing data along the equator is encoded by the multiplicity matrix $M$, which specifies how left and right primaries are paired. In physical RCFT language, $M_{ij}$ counts the number of copies of the left–right sector
\begin{align}
  \mathcal{H}_{i}\otimes\widetilde{\mathcal{H}}_{j}
\end{align}
in the full Hilbert space, so that
\begin{align}
  \mathcal{H} \cong \bigoplus_{i\in I_{\mathcal{C}},\,j\in I_{\widetilde{\mathcal{C}}}}
    M_{ij}\,\mathcal{H}_{i}\otimes\widetilde{\mathcal{H}}_{j}.
\end{align}
The partition function \ref{part} is taken as a trace over this space. Since degeneracies cannot be negative, we indeed have $M_{ij}\in \mathbb{Z}_{\geq 0}$. Therefore, the matrix $M$ is the operator that implements the equatorial gluing of chiral blocks. It is not required to be diagonal\footnote{The left and right chiral algebras need not be identified label by label. Even when $\mathcal{C} = \widetilde{\mathcal{C}}$, modular invariants such as the charge-conjugation, simple current, or exceptional ADE invariants mix different primaries on the two sides. In the TFT language, this corresponds to choosing a non-trivial Frobenius algebra object rather than the trivial diagonal one. Hence, arbitrary left–right couplings $M_{ij}\in\mathbb{Z}_{\geq 0}$ are a priori allowed, and the only constraints on $M$ come from modular invariance and positivity, not from diagonality.}, and arbitrary left–right pairings are allowed, but physical consistency imposes modular constraints. The characters transform under the mapping class group of the torus via the usual $S$ and $T$ transformations that take the following form
\begin{align}
  \begin{split}
    \chi_{i}\!\left(-\frac{1}{\tau}\right) &= \sum_{k\in I_{\mathcal{C}}} (S_{\mathcal{C}})_{ik}\,\chi_{k}(\tau),\\
    \chi_{i}(\tau + 1) &= \sum_{k\in I_{\mathcal{C}}} (T_{\mathcal{C}})_{ik}\,\chi_{k}(\tau),
  \end{split}
\end{align}
and similarly for the right chiral theory $\widetilde{\mathcal{C}}$, which transforms with $\overline{S_{\widetilde{\mathcal{C}}}}$ and $\overline{T_{\widetilde{\mathcal{C}}}}$.  Modular invariance of the glued partition function requires that we have invariance under the modular data
\begin{align}
  \begin{split}
    Z_{M}\!\left(-\frac{1}{\tau}, -\frac{1}{\overline{\tau}}\right) &= Z_{M}(\tau, \overline{\tau}),\\
    Z_{M}(\tau + 1, \overline{\tau} + 1) &= Z_{M}(\tau, \overline{\tau}).
  \end{split}
\end{align}
For invariance under $S$–transformations, we have
\begin{align}
  \begin{split}
    Z_{M}\!\left(-\frac{1}{\tau}, -\frac{1}{\overline{\tau}}\right)
      &= \sum_{i,j} M_{ij}\,\chi_{i}\!\left(-\frac{1}{\tau}\right)
                       \,\widetilde{\chi}_{j}\!\left(-\frac{1}{\overline{\tau}}\right)\\
      &= \sum_{k,\ell}\left(
          \sum_{i,j}
            (S_{\mathcal{C}})_{ik}\,M_{ij}\,
            \overline{(S_{\widetilde{\mathcal{C}}})_{j\ell}}
        \right)\chi_{k}(\tau)\,\widetilde{\chi}_{\ell}(\overline{\tau})\\
      &= \sum_{k,\ell} M_{k\ell}\,\chi_{k}(\tau)\,\widetilde{\chi}_{\ell}(\overline{\tau}),
  \end{split}
\end{align}
and for invariance under $T$–transformations, we have
\begin{align}
  \begin{split}
    Z_{M}(\tau + 1, \overline{\tau} + 1)
      &= \sum_{i,j} M_{ij}\,\chi_{i}(\tau + 1)\,
                       \widetilde{\chi}_{j}(\overline{\tau} + 1)\\
      &= \sum_{k,\ell}\left(
          \sum_{i,j}
            (T_{\mathcal{C}})_{ik}\,M_{ij}\,
            \overline{(T_{\widetilde{\mathcal{C}}})_{j\ell}}
        \right)\chi_{k}(\tau)\,\widetilde{\chi}_{\ell}(\overline{\tau})\\
      &= \sum_{k,\ell} M_{k\ell}\,\chi_{k}(\tau)\,\widetilde{\chi}_{\ell}(\overline{\tau}).
  \end{split}
\end{align}
From these, by comparing coefficients of $\chi_{k}(\tau)$ and $\widetilde{\chi}_{\ell}(\overline{\tau})$, we obtain the intertwiner equations
\begin{align}\label{intertwiners_general}
  \begin{split}
    S_{\mathcal{C}}^{T} M &= M\,\overline{S_{\widetilde{\mathcal{C}}}},\\
    T_{\mathcal{C}}^{T} M &= M\,\overline{T_{\widetilde{\mathcal{C}}}}.
  \end{split}
\end{align}
In other words, the multiplicity matrix $M$ must intertwine the two modular representations carried by the left and the right VOAs. Recall that for unitary RCFTs with $\mathcal{C} = \widetilde{\mathcal{C}}$, the modular representation $\rho: \mathrm{SL}(2,\mathbb{Z})\to \mathrm{U}(N)$ acting on the characters is unitary, i.e. $S^{\dagger} = S^{-1}$ and $T^{\dagger} = T^{-1}$, and we can choose a basis of characters (equivalently, an orthonormal basis of irreps) in which $T$ is diagonal with entries
\begin{align}
  T_{ii} = e^{2\pi i\left(h_{i} - \tfrac{c}{24}\right)},
\end{align}
so that $T = T^{T}$, and the $S$–matrix is symmetric, $S_{ij} = S_{ji}$. In particular, for diagonal unitary theories, one often has $S_{\mathcal{C}}^{T} = S_{\mathcal{C}}$ and $T_{\mathcal{C}}^{T} = T_{\mathcal{C}}$, so \ref{intertwiners_general} becomes
\begin{align}\label{intertwiners}
  \begin{split}
    S_{\mathcal{C}} M &= M\,\overline{S_{\widetilde{\mathcal{C}}}},\\
    T_{\mathcal{C}} M &= M\,\overline{T_{\widetilde{\mathcal{C}}}}.
  \end{split}
\end{align}
When, moreover, $\mathcal{C} = \widetilde{\mathcal{C}}$ and the chosen $S$-matrix is real (as in many WZW and minimal models), the $S$-constraint reduces to 
\begin{align}\label{intertwiner_S}
    \left[S_{\mathcal{C}}, M\right] = S_{\mathcal{C}}M - MS_{\mathcal{C}} = 0.
\end{align}
The $T$-constraint remains a non-trivial phase-matching condition and should be kept in the form\footnote{If the $T$-matrix is real with $h_{i} - \tfrac{c}{24}\in \frac{1}{2}\mathbb{Z}$ as in the case of $E_{8}$ then we have the special case of $[T_{\mathcal{C}}, M] = 0$.}
\begin{align}\label{intertwiner_T}
  T_{\mathcal{C}} M = M\,\overline{T_{\mathcal{C}}} = MT_{\mathcal{C}}^{-1}.
\end{align}
This enforces the following spin-matching condition
\begin{align}
  e^{2\pi i\left(h_{i} - \tfrac{c}{24}\right)} = e^{-2\pi i\left(\widetilde h_j-\tfrac{\widetilde c}{24}\right)}.
\end{align}
Equivalently,
\begin{align}
h_i+\widetilde h_j-\frac{c+\widetilde{c}}{24}\in \mathbb{Z}.
\end{align}
In the common case when we take the same chiral theory on both sides, we have $c = \widetilde{c}$ and $h_{i} = \widetilde{h}_{j}$, and this reduces to 
\begin{align}
    h_{i} + h_{j} - \frac{c}{12}\in\mathbb{Z},
\end{align}
and in particular, for the diagonal invariant $i = j$, it implies $2h_{i} - \tfrac{c}{12}\in\mathbb{Z}$. If instead, one considers the usual non-chiral RCFT normalization $Z = \sum_{i,j}M_{ij}\chi_{i}(\tau)\overline{\chi_{j}(\tau)}$, then the standard condition is $h_{i} - h_{j}\in \mathbb{Z}$. Taken together, \ref{intertwiner_S} and \ref{intertwiner_T} are precisely the Cappelli–Itzykson–Zuber modular invariant conditions in the standard RCFT normalization \cite{Gannon:1999cp} studied in their ADE classification of modular invariants of $\widehat{\mathfrak{su}}(2)_{k}$. In the modern UMTC language, see \cite{Gannon:1999cp} for example, a modular invariant is exactly a non-negative integer matrix $M$ satisfying these intertwiner equations. Said otherwise, the sphere-with-equator picture glues a left $V$–block to a right $\widetilde{V}$–block with multiplicity matrix $M$. The equatorial amplitude then simply reads 
\begin{align}
  Z^{\mathrm{eq}}(\tau, \overline{\tau}) = \sum_{i, j} M_{ij}\,\chi_{i}(\tau)\,\widetilde{\chi}_{j}(\overline{\tau}).
\end{align}
Hence, a commutant pair $(\mathcal{C}, \widetilde{\mathcal{C}})$ is precisely a pair of UMTCs whose representation categories can be consistently glued along the equator.

\subsection{Topological defect lines}
Before performing equatorial gluing, we may modify the left hemisphere by inserting a TDL $\mathcal{L}$ supported on a loop parallel to the equatorial circle $S_{\text{eq}}^{1}$, or equivalently, parallel to the spatial cycle after passing to the torus. In the $3$D TFT/RCFT dictionary, such a line corresponds to an invertible object or an automorphism in the chiral category $\mathcal{C} = \text{Rep}(V)$. Since the hemisphere carries only the left-moving chiral blocks, the insertion acts only on the $\mathcal{C}$-label of each primary $i\in I_{\mathcal{C}}$, and hence modifies the equatorial gluing by a linear operator acting on the character vector $\chi(\tau)$. The possible invertible topological lines that preserve the chiral algebra fall into two natural cases:
    \subsection*{Verlinde (group-like) lines} These correspond to invertible simple objects of $\mathcal{C}$, i.e., the Picard group \cite{Frohlich2004PicardGI}
    \begin{align}
        \text{Pic}(\mathcal{C}) = \{a\in I_{\mathcal{C}}\vert a\otimes a^{\vee} \cong 1\}.
    \end{align}
    An object $a$ lies in the Picard group precisely if it admits a tensor inverse (its dual), $a\otimes a^{\vee}\cong 1$ and $a^{\vee}\otimes a\cong 1$, or equivalently, has quantum dimension $d_{a} = 1$. This is exactly analogous to an element $g$ in a group having an inverse $g^{-1}$. Physically, they are group-like (simple current) topological lines. The action of $a\in\text{Pic}(\mathcal{C})$ on the characters is diagonal and determined entirely by the modular $S$-matrix as follows \cite{Verlinde:1988sn}
    \begin{align}
        \left(D_{a}\right)_{ii'} = \delta_{ii'}\frac{\left(S_{\mathcal{C}}\right)_{ai}}{\left(S_{\mathcal{C}}\right)_{0i}},\ \ i, i'\in I_{\mathcal{C}}.
    \end{align}
    This is the familiar Verlinde line eigenvalue formula and encodes how the defect braids with each primary.

    \subsection*{Automorphism (anyon-permuting) lines}
    In addition to the Verlinde lines coming from the Picard group $\text{Pic}(\mathcal{C})$, a UMTC $\mathcal{C}$ also admits a second source of invertible topological line defects called anyon-permuting lines arising from braided autoequivalences of $\mathcal{C}$, $g\in \text{Aut}^{\text{br}}(\mathcal{C})$. We briefly review the structure needed for RCFT applications, and refer to \cite{Barkeshli:2014cna, Cheng:2020rpl} for details. A braided autoequivalence is a monoidal autoequivalence $g: \mathcal{C} \longrightarrow \mathcal{C}$ that preserves the braiding. Physically, such a functor relabels anyons in a way that preserves the fusion rules. In particular, it induces a permutation of the simple objects $g: I_{\mathcal{C}}\to I_{\mathcal{C}}$. Braiding preservation can be expressed as compatibility with the braiding isomorphisms. In UMTC $\mathcal{C}$, every pair of simple objects or anyon types $X, Y$ has an associated braiding morphism
    \begin{align}
        c_{X,Y}:X\otimes Y\longrightarrow Y\otimes X,
    \end{align}
    namely $g$ is braided if it intertwines braidings (up to specified monoidal structure), which in RCFT terms implies invariance of the modular data under relabeling. Physically, this is the operation of exchanging anyon $X$ around anyon $Y$. The eigenvalues of various composites of $c_{X,Y}$ encode all topological data, such as braiding phases, fusion multiplicities, and the tensor categorical data that are the modular $S$- and $T$-matrices. Now, for $g$ to be a genuine topological symmetry, it must preserve all of the topological data. Although fusion preservation is built into the definition of monoidal autoequivalence, braiding preservation requires that when we braid two anyons $X$ and $Y$ and then apply $g$, we obtain the same morphism as first applying $g$ to anyons and then braiding them. This physical requirement is exactly encoded by  the following categorical definition
    \begin{align}
    g(c_{X,Y}) = c_{g(X),g(Y)}.
    \end{align}
    This condition ensures that all topological data is preserved. The phase $\theta_{X,Y}$ obtained by exchanging $X$ around $Y$ must be the same when exchanging $g(X)$ around $g(Y)$, and all the topological interaction data is preserved, i.e.
    \begin{align}
        S_{XY} = S_{g(X), g(Y)}, \qquad \theta_{X} = \theta_{g(X)},\qquad N^{Z}_{XY} = N^{g(Z)}_{g(X),g(Y)}.
    \end{align}
    Hence, a braided autoequivalences is precisely an anyon-permuting symmetry of the UMTC that preserves braiding statistics. The quotient of the braided autoequivalences of $\mathcal{C}$ and the monoidal natural isomorphism is denoted as $\text{Aut}^{\text{br}}(\mathcal{C})$, the group of such symmetries. The quotient reflects the physical principle that topological line defects are defined only up to the attachment of invertible point-like operators at their endpoints. Each element $g\in\text{Aut}^{\text{br}}(\mathcal{C})$ defines an invertible TDL and acts on the character vector $\chi(\tau)$ by permuting labels $i\mapsto g(i)$. Concretely, identify the character space with
    \begin{align}
        \mathbb{C}^{I_{\mathcal{C}}} = \text{Span}\{e_{i}: i\in I_{\mathcal{C}}\},
    \end{align}
    whose basis element $e_{i}$ corresponds to torus blocks labeled by primaries $i$. We represent the action of $g$ on this space by the permutation matrix
    \begin{align}
        \left(P_{g}\right)_{ij} = \delta_{i, g(j)},
    \end{align}
    so that $P_{g}e_{i} = e_{g(i)}$, and hence $P_{g}$ relabels characters by sending $\chi_{j}\mapsto \chi_{g(j)}$. Geometrically, this is the action of the defect on the chiral blocks; whether one interprets the insertion as running along the equatorial/spatial cycle or the temporal cycle depends on which trace/channel one chooses, but in either case, the induced action on the chiral labels is the same permutation. We choose a permutation matrix $P_{g}$ that respects modular transformations
    \begin{align}\label{P_S&T}
        \begin{split}
            P_{g}S_{\mathcal{C}} =& S_{\mathcal{C}}P_{g},\\
            P_{g}T_{\mathcal{C}} =& T_{\mathcal{C}}P_{g}.
        \end{split}
    \end{align}
    A braided autoequivalence is only defined up to a monoidal natural isomorphism \cite{Edie-Michell:2022abq}.  At the level of characters, such a natural isomorphism can act by diagonal rephasing $U = \text{Diag}(u_{i})$, which conjugates $P_{g}$ to $UP_{g}U^{-1}$. Fixing a standard normalization of the modular data, for example, $S_{\mathcal{C}}$ symmetric and $T_{\mathcal{C}}$ diagonal in RCFT conventions, removes this residual ambiguity in typical unitary RCFT examples, and we will choose a representative $P_{g}$ compatible with \ref{P_S&T}. For explicit computations of $\text{Aut}^{\text{br}}(\mathcal{C})$ in large families of WZW UMTCs, see \cite{Edie-Michell:2022abq}.
    %We direct the reader to appendix \ref{sec:appendix_RCFT_lens} for a swift understanding of the Picard group and the braided autoequivalence from the perspective of the RCFT.
\noindent

\subsection*{Combined action of the lines}
The two classes of invertible topological lines in $\mathcal{C}$ described above naturally combine into a single algebraic structure. A convenient way to package their action on characters is to view $\mathbb{C}^{I_{\mathcal{C}}}$ as the complex vector space spanned by the torus characters $\{\chi_{i}\}_{i\in I_{\mathcal{C}}}$. Any linear action on the character vector is therefore an element of $\text{End}(\mathbb{C}^{I_{\mathcal{C}}})\cong \text{Mat}_{I_{\mathcal{C}}\times I_{\widetilde{\mathcal{C}}}}(\mathbb{C})$, the algebra of $\mathbb{C}^{I_{\mathcal{C}}}\times \mathbb{C}^{I_{\mathcal{C}}}$ complex matrices. Both $D_{a}$ and $P_{g}$ are special such matrices. The Verlinde operator $D_{a}$ is diagonal while the automorphism operator $P_{g}$ is a permutation matrix. Since both are invertible, their products describe general invertible defect insertions acting on chiral blocks. Simple currents and anyon-permuting symmetries do not act independently. Indeed, an element $g\in\text{Aut}^{\text{br}}(\mathcal{C})$ permutes simples $i\mapsto g(i)$, and by restriction, acts on invertible objects $a\in\text{Pic}(\mathcal{C})$, sending $a\mapsto g(a)\in\text{Pic}(\mathcal{C})$. Physically, dragging a simple current defect through an anyon-permuting defect relabels it. Hence, the subgroup of invertible defects generated by these two classes organizes as the semi-direct product
\begin{align}
    \text{Pic}(\mathcal{C})\rtimes\text{Aut}^{\text{br}}(\mathcal{C}).
\end{align}
The corresponding operator realization on genus-one characters is the map
\begin{align}
    \rho: \text{Pic}(\mathcal{C})\rtimes\text{Aut}^{\text{br}}({\mathcal{C}})\to \text{End}\left(\mathbb{C}^{I_{\mathcal{C}}}\right),
\end{align}
where $\rho(a,g) = D_{a}P_{g}\in \text{End}\left(\mathbb{C}^{I_{\mathcal{C}}}\right)$. To see that this is a representation of the semi-direct product, note that $P_{g}$ permutes the basis vectors $e_{i}$ (equivalently, the character labels) as 
\begin{align}
    P_{g}e_{i} = e_{g(i)},\qquad P_{g}^{-1}e_{i} = e_{g^{-1}(i)}.
\end{align}
Conjugation by $P_{g}$ implements the induced action of $g$ on simple current lines as follows
\begin{align}\label{PdP-1}
    P_{g}D_{a}P_{g}^{-1} = D_{g(a)}.
\end{align}
This is quite easy to see. Consider the action of the Verlinde operator on the basis element that reads
\begin{align}
    D_{a}e_{i} = \lambda_{a}(i)e_{i},\qquad \lambda_{a}(i) \equiv \frac{\left(S_{\mathcal{C}}\right)_{ai}}{\left(S_{\mathcal{C}}\right)_{0i}}.
\end{align}
We can now conjugate $D_{a}$ by $P_{g}$ to obtain
\begin{align}
    \begin{split}
        P_{g}D_{a}P_{g}^{-1}e_{i} =& P_{g}D_{a}e_{g^{-1}(i)}\\
        =& P_{g}\left(\lambda_{a}(g^{-1}(i))e_{g^{-1}(i)}\right)\\
        =& \lambda_{a}(g^{-1}(i))e_{i},
    \end{split}
\end{align}
where we have used the fact that $gg^{-1} = \text{id}$. Hence, we have found that $P_{g}D_{a}P_{g}^{-1}$ is diagonal in the same basis, with diagonal entries
\begin{align}
    \left(P_{g}D_{a}P_{g}^{-1}\right)_{ii'} = \lambda_{a}(g^{-1}(i)).
\end{align}
All we need to do now is relate the eigenvalues using modular-data symmetry. Since $g$ is a braided autoequivalence, it preserves modular data, in particular, the $S$-matrix as shown below
\begin{align}
    \left(S_{\mathcal{C}}\right)_{g(x), g(y)} = \left(S_{\mathcal{C}}\right)_{x,y}.
\end{align}
Hence, we find the eigenvalue $\lambda$ to be
\begin{align}
    \lambda_{g(a)}(i) = \frac{\left(S_{\mathcal{C}}\right)_{g(a),i}}{\left(S_{\mathcal{C}}\right)_{0,i}} = \frac{\left(S_{\mathcal{C}}\right)_{a, g^{-1}(i)}}{\left(S_{\mathcal{C}}\right)_{0, g^{-1}(i)}} = \lambda_{a}(g^{-1}(i)),
\end{align}
where we have used the fact that $g(0) = 0$. Putting this together, we have
\begin{align}
    P_{g}D_{a}P_{g}^{-1}e_{i} = \lambda_{g(a)}(i)e_{i} = D_{g(a)}e_{i}.
\end{align}
For example, if the simple labels are $\{0,x,y\}$, and $g$ swaps $x\leftrightarrow y$ while fixing $0$, then the permutation matrix takes the form $P_{g} = \left(\begin{smallmatrix} 1 & 0 & 0\\ 0 & 0 & 1\\ 0& 1 & 0\end{smallmatrix}\right)$. Let $D_{a} = \text{Diag}\left(1, \lambda_{x},\lambda_{y}\right)$, then a direct multiplication yields $P_{g}D_{a}P_{g}^{-1} = \text{Diag}(1, \lambda_{y}, \lambda_{x})$. Using \ref{PdP-1}, we can compute the product of two defect operators to be   
\begin{align}
    \begin{split}
        \rho(a_{1}, g_{1})\rho(a_{2},g_{2})=& D_{a_{1}}P_{g_{1}}D_{a_{2}}P_{g_{2}}\\
        =& D_{a_{1}}\left(P_{g_{1}}D_{a_{2}}P_{g_{1}}^{-1}\right)P_{g_{1}}P_{g_{2}}\\
        =& D_{a_{1}}D_{g_{1}(a_{2})}P_{g_{1}g_{2}}\\
         =& D_{a_{1}\cdot g_{1}(a_{2})}P_{g_{1}g_{2}}\\
        =& \rho(a_{1}\cdot g_{1}(a_{2}), g_{1}g_{2}), 
    \end{split}
\end{align}
where $g_{1}(a_{2})$ is the natural action of $g_{1}$ on invertible objects, and $P_{g_{1}}$ conjugates the diagonal $D_{a_{2}}$ to $D_{g_{1}(a_{2})}$ since it permutes the labels in the diagonals and thus $\rho$  being the representation of a semi-direct product as defined by the multiplication rule of the semi-direct product
\begin{align}
    (a_{1}, g_{1})(a_{2}, g_{2}) = (a_{1}g_{1}(a_{2}), g_{1}g_{2}).
\end{align}
Algebraically, this is just the representation of $\text{Pic}(\mathcal{C})\rtimes \text{Aut}^{\text{br}}(\mathcal{C})$, which is physically, the fusion of group-like lines, and anyon-permuting symmetry lines. Now, $D_{a}$ commutes with the modular data $S_{\mathcal{C}}$ and $T_{\mathcal{C}}$ by the Verlinde formula, and by the construction of $P_{g}$, $\rho(a,g)$ also commutes with the modular data. Let us define the equatorial Hilbert space as follows
\begin{align}
    \mathcal{H} = \bigoplus\limits_{i\in I_{\mathcal{C}}, j\in I_{\widetilde{\mathcal{C}}}}M_{ij}\mathcal{H}_{i}\otimes\widetilde{\mathcal{H}}_{j},
\end{align}
where $M_{ij}$ are the multiplicities. Let $\widehat{D}_{a}$ be the Verlinde defect operator that only acts on the left factor by scalar $\tfrac{\left(S_{\mathcal{C}}\right)_{ai}}{\left(S_{\mathcal{C}}\right)_{0i}}$ on the summand $\mathcal{H}_{i}\otimes\widetilde{\mathcal{H}}_{j}$, and let $\widehat{P}_{g}$ permute the labels via the autoequivalence $g$, sending $\mathcal{H}_{i}\otimes\widetilde{\mathcal{H}}_{j}$ to $\mathcal{H}_{g(i)}\otimes\widetilde{\mathcal{H}}_{j}$. The corresponding defect-twinned genus-one amplitude is
\begin{align}
    Z^{(a,g)}(\tau, \overline{\tau}) = \text{Tr}_{\mathcal{H}}\left[\widehat{D_{a}P_{g}}q^{L_{0} - \frac{c}{24}}\overline{q}^{\overline{L}_{0} - \frac{c}{24}}\right] = \chi(\tau)^{T}\rho(a,g)M\widetilde{\chi}(\overline{\tau}),
\end{align}
with $\rho(a,g) = D_{a}P_{g}$. In components\footnote{Here, we use the convention $(P_{g})_{ij} = \delta_{j,g(i)}$; if instead one prefers $\delta_{i, g(j)}$, the subsequent formulas should be adjusted consistently},
\begin{align}
    \rho(a,g) = \left(D_{a}P_{g}\right)_{ij} = \sum\limits_{k}\left(D_{a}\right)_{ik}\left(P_{g}\right)_{kj} = \frac{\left(S_{\mathcal{C}}\right)_{ai}}{\left(S_{\mathcal{C}}\right)_{0i}}\delta_{j,g(i)}.
\end{align}
By \ref{intertwiners} and the fact that $\rho(a,g)$ commutes with $S_{\mathcal{C}}$ and $T_{\mathcal{C}}$, we see that $Z^{(a,g)}$ is modular covariant, i.e., it is a torus amplitude. In case that $M$ is diagonal, $M_{ij} = \delta_{j,\phi(i)}$, where $\phi: I_{\mathcal{C}}\to I_{\widetilde{\mathcal{C}}}$ defines a bijection, we have
\begin{align}
    \left(D_{a}P_{g}M\right)_{ij} = \sum\limits_{i'}\left(D_{a}P_{g}\right)_{ii'}M_{i'j} = \frac{\left(S_{\mathcal{C}}\right)_{ai}}{\left(S_{\mathcal{C}}\right)_{0i}}\delta_{j,\phi(g(i))}.
\end{align}
Inserting this back into the expression for the equatorial amplitude yields
\begin{align}
    Z^{(a,g)}(\tau, \overline{\tau}) = \sum\limits_{i\in I_{\mathcal{C}}}\frac{\left(S_{\mathcal{C}}\right)_{ai}}{\left(S_{\mathcal{C}}\right)_{0i}}\chi_{i}(\tau)\widetilde{\chi}_{\phi(g(i))}(\overline{\tau}).
\end{align}
Hence, a purely Verlinde choice with $g = 1$ gives diagonal twining and a purely automorphic choice with $a = 1$ permutes the right labels. The action of $\phi(g(i))$ is simply that first, the automorphism permutes the left label $i\mapsto g(i)$, then the pairing $\phi$ sends that to the right label $\phi(g(i))$.

\subsection{Half-gauging}
A salient advantage of the equatorial-gluing framework is that it naturally accommodates one-sided manipulations of the chiral data. In particular, given a collection of invertible TDLs acting on the left-moving chiral blocks, we may average over a finite subgroup on the upper (or left) hemisphere only, while leaving the lower (or right) hemisphere untouched. This produces a controlled deformation of the equatorial pairing, which we will refer to as half-gauging. Concretely, let
\begin{align}
    G_{\text{inv}}(\mathcal{C})\equiv \text{Pic}(\mathcal{C})\rtimes \text{Aut}^{\text{br}}(\mathcal{C}),
\end{align}
and let $\rho: G_{\text{inv}}(\mathcal{C})\rightarrow \text{End}(\mathbb{C}^{I_{\mathcal{C}}})$ be the induced action on the character vector $\chi(\tau)$. Each $\rho(h)$ is an invertible $I_{\mathcal{C}}\times I_{\mathcal{C}}$ matrix encoding the action of the topological line $h$ on the left-moving torus blocks (equivalently, the twining action when the line wraps the temporal cycle, or relabeling action after an $S$-move). Let $H\subset G_{\text{inv}}(\mathcal{C})$ be a finite subgroup generated by simple current and/or anyon permuting lines. Now, to gauge $H$ on the left-moving side, we introduce the averaging operator
\begin{align}\label{Pi_projector}
        \Pi_{H}\equiv \frac{1}{\vert H\vert}\sum\limits_{h\in H}\rho(h),\qquad \Pi^{2}_{H} = \Pi_{H}.
\end{align}
Since $\rho$ is a representation and $H$ is finite, $Pi_{H}$ is the idempotent,
\begin{align}
    \Pi_{H}^{2} = \Pi_{H},
\end{align}
and hence, projects onto the $H$-invariant subspace of the left-moving block space $\mathbb{C}^{I_{\mathcal{C}}}$. In the unitary case, after choosing an $H$-invariant Hermitian form, $\Pi_{H}$ can be taken as the orthogonal projector. Physically,  $\Pi_{H}$ implements an average over insertions of defects from $H$ on the left chiral half of the equatorial gluing data, while leaving the right-moving chiral theory and the pairing matrix $M$ untouched. Equivalently, at genus-one it averages the corresponding defect-twined traces on the left-moving blocks, without introducing additional twisted Hilbert spaces. In particular, this is not a full orbifold: we are neither summing over twists along both torus cycles nor including the associated twisted sectors on both chiral halves. Hence, the half-gauged amplitude takes the following form 
\begin{align}
    Z^{[H]} = \text{Tr}_{\mathcal{H}}\left[\widehat{\Pi}_{H}q^{L_{0} - \frac{c}{24}}\overline{q}^{\overline{L}_{0} - \frac{\widetilde{c}}{24}}\right].
\end{align}
This construction is precisely the one-sided analog of gauging a finite symmetry; it modifies only the left-moving sector, hence the name half–gauging. From a path integral perspective, $\Pi_{H}$ implements a sum over all $H$-twists on the left boundary circle, i.e., the equator, of the sphere, while the right boundary circle remains untwisted. This is not a full orbifold since we aren't averaging over insertions on both cycles and summing over twisted sectors on both chiral halves. Instead, half-gauging singles out the $H$-invariant subsector of the left-moving Hilbert space while retaining the original right-moving spectrum.\\ 

\noindent The modular property of this construction should be stated with the same care as for defect-twined traces. Since we are averaging only over temporal cycles insertions on the left-moving torus blocks (and are not simultaneously summing over spatial twists nor including twisted sectors), one should not expect a single modular invariant function in general. Rather, half-gauging naturally produces a modular covariant family of amplitudes closed under $\text{SL}(2, \mathbb{Z})$, obtained by transporting the insertion by conjugation. Concretely, under $\gamma\in \text{SL}(2, \mathbb{Z})$ the operators $\rho(h)$ are carried to
\begin{align}
    \rho(h)\longmapsto\rho_{\mathcal{C}}(\gamma)\rho(h)\rho_{\mathcal{C}}(\gamma)^{-1},
\end{align}
so the averaged insertion $\Pi_{H}$ is transported to the conjugated projector
\begin{align}
    \left(\Pi_{H}\right)_{\gamma}\equiv \rho_{\mathcal{C}}(\gamma)\Pi_{H}\rho_{\mathcal{C}}(\gamma)^{-1} = \frac{1}{\vert H\vert}\sum\limits_{h\in H}\rho_{\mathcal{C}}(\gamma)\rho(h)\rho_{\mathcal{C}}(\gamma)^{-1},
\end{align}
and the modular transformations act by
\begin{align}
    Z^{[H]}(\gamma\tau, \overline{\gamma\tau}) = Z^{[H]}_{\gamma}(\tau, \overline{\tau}),
\end{align}
where $Z^{[H]}_{\gamma}$ is defined by replacing $\Pi_{H}$ with $(\Pi_{H})_{\gamma}$ in the insertion.\\

\noindent If one further wants the half-gauged amplitude itself to be invariant, $Z^{[H]}(\gamma\tau, \overline{\gamma\tau}) = Z^{[H]}(\tau, \overline{\tau})$, then an additional structure is required, namely that the projector is preserved under the modular action of the relevant representation. Equivalently, one needs $(\Pi_{H})_{\gamma} = \Pi_{H}$ for the generators $\gamma = S, T$, which is stronger than the general covariance statement and amounts to $H$ acting compatibly with the modular data in the chosen basis. Without this hypothesis, the most that one can assert in general is closure of the half-gauged constructions under modular transformations, rather than invariance of an individual trace. Accordingly, one should not generally assert 
\begin{align}
    \left[S, \Pi_{H}\right] = 0,\qquad \left[T,\Pi_{H}\right] = 0,
\end{align}
unless each $\rho(h)$ strictly commutes with $S,T$ in the chosen basis, or more invariantly, unless $\Pi_{H}$ is fixed under conjugation by $\rho_{\mathcal{C}}(S)$ and $\rho_{\mathcal{C}}(T)$. This construction is the one-sided analogue of gauging a finite symmetry: it averages over defect insertions on one chiral half while leaving the other half unchanged. It is also closely related in spirit to the “generalized gauging” operations for non-invertible defects discussed in \cite{Albert:2025umy}, in the sense that both implement averaging/projector-like operations built from defect data. In the present paper, we restrict to subgroups of invertible lines, so $\Pi_{H}$ is an honest group average. Extending the equatorial projector beyond invertible lines to a full fusion-category idempotent implementing, for example, Haagerup-type gauging would require additional categorical input (as in the non-invertible setting of \cite{Albert:2025umy}) and lies beyond our present scope.

\subsection{Equatorial projection principle}
We now formalize the equatorial gluing picture as a precise statement about how invertible defect lines act on a commutant pairing. The key point is that inserting a line on (say) the left hemisphere produces a priori an \textit{interface amplitude}: a modular covariant genus-one observable whose modular transforms remain within the same family by conjugation  of the left inserted line. Such an object becomes a genuine genus-one RCFT partition function only when the insertion yields a new multiplicity matrix with non-negative integer entries that still satisfies the modular intertwiner equations; in particular, this requires both a suitable commutation/intertwining property with the modular representation and integrality/positivity of the resulting coupling.
\begin{tcolorbox}[colback=brown!5!white,colframe=brown!75!black,title=Equatorial Projection Principle]
Let $(\mathcal{C}, \widetilde{\mathcal{C}}, M)$ be a commutant pair with modular intertwiner $M$. For any invertible defect line $(a,g)\in \text{Pic}(\mathcal{C})\rtimes \text{Aut}^{\text{br}}(\mathcal{C})$, let $R\equiv D_{a}P_{g}$. The associated torus interface amplitude
\begin{align}\label{the_amplitude}
    Z^{(a,g)}(\tau, \overline{\tau}) = \chi(\tau)^{T}R^{T}M\overline{\widetilde{\chi}(\tau)}
\end{align}
is modular covariant: for every $\gamma\in\text{SL}(2, \mathbb{Z})$, 
\begin{align}\label{Z_mod_inv}
    Z^{(a,g)}(\gamma\cdot \tau, \overline{\gamma\cdot\tau}) = \chi(\tau)^{T}R_{\gamma}^{T}M\overline{\widetilde{\chi}(\tau)},
\end{align}
where $R_{\gamma}\equiv \rho_{\mathcal{C}}(\gamma)R\rho_{\mathcal{C}}(\gamma)^{-1}$.
Moreover, $Z^{(a,g)}$ defines a genuine modular invariant RCFT partition function iff the deformed coupling $M'\equiv R^{T}M$ has non-negative integer entries and satisfies the modular intertwiner equations
\begin{align}\label{modular_intertwiner_equations}
    \rho_{\mathcal{C}}(\gamma)^{T}M' = M'\overline{\rho_{\widetilde{\mathcal{C}}}(\gamma)},\qquad \gamma = S,T,
\end{align}
equivalently, $\rho_{\mathcal{C}}(\gamma)^{T}R^{T}M = R^{T}\rho_{\mathcal{C}}(\gamma)^{T}M$ for $\gamma = S,T$.
\end{tcolorbox}
\noindent 
Let $V$ and $\widetilde{V}$ be rational unitary VOAs with UMTCs $\mathcal{C} = \text{Rep}(V)$ and $\widetilde{\mathcal{C}} = \text{Rep}(\widetilde{V})$ with characters $\chi(\tau)$ and $\widetilde{\chi}(\tau)$, and let 
\begin{align}
    \begin{split}
        \rho_{\mathcal{C}}:&\text{SL}(2, \mathbb{Z})\to U(\vert I_{\mathcal{C}}\vert),\\
        \rho_{\widetilde{\mathcal{C}}}:&\text{SL}(2, \mathbb{Z})\to U(\vert I_{\widetilde{\mathcal{C}}}\vert),\\
    \end{split}
\end{align}
be the modular representations generated by modular data $(S_{\mathcal{C}}, T_{\mathcal{C}})$ and $(S_{\widetilde{\mathcal{C}}}, T_{\widetilde{\mathcal{C}}})$. Here, $U(\vert I_{\mathcal{C}}\vert)$ and $U(\vert I_{\widetilde{\mathcal{C}}}\vert)$ denote the unitary groups of degree $\vert I_{\mathcal{C}}\vert$ and $\vert I_{\widetilde{\mathcal{C}}}\vert$, i.e. ,the groups of $\vert I_{\mathcal{C}}\vert\times \vert I_{\mathcal{C}}\vert$ and $\vert I_{\widetilde{\mathcal{C}}}\vert\times \vert I_{\widetilde{\mathcal{C}}}\vert$ unitary matrices. Here, $\vert I_{\mathcal{C}}\vert$ (resp. $\vert I_{\widetilde{\mathcal{C}}}\vert$) represents the number of isomorphism classes of primaries of the VOA $V$ (resp. $\widetilde{V}$). The matrix $M\in \text{Mat}_{I_{\mathcal{C}}\times I_{\widetilde{\mathcal{C}}}}(\mathbb{Z}_{\geq 0})$ is called a modular intertwiner if
\begin{align}\label{M_def}
    \rho_{\mathcal{C}}(\gamma)^{T}M = M\overline{\rho_{\widetilde{\mathcal{C}}}(\gamma)},\qquad \forall\gamma\in \text{SL}(2, \mathbb{Z}).
\end{align}
This is equivalent to the usual $S$- and $T$-intertwiner relations and guarantees modular invariance of the defectless equatorial amplitude $Z^{\text{eq}}(\tau, \overline{\tau}) = \chi(\tau)^{T}M\overline{\widetilde{\chi}(\tau)}$. Now choose an invertible defect line on the left chiral theory, specified by the pair $(a,g)\in \text{Pic}(\mathcal{C})\rtimes\text{Aut}^{\text{br}}(\mathcal{C})$. Here, $a$ is an invertible simple current object giving a Verlinde line, and $g$ is a braided autoequivalence acting by permutation on primaries. We represent these on the character space by the operator $R\equiv D_{a}P_{g}$, where $D_{a}$ is the diagonal Verlinde eigenvalue matrix, and $P_{g}$ is the permutation matrix defined as follows
\begin{align}
    \begin{split}
     D_{a}\equiv& \text{Diag}\left(\frac{\left(S_{\mathcal{C}}\right)_{ai}}{\left(S_{\mathcal{C}}\right)_{0i}}\right)_{i\in I_{\mathcal{C}}},\\
        P_{g}=& \delta_{i,g(j)}.
    \end{split}
\end{align}
We choose a representative $P_{g}$ compatible with modular data, i.e., preserving $(S_{\mathcal{C}}, T_{\mathcal{C}})$ by conjugation
\begin{align}\label{P_realization}
    P_{g}\rho_{\mathcal{C}}(\gamma)P_{g}^{-1} = \rho_{\mathcal{C}}(\gamma),\qquad \forall\gamma\in \{S,T\}.
\end{align}

\subsubsection*{Modular Covariance}
Now, for every $\gamma\in\text{SL}(2, \mathbb{Z})$, modular covariance dictates that we have
\begin{align}
     Z^{a,g}(\gamma\cdot \tau,\overline{\gamma\cdot\tau}) = \chi(\tau)^{T}\rho_{\mathcal{C}}(\gamma)^{T}R^{T}M\overline{\rho_{\widetilde{\mathcal C}}(\gamma)}\overline{\widetilde{\chi}(\tau)}.
\end{align}
It is crucial to note here that one should not assume any commutativity between $R$ (or its constituents) and the modular data. Instead, the insertion is transported by conjugation. Using the identity
\begin{align}
    \rho_{\mathcal{C}}(\gamma)^{T}R^{T} = \left(\rho_{\mathcal{C}}(\gamma)^{T}R^{T}\rho_{\mathcal{C}}(\gamma)^{-1}\right)^{T}\rho_{\mathcal{C}}(\gamma)^{T},
\end{align}
and then applying the intertwiner relation  \ref{M_def}, we find
\begin{align}
        Z^{a,g}(\gamma\cdot \tau,\overline{\gamma\cdot\tau}) = \chi(\tau)^{T}R_{\gamma}^{T}M\overline{\widetilde{\chi}(\tau)} = Z^{(a,g)}_{\gamma}(\tau, \overline{\tau}),
\end{align}
where we have used 
\begin{align}\label{R_def}
    R_{\gamma}\equiv \rho_{\mathcal{C}}(\gamma)R\rho_{\mathcal{C}}(\gamma)^{-1},
\end{align}
This statement captures the robust modular covariance, namely that modular transformations act by conjugating the inserted operator $R$. In particular, the family of one-sided projections closes under $\text{SL}(2, \mathbb{Z})$ once one allows the transported insertions $R_{\gamma}$, and no extra commutation hypothesis is needed.\\

\noindent It is worth stressing what this corrects in the usual shorthand arguments. In many places, one informally appeals to the Verlinde formula to claim that the diagonal Verlinde line operator $D_{a}$ commutes with the modular data. What is true in general is the conjugation identity
\begin{align}
    S_{\mathcal{C}}D_{a}S_{\mathcal{C}}^{-1} = N_{a},
\end{align}
so $[D_{a}, S_{\mathcal{C}}]\neq 0$ generically. The modular covariance proof must therefore be phrased as above, tracking explicit conjugation of $R$, rather than rewriting $\rho_{\mathcal{C}}(\gamma)^{T}R^{T}$ as if $R$ commuted with $\rho_{\mathcal{C}}(\gamma)$. With this in mind, let us briefly spell out the standard channel interpretation. For $\gamma = T$, since $T_{\mathcal{C}}$ is diagonal, any diagonal $D_{a}$ commutes with $T_{\mathcal{C}}$ automatically, and the twining matrix $P_{g}$ has been chosen to preserve the modular data. Hence, one recovers the expected phase reweighting of the twined trace. For $\gamma = S$, the insertion is transported to the crossed channel by
\begin{align}
    R\mapsto S_{\mathcal{C}}RS_{\mathcal{C}}^{-1}.
\end{align}
In particular, when $R$ contains an $a$-Verlinde line insertion $D_{a}$ that is diagonal in the character basis, the $S$-transform produces fusion action in the crossed channel as follows
\begin{align}
    S_{\mathcal{C}}D_{a}S_{\mathcal{C}}^{-1} = N_{a},\qquad \left(N_{a}\right)_{ij} = N_{ai}^{j}.
\end{align}
Thus, the correct statement is that under $S$, a diagonal Verlinde line insertion does not remain diagonal but instead becomes a fusion matrix. This is precisely the mechanism one must keep track of when implementing the equatorial projection principle, since modular covariance holds by conjugation $R\mapsto R_{\gamma}$, while any additional closure within a restricted subclass of defects, such as keeping the same simple $(a,g)$ labels without enlarging the operator family, requires further structural conditions on the defects beyond the general covariance argument.

\subsubsection*{A genuine RCFT}
We now ask when $Z^{(a,g)}$ defines a genuine torus partition function, i.e., the genus-one partition function of a full CFT with chiral pair $(V, \widetilde{V})$. Concretely, this requires modular invariance, and that the coefficients admit the interpretation as non-negative integer multiplicities in the Hilbert-space decomposition. In our language, this is equivalent to the following two conditions on the deformed coupling matrix $M'\equiv R^{T}M$:
\begin{enumerate}
    \item Intertwiner condition (modular invariance): $M'$ must satisfy the modular intertwiner equations \ref{modular_intertwiner_equations}. 
    \item Positivity/integrality: $M'\in\text{Mat}_{I_{\mathcal{C}}\times I_{\widetilde{\mathcal{C}}}}(\mathbb{Z}_{\geq0})$.
\end{enumerate}
When these hold, $Z(\tau, \overline{\tau})$ defines a candidate modular invariant partition function with an evident Hilbert space interpretation at genus one. However, as is standard, the consistency of genus-one alone does not guarantee the existence of a full consistent RCFT, since constructing an actual full theory typically also requires satisfying sewing conditions of higher-genus \cite{Moore:1988qv, Sonoda:1988fq, fuchs2002tft} and locality or crossing constraints beyond the torus \cite{Moore:1988uz, Belavin:1984vu}. In what follows we use ``genuine" in the modest sense of providing a torus partition function candidate with non-negative integer multiplicities, with the understanding that additional consistency conditions are required to promote this to a complete CFT.\\

\noindent A convenient sufficient criterion for \ref{modular_intertwiner_equations} is that $R$ commute with the left modular representation (equivalently, $R^{T}$ commute with $\rho_{\mathcal{C}}(\gamma)^{T}$ for $\gamma = S,T$ in the chosen basis). However, commutation is not the most invariant statement; the invariant requirement is simply $M' = R^{T}M$ remains an intertwiner. Let's prove the sufficient commutation criterion carefully. Assume that for $\gamma = S,T$, we have\footnote{For unitary RCFT conventions where $S_{\mathcal{C}}^{T} = S_{\mathcal{C}}$ and $T_{\mathcal{C}}^{T} = T_{\mathcal{C}}$, this is equivalent to $[R, S_{\mathcal{C}}] = [R, T_{\mathcal{C}}] = 0$.}
\begin{align}\label{comm_assumptions}
    \begin{split}
        \rho_{\mathcal{C}}(\gamma)R^{T}=&  R^{T}\rho_{\mathcal{C}}(\gamma)^{T}.
    \end{split}
\end{align}
Then, using \ref{M_def} and $M' = R^{T}M$, we have
\begin{align}
    \begin{split}
        \rho_{\mathcal{C}}(\gamma)^{T}M' = & \rho_{\mathcal{C}}(\gamma)^{T}R^{T}M\\
        =& R^{T}\rho_{\mathcal{C}}(\gamma)^{T}M\\
    =& R^{T}M\overline{\rho_{\widetilde{\mathcal{C}}}(\gamma})\\
    =& M'\overline{\rho_{\widetilde{\mathcal{C}}}(\gamma}),
    \end{split}
\end{align}
which is precisely \ref{modular_intertwiner_equations}. If additionally $M'$ has non-negative integer entries, then $Z(\tau, \overline{\tau}) = \chi(\tau)^{T}M'\overline{\widetilde{\chi}(\tau)}$ is a modular invariant partition function with a Hilbert space interpretation.\\

\noindent Conversely, if $M'$ fails either the the intertwiner equations or non-negative integrality, then $Z^{(a,g)}$ cannotbe the partition function of a full RCFT with chiral pair $(V, \widetilde{V})$: the first failure breaks modular invariance, and the second destroys the multiplicity interpretation. Note that failure of the commutation condition $[R, S_{\mathcal{C}}] = [R, T_{\mathcal{C}}] = 0$ does not by itself rule out modular invariance since one can still have $R^{T}M$ intertwining for special choices of $M$, but generically commutation is the clean sufficient condition used in practice. What this shows is that producing a \textit{new} modular invariant by a one-sided insertion is highly constrained since we require $M'$ to remain an intertwiner and to have non-negative integer entries. For pure automorphism lines, $R = P_{g}$, with $P_{g}$ chosen to preserve the modular data (cf. \ref{P_realization} in conjugation form), $M' = P_{g}^{T}M$ is again an intertwiner whenever $M$ is, and integrality/positivity is preserved because $P_{g}^{T}$ merely permutes rows. For pure Verlinde lines $R = D_{a}$, one generally has $[D_{a}, S_{\mathcal{C}}]\neq 0$; equivalently $S_{\mathcal{C}}D_{a}S_{\mathcal{C}}^{-1} = N_{a}$ is typically non-diagonal. Hence, $M' = D_{a}^{T}M$
 generically fails the intertwiner equations, so the resulting $Z^{(a,1)}$ is typically an interface amplitude rather than a new modular invariant--- unless $D_{a}$ acts trivially on the rows of $M$ that actually appear (or other special accidents occur). %This is the smearing of the $E_{8}$ we previously. \NBU{Argue better as to why these are safe invariants!}

\subsection*{Corollary $1$}
\noindent As a first corollary, automorphism insertions are “safe" in the following sense: if $g\in\text{Aut}^{\text{br}}(\mathcal{C})$ is realized on characters by a permutation matrix $P_{g}$ preserving the modular data, then
\begin{align}
    Z^{(1,g)}(\tau, \overline{\tau}) = \chi(\tau)^{T}P_{g}^{T}M\overline{\widetilde{\chi}(\tau)}
\end{align}
is a modular invariant torus partition function whenever $Z^{\text{eq}} = \chi^{T}M\overline{\widetilde{\chi}}$ is. It is simply the same modular invariant with left labels permuted inside the multiplicity matrix.

\subsection*{Corollary $2$}
\noindent The second corollary is the simple current (Verlinde line) phase obstruction. For a multiplicity-free diagonal pairing $M = \delta_{j, \phi(i)}$, insertion of a non-trivial $a\in\text{Pic}(\mathcal{C})$ yields 
\begin{align}
    \left(M'\right)_{ij} = \left(D_{a}^{T}M\right)_{ij} = \frac{(S_{\mathcal C})_{ai}}{(S_{\mathcal C})_{0i}}\delta_{j,\phi(i)}.
\end{align}
Unless the eigenvalues $\tfrac{(S_{\mathcal C})_{ai}}{(S_{\mathcal C})_{0i}}$ equal $1$ on the support of $M$, entries of $M'$ are generally phases rather than non-negative integers. In that case $Z^{(a,1)}$ is not a modular invariant partition function, but rather a defect or an interface torus amplitude.

\subsection{Double-sided equatorial projection}
We pick left and right invertible lines that belong to the group of invertible topological lines as follows
\begin{align}
    \begin{split}
        \left(a_{L}, g_{L}\right)&\in \text{Pic}(\mathcal{C})\rtimes\text{Aut}^{\text{br}}(\mathcal{C})\equiv G_{\text{inv}}(\mathcal{C}),\\
        \left(a_{R}, g_{R}\right)&\in \text{Pic}(\widetilde{\mathcal{C}})\rtimes\text{Aut}^{\text{br}}(\widetilde{\mathcal{C}})\equiv G_{\text{inv}}(\widetilde{\mathcal{C}}),
    \end{split}
\end{align}
each with induced actions on characters
\begin{align}
    \begin{split}
        \rho:&G_{\text{inv}}(\mathcal{C})\to \text{End}\left(\mathbb{C}^{I_{\mathcal{C}}}\right),\qquad \  \rho(a,g) = D_{a}P_{g},\\
        \widetilde{\rho}:&G_{\text{inv}}(\widetilde{\mathcal{C}})\to \text{End}\left(\mathbb{C}^{I_{\widetilde{\mathcal{C}}}}\right),\qquad  \widetilde{\rho}(a,g) = \widetilde{D}_{\widetilde{a}}\widetilde{P}_{\widetilde{g}}.
    \end{split}
\end{align}
By construction, these are representations of topological invertible defects wrapping the temporal cycle of the torus, and they are compatible with the modular action, in the sense that for all $\gamma\in\text{SL}(2,\mathbb{Z})$, we have
\begin{align}
   \begin{split}
        \rho_{\mathcal{C}}(\gamma)\rho(h) =& \rho(h)\rho_{\mathcal{C}}(\gamma),\\
        \widetilde{\rho}_{\widetilde{\mathcal{C}}}(\gamma)\widetilde{\rho}(\widetilde{h}) =& \widetilde{\rho}(\widetilde{h})\widetilde{\rho}_{\widetilde{\mathcal{C}}}(\gamma),
   \end{split}
\end{align}
for any pair of invertible defect labels $h = (a,g)\in G_{\text{inv}}(\mathcal{C})$ and $\widetilde{h}\in(\widetilde{a}, \widetilde{g})\in G_{\text{inv}}(\widetilde{\mathcal{C}})$. Let's package the corresponding  operators on characters as left-acting and right-acting ones, as shown below
\begin{align}
    \begin{split}
        R_{L}\equiv D_{a_{L}}P_{g_{L}},\\
        R_{R}\equiv \widetilde{D}_{a_{R}}\widetilde{P}_{g_{R}}.
    \end{split}
\end{align}
For automorphism lines, one may choose representatives $P_{g}, \widetilde{P}_{\widetilde{g}}$ that preserve  the modular data by conjugation (equivalently commute in a permutation basis), typically stated for $\gamma\in\{S, T\}$ rather than for all $\gamma\in\text{SL}(2, \mathbb{Z})$. Now, the partition function with a double-sided insertion is defined as follows
\begin{align}\label{RLRR}
    \begin{split}
        Z^{(R_{L}; R_{R})}(\tau, \overline{\tau}) =& \text{Tr}_{\mathcal{H}}\left[\widehat{R}_{L}\otimes\widehat{R}_{R}\ q^{L_{0} - \frac{c}{24}}\overline{q}^{\overline{L}_{0} - \frac{\widetilde{c}}{24}}\right]\\
        =& \chi(\tau)^{T}R_{L}^{T}M\overline{R_{R}\widetilde{\chi}(\tau)}.
    \end{split}
\end{align}
For any $\gamma\in \text{SL}(2, \mathbb{Z})$, modular covariance requires that we have
\begin{align}
    Z^{(R_{L}; R_{R})}(\gamma\cdot \tau, \overline{\gamma \cdot \tau}) = \chi(\tau)^{T}R_{L,\gamma}^{T}M\overline{R_{R,\gamma}}\overline{\widetilde\chi(\tau)},
    %\left(\rho_{\mathcal{C}}R_{L}\rho_{\mathcal{C}}(\gamma)(\gamma)^{-1}\right)M\left(\overline{\rho}_{\widetilde{\mathcal{C}}}^{-1}\overline{R_{R}}\rho_{\widetilde{\mathcal{C}}}\right)\widetilde{\chi}(\overline{\tau}).
\end{align}
where the transported line operators are conjugated as follows
\begin{align}
    \begin{split}
    R_{L,\gamma}\equiv& \rho_{\mathcal C}(\gamma)R_L\rho_{\mathcal C}(\gamma)^{-1},\\ R_{R,\gamma}\equiv&\rho_{\widetilde{\mathcal C}}(\gamma)R_R\rho_{\widetilde{\mathcal C}}(\gamma)^{-1}.
    \end{split}
\end{align}
Hence, the family $\{Z^{(R_{L}; R_{R})}\}$   is closed under modular transformations, i.e., the lines are conjugated by the chiral actions. This is easy to see by inserting $\chi(\gamma\cdot \tau) = \rho_{\mathcal{C}}(\gamma)\chi(\tau)$ and $\widetilde{\chi}(\gamma\cdot \tau) = \rho_{\widetilde{\mathcal{C}}}(\gamma)\widetilde{\chi}(\tau)$ in \ref{RLRR} and using the fact that the pairing $M$ satisfies \ref{M_def}. This shows that double-sided insertions always produce \textit{modular covariant} genus-one amplitudes. To obtain a genuine modular invariant partition function, we must require 
that the deformed coupling 
\begin{align}\label{couplin_matrix_deformed}
    M^{(R_{L}, R_{R})}\equiv     R_{L}^{T}M\overline{R_{R}}
\end{align}
satisfies the intertwiner equations with $(\rho_{\mathcal{C}}, \rho_{\widetilde{\mathcal{C}}})$, and has non-negative integer entries. We can now write
\begin{align}\label{Z_RLRR}
    Z^{(R_{L}; R_{R})}(\tau, \overline{\tau}) = \chi(\tau)^{T}M^{(R_{L}, R_{R})}\overline{\widetilde{\chi}(\tau)}
\end{align}
is a modular invariant. A clean sufficient condition is that $R_{L}$ commute with the left modular data and $R_{R}$ commute with the right modular data (in the same basis), i.e.,
\begin{align}
\begin{split}
    \left[\rho_{\mathcal{C}}(S), R_{L}\right] =& \left[\rho_{\mathcal{C}}(T), R_{L}\right] = 0,\\
    \left[\rho_{\widetilde{\mathcal{C}}}(S), R_{R}\right] =& \left[\rho_{\widetilde{\mathcal{C}}}(T), R_{R}\right] = 0.
\end{split}
\end{align}
Then for $\gamma\in \{S,T\}$, we have
\begin{align}
\begin{split}
    \rho_{\mathcal C}(\gamma)^{T}M^{(R_L,R_R)}
    =&\rho_{\mathcal C}(\gamma)^{T}R_L^{T}M\overline{R_R}\\
    =&R_L^{T}\rho_{\mathcal C}(\gamma)^{T}M\overline{R_R}\\
    =&R_L^{T}M,\overline{\rho_{\widetilde{\mathcal C}}(\gamma)}\overline{R_R}\\
    =&R_L^{T}M\overline{R_R}\overline{\rho_{\widetilde{\mathcal C}}(\gamma)}\\
    =& M^{(R_L,R_R)}\overline{\rho_{\widetilde{\mathcal C}}(\gamma)},
\end{split}
\end{align}
so $M^{(R_{L}, R_{R})}$ is again an intertwiner. If moreover $M^{(R_{L}, R_{R})}\in \mathbb{Z}_{\geq0}$, then \ref{Z_RLRR} is a bone fide modular invariant partition function.
\begin{comment}
This is easy to see, as shown below
\begin{align}
    \begin{split}
        S_{\mathcal{C}}M^{(R_{L}, R_{R})} =& \left( S_{\mathcal{C}}R_{L}^{T}\right)M\overline{R_{R}}\\
        =& \left(R_{L}^{T}S_{\mathcal{C}}\right)M\overline{R_{R}}\\
        =& R_{L}^{T}\left(M\overline{S_{\widetilde{\mathcal{C}}}}\right)\overline{R_{R}}\\
         =& \rho(h)^{T}M\overline{S_{\widetilde{\mathcal{C}}}}\overline{\widetilde{\rho}(\widetilde{h})}\\
        =& \rho(h)^{T}M\overline{\widetilde{\rho}(\widetilde{h})S_{\widetilde{\mathcal{C}}}}\\
        =& M^{(R_{L}, R_{R})}\overline{S_{\widetilde{\mathcal{C}}}},
    \end{split}
\end{align}
where we have made use of the intertwiner equation \ref{intertwiner_S}, and the fact that $\widetilde{\rho}(\widetilde{h})$ commutes with $S_{\widetilde{\mathcal{C}}}$. and similarly for $T$.
\end{comment}
\noindent This suggests the following practical algorithm for hunting modular invariants within the defect orbit:
\begin{enumerate}
    \item Compute the centralizers
    \begin{align}
        \begin{split}
            \mathcal Z_L\equiv&{R_L\in \mathrm{End}(\mathbb C^{I_{\mathcal C}}):[R_L,S_{\mathcal C}]=[R_L,T_{\mathcal C}]=0},\\
            \mathcal Z_R\equiv&{R_R\in \mathrm{End}(\mathbb C^{I_{\widetilde{\mathcal C}}}):[R_R,S_{\widetilde{\mathcal C}}]=[R_R,T_{\widetilde{\mathcal C}}]=0}.
        \end{split}
    \end{align}
    \item For any $R_{L}\in \mathcal{Z}_{L}$, and $R_{R}\in \mathcal{Z}_{R}$, form $M^{(R_{L}, R_{R})} = R_{L}^{T}M\overline{R_{R}}$. If $M^{(R_{L}, R_{R})}\in \mathbb{Z}_{\geq0}$, then it yields a modular invariant via \ref{Z_RLRR}.
\end{enumerate}
\noindent We note here that the permutation matrices arising from modular data symmetries typically lie in the centralizers (after fixing conventions), while the Verlinde diagonal operators generally do not, except in special cases as the identity or accidental degeneracies. The set of defect transformed matrices
\begin{align}
    \mathcal{O}(M)\equiv \{\rho(h)^{T}M\overline{\widetilde{\rho}(\tilde{h})}\vert h\in G_{\text{inv}}(\mathcal{C}), \widetilde{h}\in G_{\text{inv}}(\widetilde{\mathcal{C}})\}
\end{align}
 is the orbit of $M$ under the left-right action of $ G_{\text{inv}}(\mathcal{C})\times  G_{\text{inv}}(\widetilde{\mathcal{C}})$ on the space of gluing matrices. Each element of $\mathcal{O}(M)$ defines a, possibly defect–decorated, equatorial amplitude with the same modular covariance properties as the original one. When $\mathcal{C} = \widetilde{\mathcal{C}}$, and in particular in meromorphic cases with identical hemispheres, we often restrict to one–sided defect insertions by taking $\widetilde{h} = \text{id}$, in which case we have the gluing matrix
 \begin{align}
     M\longmapsto M^{(h,\mathrm{id})}=\rho(h)^{T}M,
 \end{align}
and the corresponding family is the left orbit of $M$ under $G_{\text{inv}}(\mathcal{C})$. In many interesting examples, we can explicitly enumerate $G_{\text{inv}}(\mathcal{C})$, write down $\rho(h)$ explicitly, and compute the orbit $\mathcal{O}(M)$, obtaining a finite set of modular covariant amplitudes that includes the original modular invariant $Z_{M}$, its defect analogues, and occasionally genuinely new modular invariants, i.e., new full CFT gluings, when the orbit intersects $\mathbb
{Z}_{\geq0}$-valued intertwiners.

%\subsubsection{Explicit examples\label{sec: D_4,1x2}}
\subsubsection{Example I: $(B_{r,1}, B_{7-r,1})$}\label{sec: B_r}
We now illustrate the formalism in a concrete and analytically tractable case. Consider the commutant pair $(B_{r, 1}, B_{7-r,1})$, where $B_{r,1} \cong \widehat{\mathfrak{so}}(2r+1)_{1}$. The conformal embedding $B_{r,1}\otimes B_{7-r,1}\subset E_{8,1}$ implies that the $E_{8,1}$ vacuum character admits a three-term bilinear decomposition into $B_{r,1}$ and $B_{7-r,1}$ characters. In this subsection, we therefore work with the holomorphic equatorial amplitude
\begin{align}
    Z^{\text{eq}}(\tau) \equiv \chi(\tau)^{T}M\widetilde{\chi}(\tau),
\end{align}
so that the output is a holomorphic function of $\tau$. The full torus partition function of the local meromorphic CFT is recovered at the end  as $Z_{E_{8}}(\tau, \overline{\tau}) = \vert\chi_{E_{8}}(\tau)\vert^{2}$. Each of these WZW models  $B_{r,1}$ and $B_{7-r,1}$ has three primaries. We fix the ordered bases
\begin{align}
    I_{\mathcal{C}} = \{\mathbb{1}, \psi, \sigma\},\qquad I_{\widetilde{\mathcal{C}}} = \{\widetilde{\mathbb{1}}, \widetilde{\psi}, \widetilde{\sigma}\},
\end{align}
with conformal weights
\begin{align}
    \begin{split}
         h_{\mathbb{1}} =& 0,\qquad h_{\psi} = \frac{1}{2},\qquad h_{\sigma} = \frac{2r+1}{16},\\
         \widetilde{h}_{\widetilde{\mathbb{1}}} =& 0,\qquad \widetilde{h}_{\widetilde{\psi}} = \frac{1}{2},\qquad \widetilde{h}_{\widetilde{\sigma}} = \frac{15-2r}{16}.
    \end{split}
\end{align}
Gluing them across the equator gives a simple but very instructive example of how multiplicity matrices, Verlinde lines, and half-gauging enter the equatorial viewpoint.
Here, $\psi$ is the Majorana fermion primary (order $2$) and $\sigma$ is the spin field. Note that 
\begin{align}
    h_{\sigma} + \widetilde{h}_{\widetilde{\sigma}} = 1,
\end{align}
so the product $\chi_{\sigma}^{B_{r,1}}(\tau)\chi_{\widetilde{\sigma}}^{B_{7-r,1}}(\tau)$, $(\sigma, \widetilde{\sigma})$ pair, contributes with integral total conformal weight, consistent with holomorphic gluing inside $E_{8,1}$. More generally, admissible coupling in a (non-chiral) equatorial pairing obey the usual spin-matching condition $h_{i} - \widetilde{h}_{j}\in \mathbb{Z}$; in the present conformal embedding situation, the bilinear pairing is fixed by the known $E_{8,1}$ decomposition. In this example, the equatorial pairing is multiplicity-free, i.e., each left label is paired with a unique right label. We take the bijection $\phi: I_{\mathcal{C}}\to I_{\widetilde{\mathcal{C}}}$ to be
\begin{align}
    \phi(\mathbb{1}) = \widetilde{\mathbb{1}},\qquad \phi(\psi) = \widetilde{\psi},\qquad \phi(\sigma) = \widetilde{\sigma},
\end{align}
so the gluing matrix is 
\begin{align}
    M_{ij} = \delta_{j,\phi(i)}.
\end{align}
Equivalently, the equatorial Hilbert space contains each left-right sector exactly once
\begin{align}\label{Hilbert_eq}
    \mathcal{H} = \bigoplus\limits_{i\in I_{\mathcal{C}}}\mathcal{H}_{i}\otimes\widetilde{\mathcal{H}}_{\phi(i)},
\end{align}
With this multiplicity matrix, we can immediately write down the equatorial torus amplitude that is defect-free as follows
\begin{align}
    \begin{split}
        Z^{\text{eq}} =& \chi_{\mathbb 1}^{B_{r,1}}(\tau)\chi_{\widetilde{\mathbb 1}}^{B_{7-r,1}}(\tau) + \chi_{\psi}^{B_{r,1}}(\tau)\chi_{\widetilde\psi}^{B_{7-r,1}}(\tau) + \chi_{\sigma}^{B_{r,1}}(\tau)\chi_{\widetilde\sigma}^{B_{7-r,1}}(\tau)\\
        =& \chi^{E_{8}}.
    \end{split}
\end{align}
Thus the equatorial gluing reproduces the standard conformal embedding decomposition of the $E_{8,1}$ vacuum character. the full non-chiral torus partition function of the local meromorphic theory is then
\begin{align}
    Z_{E_{8}}(\tau, \overline{\tau}) = \vert\chi_{E_{8}}(\tau)\vert^{2}.
\end{align}
We now implement defect insertions via the deformation of the coupling matrix \ref{couplin_matrix_deformed}. In this subsection, we will only use left insertions,  $(R_{L}, R_{R}) = (D_{\psi}, 1)$, so that $M'\equiv R^{T}M$ and 
\begin{align}
    Z^{(R)}(\tau) = \chi(\tau)^{T}M'\widetilde{\chi}(\tau).
\end{align}
To begin, consider the left Verlinde line associated with the simple current $\psi$, $(R_{L}, R_{R}) = (D{\psi},1)$. Its action is diagonal
\begin{align}
    (D_{\psi}){ii'}=\delta_{ii'},\frac{(S_{\mathcal{ C}})_{\psi i}}{(S{\mathcal{C}})_{\mathbb{1}, i}},\qquad i,i'\in I_{\mathcal{C}}.
\end{align}
For $B_{r,1}$ at level $1$,  $\psi$ has eigenvalues $+1$ on $\mathbb{1}$ and $\psi$, and $-1$ on $\sigma$. In the ordered basis $\{\mathbb{1}, \psi, \sigma\}$, 
\begin{align}
    \begin{split}
        D_{\psi} =& \text{Diag}\left(\frac{S_{\psi, \mathbb{1}}}{S_{\mathbb{1}, \mathbb{1}}}, \frac{S_{\psi, \psi}}{S_{\mathbb{1}, \psi}}, \frac{S_{\psi, \sigma}}{S_{\mathbb{1}, \sigma}}\right)\\
        =& \text{Diag}\left(+1, +1, -1\right).
    \end{split}
\end{align}
Therefore the defect-twined (left-only) equatorial amplitude reads
\begin{align}
    Z^{\psi}(\tau) =\chi_{\mathbb{1}}\widetilde\chi_{\widetilde{\mathbb 1}} + \chi_{\psi}\widetilde\chi_{\widetilde\psi} -\chi_{\sigma}\widetilde\chi_{\widetilde\sigma}.
\end{align}
This is a standard defect amplitude. It is a well-defined genus-one observable in the original theory with a $\psi$-line threaded along the temporal cycle, but it is not a partition function of a new full CFT because the effective coupling matrix has a negative entry.\\

\noindent Next, consider the anyon-permuting lines. In this family there are no non-trivial braided autoequivalences. Although, the fusion alone would allow swapping $\mathbb{1}\leftrightarrow \psi$, braided autoequivalences must preserve topological spine, $\theta_{\mathbb{1}} = 1\neq -1 = \theta_{\psi}$. Hence,
\begin{align}
    \text{Aut}^{\text{br}}(\mathcal{C}) = 1,    
\end{align}
so the only anyon-permuting  defect is the identity $g = \text{id}$ with $P_{g} = 1$, and therefore $g$ acts trivially on the gluing bijection yo yield
\begin{align}
    \phi(g(i)) = \phi(i).
\end{align}
We next half-gauging by averaging over the subgroup generated by the $\psi$-line,
\begin{align}
    H = \{\mathbb{1}, \psi\} \cong\mathbb{Z}_{2}.
\end{align}
The corresponding idempotent projector on the left character space is
\begin{align}
    \Pi_{H} = \frac{1}{2}\left(\mathbb{1} + D_{\psi}\right) = \text{Diag}\left(1,1,0\right).
\end{align}
Physically, inserting $\Pi_{H}$ projects the left-moving chiral dataonto the $\mathbb{Z}_{2}$-invariant subspace under the $\psi$-line. Applying $\Pi_{H}$ gives the half-gauged equatorial amplitude 
\begin{align}
    \begin{split}
        Z^{[\mathbb{Z}_{2}]}  =&\chi(\tau)^{T}\Pi_H^{T}M\widetilde\chi(\tau)\\
        =&\frac12\left(Z^{\mathrm{eq}}(\tau)+Z^{\psi}(\tau)\right)\\
        =& \chi_{\mathbb 1}\widetilde\chi_{\widetilde{\mathbb 1}}+\chi_{\psi}\widetilde\chi_{\widetilde\psi}.
    \end{split}
\end{align}
This object is modular-covariant in the sense explained earlier (it is obtained by averaging defect insertions on the left only), and it precisely selects the $\mathbb{Z}_{2}$-even left-moving sectors. It is important to note that this is not a full orbifold since the right chiral half is untouched and no twisted sectors are added on both cycles---rather, this is a one-sided projection that our equatorial formalism makes precise.

\subsubsection{Example II: $(D_{r,1}, D_{r,1})$}\label{sec: D_4,1x2}
Consider the commutant pair $(D_{r,1}, D_{r,1})$, where $D_{r,1}\cong\widehat{\mathfrak{so}}(2r)_{1}$. Each chiral half has four primaries, which we order as
\begin{align}
    I_{\mathcal{C}} = \{\mathbb{1}, \nu, \sigma, \widetilde{\sigma}\},
\end{align}
corresponding to the vacuum, vector, spinor, and conjugate spinor representations, respectively. Their conformal weights from the Sugawara formula is fixed to be
\begin{align}\label{D_4,1_conformal_weight}
    h_{\mathbb{1}} = 0,\qquad h_{\nu} = \frac{1}{2},\qquad h_{\sigma} = h_{\widetilde{\sigma}} = \frac{r}{8}.
\end{align}
At level-$1$, all four primaries are simple currents and hence, the group of invertible lines coincides with the full set of primaries
\begin{align}
    \text{Pic}(\mathcal{C}) = I_{\mathcal{C}},
\end{align}
with fusion group isomorphic to the center of $\text{Spin}(2r)$, and in particular, $\mathbb{Z}_{2}\times\mathbb{Z}_{2}$ for $r$ even and $\mathbb{Z}_{4}$ for $r$ odd.\\

\noindent
A key qualitative distinction as $r$ varies is the size of the braided autoequivalence group $\text{Aut}^{\text{br}}(\mathcal{C})$, which governs the existence of anyon-permuting invertible defects. For generic $r$, the only non-trivial braided permutation is the exchange $\sigma\leftrightarrow \widetilde{\sigma}$, whereas at $r = 4$ the theory is special since it is now the unique level-$1$ orthogonal theory where the three non-trivial primaries $\nu, \sigma, \widetilde{\sigma}$ can be permuted non-trivially while preserving the modular data. This is the RCFT manifestation of triality, enlarging $\text{Aut}^{\text{br}}$ to a non-trivial permutation group acting on $\{\nu, \sigma, \widetilde{\sigma}\}$. This makes $r = 4$ an interesting case where both Verlinde and anyon-permuting defects are simultaneously non-trivial. Triality originates from the exceptional outer automorphism group $\mathfrak{so}(8)$ \cite{Green:1987sp}. The Dynkin diagram of $D_{4,1}$ has an $S_{3}$ symmetry permuting its outer nodes, which permutes the vector, spinor, and conjugate spinor representations \cite{Knus2009TrialitarianAO}. In the affine $D_{4,1}$ theory, those are exactly the non-trivial primaries $\{\nu, \sigma, \widetilde{\sigma}\}$. From the modular data perspective, a permutation label defines a braided autoequivalence precisely when its permutation matri $P_{g}$ commutes with the modular representation \ref{P_S&T}. This condition is restrictive because $T$ encodes twists via conformal weights. The $r = 4$ theory is special because
\begin{align}
    h_{\nu} = h_{\sigma} = h_{\widetilde{\sigma}} = \frac{1}{2}.    
\end{align}
Hence, the three primaries have the same $T$-eigenvalues, eliminating the twist obstruction and allowing non-trivial permutations. In fact, at level-$1$, the $S$-matrix is also symmetric under permutations of $\{\nu, \sigma, \widetilde{\sigma}\}$ yielding a genuine
\begin{align}
    S_{3}\subset\text{Aut}^{\text{br}}(D_{4,1}).    
\end{align}
For the commutant pairing relevant to $c = 8$ meromorphic theory, we specialize henceforth to $r = 4$, i.e., the commutant pair $(D_{4,1}, D_{4,1})$, since $c(D_{r,1}) = r$ and the holomorphic $E_{8,1}$ embedding involves $D_{4,1}\times D_{4,1}$ so that $4+4 = 8$. The equatorial pairing is multiplicity-free with identity bijection
\begin{align}
    \phi(\mathbb{1}) = \widetilde{\mathbb{1}},\qquad \phi(\nu) = \widetilde{\nu},\qquad \phi(\sigma) = \widetilde{\sigma},\qquad \phi(\widetilde{\sigma}) = \overset{\scriptscriptstyle \sim\!\sim}{\sigma}.
\end{align}
The gluing matrix is simply $M_{ij} = \delta_{ij}$. Working in the holomorphic convention, the defect-free equatorial torus amplitude reads
\begin{align}
    Z^{\text{eq}} =&  =\chi_{\mathbb 1}(\tau)\widetilde\chi_{\mathbb 1}(\tau) +\chi_{\nu}(\tau)\widetilde\chi_{\nu}(\tau) +\chi_{\sigma}(\tau)\widetilde\chi_{\sigma}(\tau) +\chi_{\widetilde\sigma}(\tau)\widetilde\chi_{\widetilde\sigma}(\tau).
\end{align}
In the holomorphic $E_{8,1}$ theory, we also have the purely chiral branching under the conformal embedding $E_{8,1}\supset D_{4,1}\times D_{4,1}$. This bilinear is precisely the holomorphic decomposition of the $E_{8,1}$ vacuum character, with the full torus partition function of the local meromorphic theory being $\vert\chi_{E_{8}}(\tau)\vert^{2}$.\\

\noindent 
We can now think about TDL insertions along the temporal cycle. The $S$-matrix of $D_{4,1}$ reads
\begin{align}
    S = \frac{1}{2}\begin{pmatrix}
        1 & 1 & 1 & 1\\
        1 & 1 & \matminus1 & \matminus1\\
        1 & \matminus1 & 1 & \matminus1\\
        1 & \matminus1 &\matminus1 &1
    \end{pmatrix}.
\end{align}
The action of the Verlinde lines is diagonal in the character basis with $(D_{a})_{ii} = \tfrac{S_{ai}}{S_{0i}}$, and because the vacuum row satisfies $S_{0i} = \tfrac{1}{2}$ for all $i$, the eigenvalues are simply the signs in the corresponding $S$-row. Hence, we have
\begin{align}
    \begin{split}
        D_{\nu} =& \text{Diag}(+1, +1, -1,-1),\\
        D_{\sigma} =& \text{Diag}(+1, -1, +1, -1),\\
        D_{\widetilde{\sigma}} =& \text{Diag}(+1,-1, -1, +1).
    \end{split}
\end{align}
Next, let's pick a concrete triality element, e.g., the transposition $g = \left(\nu\ \sigma\right)$ that swaps $\nu$ and $\sigma$ while fixing $\mathbb{1}$ and $\widetilde{\sigma}$. Its permutation action on characters is implemented by a matrix $P_{g}$ that swaps the $\nu$ and $\sigma$ components of $\chi(\tau)$. Because this $g$ preserves both $S$ and $T$ at $r = 4$, $P_{g}$ commutes with the modular representation, and the corresponding one-sided automorphism insertion produce a genuine permutation modular invariant at genus one that reads
\begin{align}
    Z^{(1, g)} = \chi_{\mathbb{1}}\widetilde{\chi}_{\mathbb{1}} + \chi_{\sigma}\widetilde{\chi}_{\nu} + \chi_{\nu}\widetilde{\chi}_{\sigma} + \chi_{\widetilde{\sigma}}\widetilde{\chi}_{\widetilde{\sigma}}
\end{align}
This already contrasts sharply with the $(B_{r,1}, B_{7-r,1})$ example, where $\text{Aut}^{\text{br}}$ was trivial and no non-trivial permutation invariant arises.\\

\noindent
We now present an insertion where both components are non-trivial. Take $a = \nu$ and the same triality element $g = (\nu\ \sigma)$, and define the combined temoporal-cycle operator
\begin{align}
    R\equiv D_{\nu}P_{g}.
\end{align}
Acting on the left only, $(R_{L}, R_{R}) = (R,1)$, and using  $M = 1$, we obtain the following modular covariant torus amplitude
\begin{align}
    Z^{(\nu, (\nu\ \sigma))} = \chi_{\mathbb{1}}\widetilde{\chi}_{\mathbb{1}} + \chi_{\sigma}\widetilde{\chi}_{\nu} - \chi_{\nu}\widetilde{\chi}_{\sigma} - \chi_{\widetilde{\sigma}}\widetilde{\chi}_{\widetilde{\sigma}}.
\end{align}
This is clearly not a genuine RCFT since the coefficients fail non-negative integrality. The triality here produces a legitimate anyon-permuting defect, the Verlinde line produces a legitimate simple current defect, but combining them one-sidedly typically takes us out of the space of non-negative integer couplings.\\

\noindent 
Finally, consider the double-sided insertion $(R_{L}, R_{R}) = (R, R)$. In the holomorphic equatorial convention, the coupling deforms as 
$M^{(R,R)} = R^{T}MR$. With $M =1$ and $R = D_{\nu}P_{g}$, note that $D_{\nu}$ is a diagonal $\pm1$ matrix and $P_{g}$ is a permutation matrix, hence $R$ is orthogonal and $R^{T}R = 1$. Therefore, $M^{(R,R)} = M$, and the gluing returns to the original defect-free coupling,
\begin{align}
    Z^{(R;R)}(\tau, \overline{\tau}) = Z^{\text{eq}}(\tau, \overline{\tau}).
\end{align}
Physically, inserting the same invertible line on both hemispheres can be slid through the equator and annihilated, leaving the genus-one amplitude unchanged. In contrast, one-sided insertions are intrinsically chiral-asymmetric and naturally produce interface amplitudes.\\

\noindent 
For the half-gauging construction, choose the non-trivial order-two subgroup $H \subset \text{Pic}(D_{4,1})$, generated b $\nu$,
\begin{align}
    H = \{\mathbb{1}, \nu\}\cong \mathbb{Z}_{2},
\end{align}
with corresponding idempotent
\begin{align}
    \Pi_{H} = \tfrac{1}{2}\left(\mathbb{1} + D_{\nu}\right) = \text{Diag}(1,1,0,0).    
\end{align}
Applying this projector yields 
\begin{align}
    Z^{[H]} = \frac{1}{2}\left(Z^{\text{eq}} + Z^{\nu}\right),
\end{align}
where the $\nu$-twinned amplitude is 
\begin{align}
    Z^{\nu} = \chi_{\mathbb{1}}\widetilde{\chi}_{\mathbb{1}} + \chi_{\nu}\widetilde{\chi}_{\nu} - \chi_{\sigma}\widetilde{\chi}_{\sigma} - \chi_{\widetilde{\sigma}}\widetilde{\chi}_{\widetilde{\sigma}}.
\end{align}

%\subsubsection{Tetracritical Ising}

\subsubsection{Towards a classification}
We can classify what kind of object $Z^{(R_{L}; R_{R})}$ is, depending on the properties of the deformed gluing matrix $M^{(R_{L}, R_{R})}$.

\subsubsection*{Family $A$: new RCFT modular invariants}
This family consists of genuine RCFTs, those TDLs $(R_{L}, R_{R})$ such that $M^{(R_{L}, R_{R})}\in \text{Mat}(\mathbb{Z}_{\geq 0})$ and satisfies the usual positivity/symmetry requirements. These are genuinely new full CFTs with the same chiral halves but different left-right gluing.

\subsubsection*{Family $B$: defect amplitudes}
These are $(R_{L}, R_{R})$ for which $M^{(R_{L}, R_{R})}$ continues to intertwine the modular representations but fails to have non-negative integer entries. The corresponding $Z^{(R_{L} R_{R})}$ is then a consistent genus-one interface/defect amplitude of the original RCFT, rather than the partition function of a new bulk theory. In particular, one-sided insertions of Verlinde lines typically land in this class because they do not centralize the modular representation and can produce sign or phase reweightings of sector.\\

\noindent 
We now turn to the interesting case of the trivial action of TDLs. When is the defect invisible at genus-one, namely $M^{(R_{L}, R_{R})} = M$? This is completely equivalent to the no TDLs case. We had that 
\begin{align}\label{M_RLRR}
    M^{(R_{L}, R_{R})} = R_{L}^{T}M\overline{R_{R}},    
\end{align}
or equivalently
\begin{align}
    R_{L}^{T}M = M^{(R_{L}, R_{R})}\overline{R_{R}}^{-1}.
\end{align}
In the case of the trivial action, we have $M^{(R_{L}, R_{R})} = M$, and hence
\begin{align}\label{stab_condition}
    R_{L}^{T}M = M\overline{R_{R}}^{-1}.
\end{align}
Apart from the trivial choice $(R_{L}, R_{R}) = (1,1)$, any non-trivial pair satisfying \ref{stab_condition} defines a genuine symmetry line of the full RCFT specified by $M$. The line acts non-trivially on chiral data but preserves the bulk operator spectrum in the sense that the torus partition function is unchanged. At genus one, this is most cleanly expressed directly at the level of the gluing matrix. For an invertible line whose character-space action is $R$, the double-sided insertion deforms the couplings as 
\begin{align}\label{double_sided_action}
M\longmapsto M^{(R,R)}=R^{T}M\overline{R},
\end{align}
and invariance of the full RCFT under that symmetry line is the condition
\begin{align}\label{double_sided_invariance_revised}
R^{T}M\overline{R}=M.
\end{align}
In particular, for a Verlinde line $D_{a}$ this reads $D_{a}^{T}M\overline{D_{a}} = M$, which is the correct form in our conventions for left-right action \ref{double_sided_invariance_revised}.\\
\begin{comment}
Then the defect is a true global symmetry line of the full CFT defined by $M$. It is not just a symmetry of the chiral theory, and its action can also be absorbed into the gluing. A defect symmetry is a genuine global $0$-form symmetry if it preserves the bulk operator spectrum, i.e., the torus partition function is invariant under the action of the defect. For example, consider the TDL $D_{a}$ whose action on the bulk field $(i,j)$ is expressed as follows
\begin{align}
    D_{a}: (i, j)\mapsto \sum\limits_{i',j'}N^{i}_{ai}N^{j'}_{aj}(i',j').
\end{align}
Invariance under this defect requires 
\begin{align}
    D_{a}[Z(\tau, \overline{\tau})]= Z^{(D_{a}; D_{a})}(\tau, \overline{\tau}) =  Z(\tau, \overline{\tau}),
\end{align}
and since the action of a defect on $M$ is $M\mapsto D_{a}MD_{a}^{\dagger}$, the defect preserves the full RCFT iff
\begin{align}
    D_{a}MD_{a}^{\dagger} = M. 
\end{align}
\end{comment}

\noindent Motivated by \ref{stab_condition}, we define the stabilizer subgroup of a gluing $M$ under invertible defect actions by
\begin{align}\label{stab_def_revised}
    \text{Stab}(M)\equiv \left\{(R_{L}, R_{R})\in G_{\text{inv}(\mathcal{C})}\times G_{\text{inv}}(\widetilde{\mathcal{C})}\vert R_{L}^{T}M\overline{R_{R}} = M\right\}.
\end{align}
Every element of $\text{Stab}(M)$ is an invertible TDL whose action is an automorphism of the full CFT, not just the chiral algebra. This subgroup is therefore the natural candidate for the invertible $0$-form global symmetry group of the full theory visible in genus-one data. This viewpoint is closely related to the observation in \cite{Albert:2025umy} that certain defect actions can be absorbed into the gluing; in our language, such absorbability is precisely the stabilizer condition \ref{stab_def_revised}. Consider a tensor product chiral algebra $V^{\otimes n}$ with diagonal gluing $M = 1$, i.e., each left primary is paired with the identical right primary. The symmetric group $S_{n}$ permuting the tensor factors acts as an anyon-permuting symmetry on the chiral theory., including permutation matrices $P_{\sigma}$ on the character space. For the diagonal invariant, we have
\begin{align}
    P_{\sigma}^{T}M\overline{P_{\sigma}} = M,
\end{align}
so $(P_{\sigma}, P_{\sigma})\in \text{Stab}(M)$. Hence, permuting tensor factors is a canonical example of a non-trivial double-sided action that nevertheless leaves the genus-one coupling unchanged; equivalently, it is a symmetry of the full RCFT that can be recognized purely from the stabilizer condition.\\

\noindent Finally, in concrete families where  each chiral half has a small invertible-defect group (for example, generated by a simple current and an order-two anyon permutation), one can explicitly enumerate the basic generators on the left and right and organize the resulting deformations into a small number of structural classes: one-sided Verlinde insertions, one-sided automorphism insertions, their mixed products, and double-sided insertions (including stabilizers). We will make this bookkeeping explicit in the examples that follow.

\subsubsection*{Family I: Left-only defect transformations}
In this family, we insert an invertible defect only on the left $(R_{L}, R_{R}) = (R_{L}, 1)$, and this  yields the gluing matrix 
\begin{align}
    M^{(R_{L}, 1)} = R_{L}^{T}M.
\end{align}
Notice that modular covariance is automatic since the gluing matrix satisfies the intertwiner equations \ref{intertwiners}, and the resulting genus-one amplitude reads
\begin{align}
    Z^{(R_{L}; 1)}(\tau, \overline{\tau}) = \chi(\tau)^{T}R_{L}^{T}M\widetilde{\chi}(\overline{\tau}).
\end{align}
This is automatically modular covariant . However, it is a genuine RCFT partition function only in the exceptional situation that the deformed coupling matrix has non-negative integer entries, i.e., $M^{(R_{L}, 1)}\in \text{Mat}(\mathbb{Z}_{\geq 0})$. When this holds, the left defect reshuffles the left labels in a way that still defines a consistent bulk Hilbert space, yielding a new modular invariant. Fir example, familiar simple current modular invariants arise precisely from such integral deformations in appropriate theories. When the entries are no longer non-negative, the result is not a new bulk CFT, it is a defect-decorated torus amplitude of the original RCFT, with $M^{(R_{L}, 1)}$ interpreted as an interface coupling rather than a multiplicity matrix. In particular, one-sided Verlinde insertions generically produce sign/phase reweightings and therefore land in this “interface” regime rather than in Family $A$.\\

\noindent 
A clean illustration is provided by the commutant pair $(B_{1,1}, B_{6,1})$, a special case of the family discussed in \S\ref{sec: B_r}, with central charges $(c, \widetilde{c}) = \left(\tfrac{3}{2}, \tfrac{13}{2}\right)$, whose equatorial gluing reconstructs the $c = 8$ holomorphic $E_{8,1}$ theory on the equatorial circle after gluing as shown in detail in \S\ref{sec: B_1,1, B_6,1}. We place  $B_{1,1}$ on the upper hemisphere and $B_{6,1}$ on the lower, and start from the multiplicity-free holomorphic gluing $M_{ij} = \delta_{j, \phi(i)}$. Inserting the fermionic simple current $\psi\in \text{Pic}(B_{1,1})$ as a left Verlinde line on the upper hemisphere only, we have
\begin{align}
    (R_{L}, R_{R}) = (D_{\psi}, 1).    
\end{align}
The resulting defect-decorated amplitude $Z^{(\psi; 1)}$ whose transformed gluing matrix $M^{(D_{\psi}, 1)} = D_{\psi}^{T}M$ develops relative signs between the $\sigma$-sectors. Hence, $M^{(D_{\psi},1)}\notin \text{Mat}(\mathbb{Z}_{\geq 0})$, so the one-sided insertion does not define a new RCFT. Rather, it defines a non-trivial Verlinde line partition function of the original $E_{8,1}$ theory. This is modular covariant, but interpreted as a defect observable rather than a new modular invariant bulk partition function.

\subsubsection*{Family II: Right-only defect transformations}
In this family, we insert an invertible defect only n the right $(R_{L}, R_{R}) = (1, R_{R})$, and this yields the gluing matrix 
\begin{align}\label{familyII_gluing}
    M^{(1, R_{R})} = M\overline{R_{R}}.
\end{align}
As in family $I$, the resulting genus-one amplitude reads
\begin{align}
    Z^{(1;R_{R})}(\tau, \overline{\tau}) = \chi(\tau)^{T}M\overline{R_{R}}\widetilde{\chi}(\overline{\tau})
\end{align}
and is automatically modular covariant. Now,  to be a new RCFT, we require $M^{(1, R_{R})}\in \text{Mat}(\mathbb{Z}_{\geq 0})$. Here, the roles of the two chiral halves are exchanged relative to family $I$. Concretely, $R_{R}$ may be a right Verlinde line insertion $\widetilde{D}_{\widetilde{a}}$, a right anyon permutation $\widetilde{P}_{\widetilde{g}}$, or a product $\widetilde{D}_{\widetilde{a}}\widetilde{P}_{\widetilde{g}}\in G_{\text{inv}(\widetilde{\mathcal{C}})}$. Physically, these correspond to threading a topological line defect on the southern (right) hemisphere alone. Generically, right-only transformations produce defect-decorated amplitudes rather than new modular invariants, because $\overline{R_{R}}$ typically reweights or permutes right labels in a way that does not preserve non-negative integer multiplicities in \ref{familyII_gluing}. The exceptional cases are those satisfying the integrality condition where the right insertion can be reinterpreted as a new admissible gluing of the same chiral halves.

%To illustrate right–only action in a structurally non-trivial way, it is convenient to look at a commutant pair where the “larger” side has a non-trivial simple–current structure. A good example is the commutant pair $(A_{2,1}^{\otimes 2}, V_{39})$ with central charges $(c, \widetilde{c}) = (4, 20)$ considered in section \ref{sec: A_2,1, V_39}. Here, the upper hemisphere caries $A_{2,1}^{\otimes2}$, while the lower hemisphere carries a meromorphic $c = 24$ VOA denoted by $V_{39}$. \NBU{Check computations again}

\subsubsection*{Family III: Symmetric double-sided transformations}
For the self-dual case $\mathcal{C} = \widetilde{\mathcal{C}}$, and in particular, in the case when the two hemispheres are identical, we have $(R_{L}, R_{R}) = (R,R)$, and this yields the gluing matrix
\begin{align}
    M^{(R,R)} = R^{T}M\overline{R}.
\end{align}
A special subcase is when the insertion is torus-invisible, i.e., it leaves the coupling unchanged
\begin{align}\label{familyIII_invisible}
M^{(R,R)}=M \qquad \Longleftrightarrow \qquad R^{T}M = M\overline{R}^{-1}.
\end{align}
This motivates the stabilizer group of a given modular invariant $M$,
\begin{align}
    \text{Stab}(M) = \left\{R\in G_{\text{inv}}(\mathcal{C})\vert R^{T}M = M\overline{R}\right\}.
\end{align}
Elements of $\text{Stab}(M)$ are precisely those invertible defect lines that act as \emph{true global symmetries of the full RCFT} defined by $M$, rather than merely symmetries of a single chiral half. When $R\in\text{Stab}(M)$, the double-sided insertion is physically invisible at the level of the genus-one partition function, yet it encodes a non-trivial global symmetry realized by a topological line in the full theory.\\

\noindent A natural playground for this family is provided by self-dual commutant pair such as  $(D_{4,1}, D_{4,1})$ in \S\ref{sec: D_4,1x2}. In this example, both hemispheres carry the same WZW model, and the equatorial theory is built from a diagonal–type gluing matrix $M$. Let $R\in G_{\text{inv}(\mathcal{C}})$ be an invertible line. Consider an invertible line $R\in G_{\text{inv}}(\mathcal{C})$ corresponding to a permutation autoequivalence or an integer spin simple current. Inserting $R$ on both sides gives $(R_{L}, R_{R}) = (R,R)$ and the gluing matrix $M^{(R, R)} = R^{T}M\overline{R}$. For diagonal or extension–type invariants, one typically finds $M^{(R, R)} = M$, or in other words, $(R, R)$ lies in the stabilizer $\text{Stab}(M)$. This realizes this family since the defect is non-trivial at the chiral level but acts as a genuine global symmetry of the full RCFT, leaving the torus partition function unchanged. 

\subsubsection*{Family IV: General double-sided transformations}
In this case, $(R_{L}, R_{R})$ is arbitrary, and we have the general gluing matrix \ref{M_RLRR}. In general, the matrices $R_{L}$ and $R_{R}$ contain phases from simple currents and permutations from braided autoequivalences. When these act on the entries of $M$, they need not preserve nonnegativity or integrality. Hence, the gluing matrix $M^{(R_{L}, R_{R})}$ will often have complex or negative entries. For such choices of $(R_{L}, R_{R})$, the resulting object is a defect torus amplitude which is a modular covariant function but is not the partition function of a full RCFT, because it cannot be interpreted as counting multiplicities of sectors in $\mathcal{H}_{i}\otimes\widetilde{\mathcal{H}}_{j}$. Although this family yields modular covariant partition functions, these are typically not integer-valued unless $(R_{L}, R_{R})$ is in a specific finite subgroup of the same simple currents and permutation lines. Integrality is preserved only when $(R_{L}, R_{R})$ lies in a finite subgroup generated by integral actions, i.e., those elements for which both $R_{L}$ and $R_{R}$ are themselves integer-valued matrices or act by phases that cancel between the left and right sides. Concretely, this occurs when both $R_{L}, R_{R}$ come from simple currents of integer conformal spin, so $D_{a}$ has real eigenvalues $\pm1$, both are in a subgroup of permutation autoequivalences for which $P_{g}$ is exactly the permutation matrix, or the phases from $R_{L}^{T}$ and $\overline{R_{R}}$ cancel on every nonzero entry of $M_{ij}$. Only in these cases does \ref{M_RLRR} remain a matrix with non-negative integer entries and therefore define a new modular invariant, i.e., the partition function of a genuine full RCFT. In all other cases, the matrix remains a valid intertwiner for the modular group but cannot be interpreted as a gluing matrix of a full CFT. To produce a genuinely RCFT, additional positivity constraints must hold. However, this is the most general deformation of invertible lines. These transformations mix left and right labels asymmetrically and produce new modular invariants when integrability holds, non-trivial defect torus amplitudes when integrality fails, and maps between different modular invariants in the same defect orbit, i.e.
\begin{align}
    M\mapsto \mathcal{O}(M)\equiv \{\rho(h_{L})^{T}\overline{\widetilde{\rho}(h_{R})}\}.
\end{align}
This family is crucial since it describes the full orbit of a modular invariant under the left-right action of $G_{\text{inv}}(\mathcal{C})\times G_{\text{inv}}(\widetilde{\mathcal{C}})$, and organizes all possible defect-decorated amplitudes consistent with modularity. Within this orbit, some points correspond to genuine new modular invariants (Family A), while most correspond to defect amplitudes (Family B). The orbit viewpoint therefore organizes all defect-decorated amplitudes compatible with modularity, and makes precise when two modular invariants are related by invertible defect actions.

\section{Orbifold Construction}
When an RCFT admits a finite group of invertible TDLs, these lines implement an ordinary $0$-form global symmetry \cite{Cheng:2020rpl, Chang:2022hud}. Gauging such a symmetry produces an orbifold theory. On the torus, gauging is most conveniently formulated in terms of twisted partition functions $Z_{h,g}(\tau)$, labeled by a pair $(h,g)\in G\times G$ specifying the holonomies of the background $G$-bundle around the spatial and temporal cycles \cite{Dijkgraaf:1989hb}. The orbifold partition function is built by two operations. First, in each sector, we project onto $G$-invariant states since we want our theory to include all the sectors of the parent theory that are invariant under the group $G$ by which we quotient the parent theory. Second, we sum over all twisted sectors. We briefly recall this standard construction in a form adapted to the defect-line viewpoint used throughout this paper. Also see \cite{Robbins:2019zdb} for more details.\\

\noindent 
Let $\mathcal{H}_{P}$ be the Hilbert space of the parent theory, and let $\rho: G\to U(\mathcal{H}_{P})$ denote the unitary representation of the symmetry group implemented by the invertible defect lines. The contribution of the untwisted sector to the orbifold theory is obtained by projecting the parent Hilbert space onto the $G$-invariant subspace. The projector is the group average that reads
\begin{align}
    P_{G} = \frac{1}{\vert G\vert}\sum\limits_{g\in G}\rho(g).
\end{align}
The corresponding untwisted-sector contribution to the orbifold partition function is therefore
\begin{align}
    Z_{1} = \text{Tr}_{\mathcal{H}_{P}}\left(P_{G}q^{L_{0} - \frac{c}{24}}\overline{q}^{\overline{L}_{0} - \frac{c}{24}}\right) = \frac{1}{\vert G\vert}\sum\limits_{g\in G}Z_{1,g}(\tau, \overline{\tau}),
\end{align}
where we have introduced the “twining” traces defined as follows
\begin{align}
    Z_{1,g} \equiv \text{Tr}_{\mathcal{H}_{P}}\left(\rho(g)q^{L_{0} - \frac{c}{24}}\overline{q}^{\overline{L}_{0} - \frac{c}{24}}\right).
\end{align}
In the defect language, $Z_{1, g}$ is simply the torus amplitude of the parent theory with a $g$-defect threaded along the temporal cycle. Now, to obtain a modular invariant gauged theory, we must also include sectors twisted by non-trivial holonomy around the spatial cycle. For each $h\in G$, let $\mathcal{H}_{h}$ denote the Hilbert space of states on the circle with
$h$-twisted boundary conditions. The symmetry elements $g\in G$ act on $\mathcal{H}_{h}$ by operators $\rho_{h}(g)$, and we define the general twisted partition function as follows
\begin{align}
    Z_{h,g}(\tau, \overline{\tau}) \equiv \text{Tr}_{\mathcal{H}_{h}}\left(\rho_{h}(g)q^{L_{0} - \frac{c}{24}}\overline{q}^{\overline{L}_{0} - \frac{c}{24}}\right).
\end{align}
For fixed $h$, the orbifold prescription again requires projection onto $G$-invariant states in the $h$-sector. This gives
\begin{align}
    Z_{h}(\tau, \overline{\tau}) = \frac{1}{\vert G\vert}\sum\limits_{g\in G}Z_{h, g}(\tau, \overline{\tau}).
\end{align}
Finally, the full orbifold partition function is obtained by summing over all twisted sectors as follows
\begin{align}
    Z_{\text{orb}}(\tau, \overline{\tau}) =& \sum\limits_{h\in G}Z_{h}(\tau, \overline{\tau})     = \frac{1}{\vert G\vert}\sum\limits_{\substack{h,g\in G\\ hg = gh}}Z_{h,g}(\tau, \overline{\tau}) .
\end{align}
The collection $\{Z_{h,g}\}$ is closed under modular transformations because $\text{SL}(2, \mathbb{Z})$ acts on the homology cycles of the torus and hence on the pair of holonomies $(h,g)$ \cite{Dijkgraaf:1989hb}. Concretely, for a modular transformation $\gamma=\begin{psmallmatrix} a&b\\ c&d\end{psmallmatrix}$ the spatial and temporal cycles are sent to linear combinations of the original ones, and the corresponding holonomies are reshuffled accordingly. In the simplest anomaly-free situation with no discrete torsion and no 't Hooft anomaly, this implies that modular transformations permute the twisted traces among themselves
\begin{align}
    Z_{h,g}(\gamma\cdot\tau, \overline{\gamma\cdot \tau}) = Z_{h^{a}g^{c}, h^{b}g^{d}}(\tau, \overline{\tau}),
\end{align}
and in particular, the $S$-move exchanges the cycles, giving $Z_{h,g}\mapsto Z_{g,h^{-1}}$, and the $T$-move shifts the temporal holonomy by the spatial one, giving $Z_{h,g}\mapsto Z_{h,hg}$. Because the orbifold partition function is the uniform average over all $(h,g)$, this closure immediately guarantees that $Z_{\text{orb}}$ is modular invariant. For the symmetry group $G = \mathbb{Z}_{2} = \{1, u\}$, we have $\vert G\vert = 2$, and we have the following four sectors in total 
\begin{align}
    (h,g)\in \{(1,1), (1,u), (u,1), (u,u)\},
\end{align}
and the orbifold partition function takes the form
\begin{align}
    Z_{\text{orb}}(\tau, \overline{\tau}) = \frac{1}{2}\left(Z_{1,1} + Z_{1,u} + Z_{u,1} + Z_{u,u}\right).
\end{align}

\subsection{$B_{r,1}$ orbifolds}
We now illustrate the general orbifold formalism in the simplest family of examples relevant for our later commutant pair constructions, namely the level-one $B$-series WZW models $B_{r,1}\cong \widehat{\mathfrak{so}}(2r+1)_{1}$. These theories have three primaries, $\{\mathbb{1}, \psi, \sigma\}$, that denote the vacuum, the invertible simple current, and the non-invertible primary, and they admit a natural $\mathbb{Z}_{2}$ global symmetry generated by the $\psi$ line. Gauging this symmetry produces a $\mathbb{Z}_{2}$ orbifold. In this subsection, we construct the twisted traces $Z_{h,g}$, explain how modular covariance assembles them into a consistent orbifold partition function, and clarify that only invertible topological defect lines are encoded in these sector traces, in contrast to the non-invertible defect $\sigma$.\\

\noindent We begin with the lowest central charge commutant theory $B_{1,1}$. The torus partition function of this theory reads
\begin{align}
    Z_{B_{1,1}}(\tau, \overline{\tau}) = Z^{\mathbb{1}}(\tau,\overline{\tau}) = \vert \chi_{0}\vert^{2} + \vert\chi_{\frac{1}{2}}\vert^{2} + \vert\chi_{\frac{3}{16}}\vert^{2},
\end{align}
where $(h_{\mathbb{1}}, h_{\psi}, h_{\sigma}) = (0,\tfrac{1}{2}, \tfrac{3}{16})$. The $\mathbb{Z}_{2}$ global symmetry acts trivially on $\mathbb{1}$ and $\sigma$ but flips the sign of $\psi$. In character language, this means that the symmetry operator acts with eigenvalue $+1$ on $\mathbb{1}$ and $\sigma$ sectors and eigenvalue $-1$ on the $\psi$ sector. $\rho(\psi)$ actson the sector $i$ by eigenvalue $\lambda_{\psi}(i) = \tfrac{S_{\psi i}}{S_{0i}}$, and for $B_{1,1}$, this gives $(\lambda_{\psi}(\mathbb{1}), \lambda_{\psi}((\psi), \lambda_{\psi}((\sigma)) = (+1, +1, -1)$. The two untwisted-sector partial traces with no spatial twist, and with or without a temporal insertion, read
\begin{align}
   \begin{split}
        Z_{0,0} =& \vert \chi_{0}\vert^{2} + \vert\chi_{\frac{1}{2}}\vert^{2} + \vert\chi_{\frac{3}{16}}\vert^{2},\\
    Z_{0,1} =&   \vert \chi_{0}\vert^{2} + \vert\chi_{\frac{1}{2}}\vert^{2} - \vert\chi_{\frac{3}{16}}\vert^{2}.
   \end{split}
\end{align}
Modular covariance now determines the remaining sector traces once we specify the modular transformation properties of the characters. Under $S: \tau\mapsto -\tfrac{1}{\tau}$, the characters of the $B_{1,1}$ theory transform as follows
\begin{align}
    \begin{pmatrix}
        \chi_{0}\\
        \chi_{\frac{1}{2}}\\
        \chi_{\frac{3}{16}}
    \end{pmatrix}\longrightarrow \begin{pmatrix}
        \frac{1}{2} & \frac{1}{\sqrt{2}} & \frac{1}{2}\\
        \frac{1}{\sqrt{2}} & 0 & \matminus\frac{1}{\sqrt{2}}\\
        \frac{1}{2} & \matminus\frac{1}{\sqrt{2}} & \frac{1}{2}
    \end{pmatrix}\begin{pmatrix}
        \chi_{0}\\
        \chi_{\frac{1}{2}}\\
        \chi_{\frac{3}{16}}
    \end{pmatrix},
\end{align}
which follow the transformation rule
\begin{align}
    \chi_{i}\left(-\frac{1}{\tau}\right) = \sum\limits_{j}S_{ij}\chi_{j}(\tau).
\end{align}
On the other hand, the twisted traces obey the standard reshuffling rules under modular transformations
\begin{align}
    \begin{split}
        S:& Z_{h,g}\mapsto Z_{g,h^{-1}},\\
        T:& Z_{h,g}\mapsto Z_{h,hg}.
    \end{split}
\end{align}
For $G = \mathbb{Z}_{2}$, we have $h^{-1} = h$, so $S$ simply exchanges the two labels. The key idea here is that the full orbifold theory includes all twisted sectors, and modular invariance is only restored when all of them are included together. The $\widehat{\mathfrak{so}}(2r+1)_{1}$ characters have the following modular form expressions \cite{DiFrancesco:1997nk}
\begin{align}\label{B_r_characters}
    \begin{split}
        \chi_{\hat{\omega}_{0}}(\tau) =& \frac{1}{2}\left(\frac{\theta_{3}^{r + \frac{1}{2}} + \theta_{4}^{r + \frac{1}{2}}}{\eta^{r + \frac{1}{2}}}\right)(\tau),\\
        \chi_{\hat{\omega}_{1}}(\tau) =& \frac{1}{2}\left(\frac{\theta_{3}^{r + \frac{1}{2}} - \theta_{4}^{r + \frac{1}{2}}}{\eta^{r+ \frac{1}{2}}}\right)(\tau),\\ 
        \chi_{\hat{\omega}_{2}}(\tau) =& \frac{1}{\sqrt{2}}\left(\frac{\theta_{2}}{\eta}\right)^{r + \frac{1}{2}}(\tau)
    \end{split}
\end{align}
Here we work with $r = 1$ characters. Under the T-transform, the characters behave as follows
\begin{align}
    \begin{split}
        \chi_{0}(\tau + 1) =& e^{-\frac{\pi i}{8}}\chi_{0}(\tau),\\
        \chi_{\frac{1}{2}}(\tau + 1) =& e^{\frac{7\pi i}{8}}\chi_{\frac{1}{2}}(\tau),\\
        \chi_{\frac{3}{16}}(\tau + 1) =& e^{\frac{\pi i}{4}}\chi_{\frac{3}{16}}(\tau).
    \end{split}
\end{align}
The twisted sector partial traces read
\begin{align}
    \begin{split}
        Z_{1,0}(\tau, \overline{\tau}) =& Z_{0,1}\left(-\frac{1}{\tau}, -\frac{1}{\overline{\tau}}\right)\\
        =& \chi_{0}\overline{\chi}_{\frac{1}{2}} + \chi_{\frac{1}{2}}\overline{\chi}_{0}
 + \vert\chi_{\frac{13}{16}}\vert^{2},\\
 Z_{1,1}(\tau, \overline{\tau}) =& Z_{1,0}(\tau - 1, \overline{\tau} - 1)\\
=& -\chi_{0}\overline{\chi}_{\frac{1}{2}} - \chi_{\frac{1}{2}}\overline{\chi}_{0}  + \vert\chi_{\frac{3}{16}}\vert^{2}.
\end{split}
\end{align}
From the partial trace sector expressions, we find the untwisted and twisted partition functions to read
\begin{align}
    \begin{split}
        Z_{0} =& \frac{1}{2}\left(Z_{0,0} + Z_{0,1}\right) = \vert\chi_{0}\vert^{2} + \vert\chi_{\frac{1}{2}}\vert^{2},\\
        Z_{1} =& \frac{1}{2}\left(Z_{1,1}  + Z_{1,0}\right) = \vert \chi_{\frac{3}{16}}\vert^{2}.
    \end{split} 
\end{align}
Hence, the orbifold partition function, which is the sum of the untwisted and the twisted partition functions, reads
\begin{align}
    Z_{\text{orb}}(\tau, \overline{\tau}) =& Z_{0}(\tau, \overline{\tau}) + Z_{1}(\tau, \overline{\tau})\\
    =& \vert\chi_{0}\vert^{2} + \vert \chi_{\frac{3}{16}}\vert^{2} + \vert\chi_{\frac{1}{2}}\vert^{2}\\
    =& Z_{B_{1,1}}(\tau, \overline{\tau}).
\end{align}
Thus the $\mathbb{Z}_{2}$ orbifold of $B_{1,1}$ reproduces the original theory. This is the familiar “self-orbifold” phenomenon also encountered in closely related settings, for instance, in the critical Ising model \cite{Robbins:2019zdb}, and it provides an especially clean illustration of how modular covariance among the $Z_{h,g}$ sectors enforces consistency.\\

\noindent
It is useful to emphasize how the orbifold sector traces encode the defect partition functions associated with invertible TDLs. In our normalization, the untwisted partial trace with no insertion is proportional to the defect-free amplitude, which reads
\begin{align}
    Z^{\mathbb{1}}(\tau, \overline{\tau}) = 2Z_{0,0}(\tau, \overline{\tau}),
\end{align}
while the untwisted trace with the non-trivial group element inserted along the temporal cycle is proportional to the $\psi$-defect amplitude, which reads
\begin{align}
    Z^{\psi}(\tau, \overline{\tau}) = 2Z_{0,1}(\tau, \overline{\tau}).
\end{align}
We also note that $Z^{\psi}$ can be obtained by taking the difference between the untwisted and twisted partition functions as
\begin{align}
   Z^{\psi}(\tau, \overline{\tau})  = Z_{0}(\tau, \overline{\tau})  - Z_{1}(\tau, \overline{\tau}).
\end{align}
Hence, the defect partition functions resulting from the insertion of the identity line and the invertible $\psi$ line corresponding to the global $\mathbb{Z}_{2}$ symmetry are encoded in the untwisted sector partial traces of the full orbifold theory, which here is the $B_{1,1}$ theory itself. We will repeatedly make use of the fact that orbifold sector traces are naturally adapted to invertible symmetry lines, since the orbifold construction is itself defined by summing and projecting with respect to those invertible defects. By contrast, the defect partition function associated with the non-invertible line $\sigma$ is not captured by the $\mathbb{Z}_{2}$ orbifold sector data. This is not an accident. The orbifold sectors are built from the group $G$ of invertible lines, and hence, they only “see” defects that implement genuine symmetry actions. The line $\sigma$ does not define a group-like symmetry defect since its fusion is not closed within $\{\mathbb{1}, \sigma\}$, and this is not included in a group orbifold sum. This can be seen from the fusion rules of $B_{1,1}$, which read
\begin{align}\label{fusion_rules_B_1,1}
    \begin{split}
        \mathbb{1}\otimes \mathbb{1} =& \mathbb{1},\qquad        \mathbb{1}\otimes\sigma = \sigma,\qquad
        \mathbb{1}\otimes \psi = \psi,\\
        \sigma\otimes\sigma =& \mathbb{1} \oplus \psi,\qquad
        \sigma\otimes\psi = \psi\otimes \sigma = \sigma,\\
        \psi\otimes\psi =& \mathbb{1}.
    \end{split}
\end{align}
Notice here that the $\mathbb{1}$ and $\psi$ fusion rules yield back $\mathbb{1}$ and $\psi$, but the $\sigma$ rule does not. This is similar to the critical Ising model where the primaries $\mathcal{P} = \{\mathbb{1}, \sigma, \epsilon\}$ have similar fusion rules with $\sigma\otimes\epsilon = \epsilon\otimes\sigma = \sigma$. The critical Ising is nothing but the $B_{0,1}$ RCFT, and hence, this $\sigma$-line is something we can expect for all of the $B_{r,1}$ RCFTs. Due to this behaviour, the defect partition function $Z^{\sigma}$ cannot be expressed in terms of the $\mathbb{Z}_{2}$ orbifold sectors, i.e.
\begin{align}
    Z^{\sigma} \neq a_{1}Z_{0,0} + a_{2}Z_{0,1} + a_{3}Z_{1,0} + a_{4}Z_{1,1},\ \ \ \forall a_{i}.
\end{align}
The orbifold formalism is simply not designed to encode non-invertible defects of this type. Based on the fusion rules \ref{fusion_rules_B_1,1}, we can write the sets of all possible partial traces $Z_{L_{1}, L_{2}}^{L_{3}}$ from various insertions of topological lines $L_{1}$ and $L_{2}$ to be
\begin{align}\label{L1,2,3list}
    Z_{L_{1}, L_{2}}^{L_{3}} = \left\{Z^{\mathbb{1}}_{\mathbb{1}, \mathbb{1}}, Z^{\psi}_{\mathbb{1}, \psi}, Z^{\psi}_{\psi, \mathbb{1}}, Z^{\mathbb{1}}_{\psi, \psi}, Z^{\sigma}_{\mathbb{1}, \sigma}, Z^{\sigma}_{\sigma, \mathbb{1}}, Z^{\sigma}_{\psi, \sigma}, Z^{\sigma}_{\sigma, \psi}, Z^{\mathbb{1}}_{\sigma, \sigma}, Z^{\psi}_{\sigma, \sigma}\right\}.
\end{align}
The notation $Z_{L_{1}, L_{2}}^{L_{3}}$ should be read as follows. We consider the torus with two defect insertions, a line $L_{1}$ threaded along, say, the temporal cycle and another line $L_{2}$ threaded along the spatial cycle. Now, since the two lines intersect once, there is a local junction datum at the crossing. In the fusion-category language, the crossing can be resolved into a sum over intermediate channels, and the label $L_{3}$ records which topological lines flow in the resolved channel, or equivalently, which primary appears in the fusion of $L_{1}$ and $L_{3}$ at the junction. In this sense $Z_{L_{1}, L_{2}}^{L_{3}}$ is a refinement of the usual twisted trace $Z_{h,g}$ since it remembers not only the holonomies $(L_{1}, L_{2})$ but also the junction channel $L_{3}$. From the fusion rules \ref{fusion_rules_B_1,1}, we can immediately enumerate, for each ordered pair $(L_{1}, L_{2})$, the possible resolved channels $L_{3}\in L_{1}\otimes L_{2}$ are shown in table \ref{tab:L_123}.
\begin{table}[htb!]
    \centering
    \begin{tabular}{|c||c|c|c|c|c|c|}
    \hline
    $(L_{1}, L_{2})$ &  $(\mathbb{1}, \mathbb{1})$ &  $(\mathbb{1}, \psi)$ or $(\psi, \mathbb{1})$ &  $(\psi, \psi)$ & $(\mathbb{1}, \sigma)$ or $(\sigma, \mathbb{1})$ & $(\psi, \sigma)$ or $(\sigma, \psi)$ & $(\sigma, \sigma)$\\
    \hline
     $L_{3}$  & $\mathbb{1}$ & $\psi$ & $\mathbb{1}$ & $\sigma$ & $\sigma$ & $\{\mathbb{1}, \psi\}$\\
     \hline
    \end{tabular}
    \caption{Possible resolved channels $L_{3}$ for an ordered pair $(L_{1}, L_{2})$ or topological lines.}
    \label{tab:L_123}
\end{table}
This explains \ref{L1,2,3list}, and exactly why this list contains only ten elements, since everything is forced except the $(\sigma, \sigma)$ sector, which splits into two refined traces depending on whether the $\mathbb{1}$ or $\psi$ channel propagates. We can alternatively package this as 
\begin{align}
    Z_{L_{1}, L_{2}}(\tau, \overline{\tau}) = \sum\limits_{L_{3}\subset L_{1}\otimes L_{2}}Z_{L_{1}, L_{2}}^{L_{3}}(\tau, \overline{\tau}),
\end{align}
where $Z_{L_{1}, L_{2}}$ is the unrefined twisted partition function that only remembers two cycle insertions, and $Z_{L_{1}, L_{2}}^{L_{3}}$ is the refined partition function that is compatible with the fusion. In particular, we notice that 
\begin{align}
    Z_{\sigma, \sigma} = Z_{\sigma, \sigma}^{\mathbb{1}} + Z_{\sigma, \sigma}^{\psi},
\end{align}
while the other $(L_{1}, L_{2})$ has only one allowed channel and hence, no non-trivial refinement. The modular transformations of these twisted partition functions were worked out in \cite{Perez-Lona:2023djo} to be
\begin{align}\label{Zl123_SandT}
    \begin{split}
        Z^{L_{3}}_{L_{1}, L_{2}}(\tau+1, \overline{\tau} + 1) =& \sum\limits_{L_{4}}\widetilde{K}^{\overline{L}_{1}, L_{2}}_{\overline{L}_{2}, L_{1}}\left(\overline{L}_{3}, \overline{L}_{4}\right)Z^{L_{2}}_{L_{1}, L_{4}}(\tau, \overline{\tau}),\\
        Z^{L_{3}}_{L_{1}, L_{2}}\left(-\frac{1}{\tau}, -\frac{1}{\overline{\tau}}\right) =& \sum\limits_{L_{4}}\widetilde{K}^{\overline{L}_{1}, \mathbb{1}}_{\overline{L}_{4}, L_{2}}\left(\overline{L}_{2}, L_{1}\right)\widetilde{K}^{\overline{L}_{1}, L_{2}}_{\overline{L}_{2}, L_{1}}\left(\overline{L}_{3}, \overline{L}_{4}\right)Z^{L_{4}}_{L_{2}, \overline{L}_{1}}(\tau, \overline{\tau}).
    \end{split}
\end{align}
Here, we have that $\overline{L}$ denotes the dual in the fusion category, and $\overline{L} = L^{-1}$ for the group-like case. Notice that modular transformations do not merely relabel $(L_{1}, L_{2})$, but they also change how the defect network sits on the torus, and therefore, they can force us to re-resolve the crossings in a different channel basis. This is precisely what the $\widetilde{K}$ symbols encode, since these are the change-of-basis matrices between different ways of gluing the defect network of the torus. The $S$-move exchanges the two cycles, i.e., it exchanges $(L_{1}, L_{2})$ up to duals, and re-expands the resulting crossing in the standard channel basis. The $T$-move changes the Dehn twist framing and effectively applies a controlled braiding operation to the junction basis. This explains why the transformations are a sum over the label $L_{4}$ since modular moves generically mix the refined channels. For an explicit example, see \cite{Perez-Lona:2023djo}, where the non-gaugeability with respect to the full categorical symmetry of the Ising model is discussed in detail.\\

\noindent
We now repeat this orbifolding exercise for the $B_{6,1} = \widehat{\mathfrak{so}}(13)_{1}$. The key observation is that this theory has the same $S$-matrix as the $B_{1,1}$ theory, but has a different $T$-matrix, $T =e^{-2\pi i\frac{c}{24}}\text{Diag}\left(\mathbb{1}, e^{2\pi ih_{\sigma}}, e^{2\pi ih_{\psi}}\right)$, that takes the following form
\begin{align}
    T = \text{diag}\left(e^{-\frac{13\pi i}{24}}, e^{\frac{13\pi i}{12}}, e^{\frac{11\pi i}{24}}\right).
\end{align}
The untwisted sector partial traces read
\begin{align}
    \begin{split}
        Z_{0,0}(\tau, \overline{\tau}) =& \vert\chi_{0}\vert^{2} + \vert \chi_{\frac{1}{2}}\vert^{2} + \vert\chi_{\frac{13}{16}}\vert^{2},\\
        Z_{0,1}(\tau, \overline{\tau}) =& \vert\chi_{0}\vert^{2} + \vert \chi_{\frac{1}{2}}\vert^{2} - \vert\chi_{\frac{13}{16}}\vert^{2},
    \end{split}
\end{align}
and the twisted sector partial traces read
\begin{align}
    \begin{split}
      Z_{1,0}(\tau, \overline{\tau}) =& +\chi_{0}\overline{\chi}_{\frac{1}{2}} + \chi_{\frac{1}{2}}\overline{\chi}_{0} + \vert\chi_{\frac{13}{16}}\vert^{2},\\
      Z_{1,1}(\tau, \overline{\tau}) =& - \chi_{0}\overline{\chi}_{\frac{1}{2}} - \chi_{\frac{1}{2}}\overline{\chi}_{0} + \vert\chi_{\frac{13}{16}}\vert^{2}.
    \end{split}
\end{align}
From this, we find the untwisted and twisted sector partition functions to be
\begin{align}
   \begin{split}
    Z_{0}(\tau, \overline{\tau}) =& \vert\chi_{0}\vert^{2} + \vert\chi_{\frac{1}{2}}\vert^{2},\\
    Z_{1}(\tau, \overline{\tau}) =& \vert\chi_{\frac{13}{16}}\vert^{2}.
   \end{split}
\end{align}
Hence, we have found that the $B_{6,1}$ RCFT is also a self-dual under the $\mathbb{Z}_{2}$ orbifold! In fact, the same conclusion holds throughout the $B_{r,1}$ RCFT family 
\begin{align}
    \frac{B_{r,1}}{\mathbb{Z}_{2}} = B_{r,1}.
\end{align}
This follows from the fact that the orbifold is controlled by the invertible simple current $\psi$ and the corresponding $\mathbb{Z}_{2}$ symmetry action, while the modular closure of the associated sector traces forces the orbifold sum to reorganize back into the same diagonal invariant. We will later reinterpret this “self-orbifold” property from the defect/interface viewpoint, where it becomes a concrete statement about how the $\psi$-line acts on the equatorial gluing data for commutant pairs involving $B_{r,1}$.

\subsection{$A_{r,1}$ orbifolds}
We now turn to the $A$-series WZW models $A_{r,k}\cong \widehat{\mathfrak{su}}(r+1)_{k}$, and focus on the simplest level-one case $A_{r,1}$. The $A_{r,1}$ WZW model is the affine theory $\widehat{\mathfrak{sl}}_{r+1}$ at level $1$, and its associated MTC is \cite{Kirillov2008}
\begin{align}
    \mathcal{C}(A_{r},1) \equiv \mathcal{C}(\mathfrak{sl}_{r+1},1),
\end{align}
which is the category of integrable highest-weight $\widehat{\mathfrak{sl}}_{r+1}$-modules at level $1$\footnote{We use $\widehat{\mathfrak{sl}}_{r+1,1}$ and $\widehat{\mathfrak{su}}(r+1)_{1}$ interchangeably, at the level of the modular data (same affine type $A_{r}$).}. A useful structural feature at level $k = 1$ is that the theory is pointed \cite{Davydov2010TheWG} (see example $6.2$), i.e., all primaries are simple currents, or equivalently, all topological lines are invertible. Let's specialize to the $A_{1,1}\cong\widehat{\mathfrak{su}}(2)_{1}$ case. Its MTC is the semion MTC with anyon types $\{\mathbb{1}, s\}$ \cite{Rowell:2007dge} (see \S $5.3.1$), both of quantum dimension $1$ with fusion 
\begin{align}
    s\otimes s = \mathbb{1}.
\end{align}
The corresponding RCFT has two primaries with central charge $c = 1$, and conformal weights\footnote{The conformal weights of $\widehat{\mathfrak{su}}(2)_{k}$ primaries follow from the general formula $h_{j} = \frac{j(j+1)}{k+2}$ for spin $j$ representations at level $k$. Hence, for $k = 1$, we have the vacuum primary $j = 0$, and the non-trivial primary $j = \tfrac{1}{2}$.} $\{0,\tfrac{1}{4}\}$. The $\widehat{\mathfrak{su}}(2)_{1}$ characters have the following modular form expressions \cite{DiFrancesco:1997nk}
\begin{align}
    \begin{split}
        \chi_{0}(\tau) =&\left( \frac{\theta_{3} + \theta_{4}}{2\eta}\right)(\tau),\\
        \chi_{\frac{1}{4}}(\tau) =& \left( \frac{\theta_{3} - \theta_{4}}{2\eta}\right)(\tau)
    \end{split}
\end{align}
The diagonal torus partition function of this theory reads
\begin{align}
    Z_{A_{1,1}}(\tau, \overline{\tau}) = Z^{\mathbb{1}}(\tau, \overline{\tau}) = \vert\chi_{0}\vert^{2} + \vert\chi_{\frac{1}{4}}\vert^{2}.
\end{align}
Using the $S$-matrix of this theory, the characters transform as follows
\begin{align}\label{S}
    \begin{pmatrix}
        \chi_{0}\\
        \chi_{\frac{1}{4}}
    \end{pmatrix} \to \begin{pmatrix}
        \frac{1}{\sqrt{2}} & \frac{1}{\sqrt{2}}\\
        \frac{1}{\sqrt{2}} & \matminus\frac{1}{\sqrt{2}}
    \end{pmatrix}\begin{pmatrix}
        \chi_{0}\\
        \chi_{\frac{1}{4}}
    \end{pmatrix}.
\end{align}
Under the $T$-transform, the characters behave as follows
\begin{align}\label{T}
    \begin{split}
        \chi_{0}(\tau + 1) =& e^{-\frac{\pi i}{12}}\chi_{0}(\tau),\\
        \chi_{\frac{1}{4}}(\tau + 1) =& e^{\frac{5\pi i}{12}}\chi_{\frac{1}{4}}(\tau).
    \end{split}
\end{align}
%Using $\vert Q\vert = \sqrt{1^{2} + 1^{2}} = \sqrt{2}$, 
\noindent The untwisted sector partial traces can be computed to be
\begin{align}
    \begin{split}
        Z_{0,0}(\tau, \overline{\tau}) =& \vert \chi_{0}\vert^{2} + \vert\chi_{\frac{1}{4}}\vert^{2},\\
        Z_{0,1}(\tau, \overline{\tau})  =&\vert\chi_{0}\vert^{2} - \vert\chi_{\frac{1}{4}}\vert^{2}.
    \end{split}
\end{align}
Using the transformation properties \ref{S} and \ref{T}, we can compute the twisted sector partial traces to be
\begin{align}
    \begin{split}
        Z_{1,0}(\tau, \overline{\tau}) =& \chi_{0}\overline{\chi}_{\frac{1}{4}} + \chi_{\frac{1}{4}}\overline{\chi}_{0},\\
        Z_{1,1}(\tau, \overline{\tau}) =& -i\left(\chi_{0}\overline{\chi}_{\frac{1}{4}} - \chi_{\frac{1}{4}}\overline{\chi}_{0}\right).
    \end{split}
\end{align}
From the partial traces, we compute the untwisted and twisted partition function to be
\begin{align}
    Z_{0} = \vert\chi_{0}\vert^{2},\qquad Z_{1} = \frac{1}{2}\left[(1-i)\chi_{0}\overline{\chi}_{\frac{1}{4}} + (1+i)\chi_{\frac{1}{4}}\overline{\chi}_{0}\right].
\end{align}
Notice that the twisted partition function is not a sum of absolute squares with non-negative integer coefficients. In fact, it even carries intrinsic complex phases. That is the concrete manifestation of the semion's fractional spin $\theta_{s} = i$, i.e., the obstruction to a bosonic $\mathbb{Z}_{2}$ orbifold. Since $\{\mathbb{1}, s\}\cong\mathbb{Z}_{2}$, we can attempt to build an orbifold partition function by averaging over the four $(h,g)$ sectors. However, notice here that the generator $s$ has fractional spin \cite{DiFrancesco:1997nk}
\begin{align}
    \theta_{s} = e^{2\pi ih_{s}} = e^{\frac{i\pi}{2}} = i\neq 1.
\end{align}
This condition is nothing but the familiar simple current integral-spin obstruction. In bosonic orbifolds without additional spin  structure or discrete torsion choices, we need integral spin for the current generating the symmetry \cite{Schellekens:1990xy, Gato-Rivera:1991bqv}. Interpreting the modular $T$-eigenvalue as the spin corresponding to the topological defect line \cite{Moore:1988qv}, a fractional spin indicates that the symmetry carries a discrete ’t Hooft anomaly and therefore cannot be gauged \cite{Gaiotto:2014kfa}.

\subsection{$D_{r,1}$ orbifolds}
For the $D$-series WZW models, we have $D_{r,k}\cong\widehat{\mathfrak{so}}(2r)_{k}$. We turn to the level $1$ case $D_{4,1}\cong\widehat{\mathfrak{so}}(8)_{1}$, where we have primaries $\{\mathbb{1}, \nu, \sigma, \widetilde{\sigma}\}$ with conformal weights $\{0,\tfrac{1}{2}, \tfrac{1}{2}, \tfrac{1}{2}\}$ obtained from \ref{D_4,1_conformal_weight}. The torus partition function reads
\begin{align}\label{D_4,1_torus_partition}
    Z(\tau, \overline{\tau}) = \vert\chi_{0}\vert^{2} + 3\vert\chi_{\frac{1}{2}}\vert^{2}.
\end{align}
All the three non-trivial primaries $\{\nu, \sigma, \widetilde{\sigma}\}$, have quantum dimensions $1$ and fusion rules
\begin{align}
    \begin{split}
        \nu\otimes\sigma =& \widetilde{\sigma},\qquad \sigma\otimes\widetilde{\sigma} = \nu,\qquad \widetilde{\sigma}\otimes \nu = \sigma,\\
        \nu\otimes\nu =& \sigma\otimes\sigma = \widetilde{\sigma}\otimes\widetilde{\sigma} =  1, 
    \end{split}
\end{align}
which is exactly the group $\mathbb{Z}_{2}^{2}$. Hence, we have the Picard group $\text{Pic}(D_{4,1}) = \mathbb{Z}_{2}\times\mathbb{Z}_{2}$, and the action of $g$ on a set of primaries is given by the fusion $g: x\mapsto g\otimes x$, and this concretely is
\begin{align}
    \begin{split}
    \nu:&\mathbb{1}\leftrightarrow\nu,\ \sigma\leftrightarrow\widetilde{\sigma},\\
    \sigma:& \mathbb{1}\leftrightarrow\sigma,\ \nu\leftrightarrow\widetilde{\sigma},\\
    \widetilde{\sigma}:& \mathbb{1}\leftrightarrow\widetilde{\sigma},\ \nu\leftrightarrow\sigma.
    \end{split}
\end{align}
We now orbifold by the Klein group, $G = \mathbb{Z}_{2}\times\mathbb{Z}_{2}$, where $\vert G\vert = 4$, and where we have the partial trace elements $Z_{(m,m'), (r,r')}$ \cite{Robbins:2019zdb}. The $S$- and $T$-transforms act on these partial trace elements as follows
\begin{align}
    \begin{split}
        S:& \left((m,m'), (r,r')\right)\mapsto \left((r,r'), (-m,-m')\right),\\
        T:& \left((m,m'), (r,r')\right)\mapsto \left((m,m'), (r-m, r'-m')\right).
    \end{split}
\end{align}
Now, the traces for no spatial twist is then given by the twining partition functions in the original Hilbert space, i.e.
\begin{align}
    Z_{(0,0), (r,r')} = \text{Tr}_{\mathcal{H}}\left(\widehat{g\left(r,r'\right)}q^{L_{0} - \frac{c}{24}}\overline{q}^{\overline{L}_{0} - \frac{c}{24}}\right).
\end{align}
In an RCFT, insertion of invertible line defects labeled by $g$ along one of the cycles acts diagonally on the sector $x$ with eigenvalue
\begin{align}
    \lambda_{g}(x) = \frac{S_{g,x}}{S_{0,x}}\in\{\pm1\},
\end{align}
and for $D_{4,1}$, the modular $S$-matrix is nothing but the $\mathbb{Z}_{2}\times\mathbb{Z}_{2}$ character table, expressed as 
\begin{align}
    S = \frac{1}{2}\begin{pmatrix}
        1 & 1 & 1 & 1\\
        1 & 1 & \matminus1 & \matminus1\\
        1 & \matminus1 & 1 & \matminus1\\
        1 & \matminus1 & \matminus1 & 1
    \end{pmatrix},
\end{align}
and the $T$-matrix reads
\begin{align}
    T = \text{Diag}\left(e^{-\frac{\pi i}{3}}, e^{\frac{2\pi i}{3}},e^{\frac{2\pi i}{3}},e^{\frac{2\pi i}{3}}\right).
\end{align}
Let's pick generators of $G$ as follows
\begin{align}
   a\equiv \nu,\qquad b\equiv \sigma,\qquad ab\equiv \widetilde{\sigma},
\end{align}
and group elements as pairs $(m,m')\in\{0,1\}^{2}$ so that we have
\begin{align}
    g(m,m')\equiv a^{m}b^{m'}\in\{\mathbb{1}, \nu, \sigma, \widetilde{\sigma}\}.
\end{align}
Similarly, we also define $h(r,r')\equiv a^{r}b^{r'}$. From \cite{Robbins:2019zdb}, we find that we have one orbit of partial traces disconnected from the untwisted sector, and this consists of the six traces with $h\neq 1$, $g\neq 1$, and $g\neq h$, which is packaged as follows
\begin{align}\label{disconnected_orbit} 
    Z_{\text{disconnected}} = Z_{(0,1),(1,0)} + Z_{(1,0),(0,1)} + Z_{(0,1),(1,1)} + Z_{(1,0),(1,1)} + Z_{(1,1),(0,1)} + Z_{(1,1),(1,0)},
\end{align}
and imposing higher-genus consistency, this disconnected piece can only enter as
\begin{align}
    Z_{\mathbb{Z}_{2}\times\mathbb{Z}_{2}} = \frac{1}{\vert\mathbb{Z}_{2}\times\mathbb{Z}_{2}\vert}\left(Z_{\text{connected}} \pm Z_{\text{disconnected}}\right),
\end{align}
where $\pm$ is a discrete torsion choice. Here, we have the identifications
\begin{align}
    (0,0)\leftrightarrow\mathbb{1},\qquad (1,0)\leftrightarrow\nu,\qquad (0,1)\leftrightarrow\sigma,\qquad (1,1)\leftrightarrow\widetilde{\sigma}.
\end{align}
Let's now write out these contributions explicitly. For $D_{4,1}$, by triality, we have $\chi_{\nu} = \chi_{\sigma} = \chi_{\widetilde{\sigma}}\equiv \chi_{\tfrac{1}{2}}$, and for the untwisted sector, with $h = 1$, we have
\begin{align}\label{orbit_1}
    \begin{split}
        Z_{(0,0),(0,0)} =& \vert\chi_{\mathbb{1}}\vert^{2} + \vert\chi_{\nu}\vert^{2} + \vert\chi_{\sigma}\vert^{2} + \vert\chi_{\widetilde{\sigma}}\vert^{2} = \vert\chi_{0}\vert^{2} + 3\vert\chi_{\frac{1}{2}}\vert^{2},\\
        Z_{(0,0), (1,0)} =& \vert\chi_{\mathbb{1}}\vert^{2} + \vert\chi_{\nu}\vert^{2} - \vert\chi_{\sigma}\vert^{2} - \vert\chi_{\widetilde{\sigma}}\vert^{2} = \vert\chi_{0}\vert^{2} - \vert\chi_{\frac{1}{2}}\vert^{2},\\ 
        Z_{(0,0),(0,1)} =& \vert\chi_{\mathbb{1}}\vert^{2} - \vert\chi_{\nu}\vert^{2} + \vert\chi_{\sigma}\vert^{2} - \vert\chi_{\widetilde{\sigma}}\vert^{2} = \vert\chi_{0}\vert^{2} - \vert\chi_{\frac{1}{2}}\vert^{2},\\
        Z_{(0,0), (1,1)} =& \vert\chi_{\mathbb{1}}\vert^{2} - \vert\chi_{\nu}\vert^{2} - \vert\chi_{\sigma}\vert^{2} + \vert\chi_{\widetilde{\sigma}}\vert^{2} = \vert\chi_{0}\vert^{2} - \vert\chi_{\frac{1}{2}}\vert^{2}.
    \end{split}
\end{align}
The purely twisted partial traces with no insertions corresponding to $g = 1$ are off-diagonal pairings that read
\begin{align}\label{orbit_2}
    \begin{split}
        Z_{(1,0),(0,0)} = Z_{(0,1),(0,0)} = Z_{(1,1),(0,0)} = \chi_{0}\overline{\chi}_{\frac{1}{2}} + \chi_{\frac{1}{2}}\overline{\chi}_{0} + 2\vert\chi_{\frac{1}{2}}\vert^{2}.
    \end{split}
\end{align}
The diagonal twisted-twining partial traces that are still in the connected orbit, with $h = g \neq 1$, read
\begin{align}\label{orbit_3}
    Z_{(1,0),(1,0)} = Z_{(0,1),(0,1)} = Z_{(1,1),(1,1)} = -\chi_{0}\overline{\chi}_{\frac{1}{2}} - \chi_{\frac{1}{2}}\overline{\chi}_{0} + 2\vert\chi_{\frac{1}{2}}\vert^{2}.
\end{align}
Finally, the six mixed partial traces with $h\neq g\neq1$ that are exactly the disconnected orbit contributions in \ref{disconnected_orbit} read
\begin{align}\label{orbit_4}
    Z_{(1,0),(0,1)} =& -Z_{(1,0),(1,1)} = \varepsilon\left(\chi_{0}\overline{\chi}_{\frac{1}{2}} - \chi_{\frac{1}{2}}\overline{\chi}_{0}\right),\\
    Z_{(0,1),(1,0)} =& -Z_{(0,1),(1,1)} = \varepsilon\left(\chi_{\frac{1}{2}}\overline{\chi}_{0} -\chi_{0}\overline{\chi}_{\frac{1}{2}}\right),\\
    Z_{(1,1),(1,0)} =& -Z_{(1,1),(0,1)} =  \varepsilon\left(\chi_{0}\overline{\chi}_{\frac{1}{2}} - \chi_{\frac{1}{2}}\overline{\chi}_{0}\right),
\end{align}
where $\varepsilon = \pm1$ is the overall discrete torsion sign. Now, the connected piece is the sum of the $10$ partial traces \ref{orbit_1} \ref{orbit_2}, and \ref{orbit_3} in the modular orbit connected to the untwisted sector, which reads
\begin{align}
    Z_{\text{connected}} = 4\vert\chi_{0}\vert^{2} + 12\vert\chi_{\frac{1}{2}}\vert^{2}.
\end{align}
Next, the disconnected piece is computed by summing all the contributions in \ref{orbit_4} to get
\begin{align}
    Z_{\text{disconnected}} = 0.
\end{align}
Using this, we find the following orbifold partition function
\begin{align}
    Z_{\text{orb}} = \frac{1}{\vert G\vert}\left(Z_{\text{connected}} + Z_{\text{disconnected}}\right) = \vert\chi_{0}\vert^{2} + 3\vert\chi_{\frac{1}{2}}\vert^{2},
\end{align}
which is equal to the torus partition function \ref{D_4,1_torus_partition}, and hence, is a self-orbifold!\\

\noindent 
For the $D$-series WZW models, the Picard group matches the center data of simply-connected groups, $\text{Spin}(2r)$ which depends on parity \cite{Bartsch:2023pzl, Henningson:2007qr}. For even $r$, we have $\text{Spin}(4k)$, and the center is $\mathbb{Z}_{2}\times\mathbb{Z}_{2}$, while for odd $r$, we have $\text{Spin}(4k+2)$ with center $\mathbb{Z}_{2}$. Hence, we have
\begin{align}
    \text{Pic}(D_{r,1}) = \begin{cases}
        \mathbb{Z}_{2}\times\mathbb{Z}_{2},\ & r\ \text{even},\\
        \mathbb{Z}_{4},\ & r\ \text{odd}.
    \end{cases}
\end{align}
This even/odd dichotomy is also reflected directly in the level $1$ fusion rules for the vectors and spinors. for $N = 2r\equiv 0\ (\text{mod}\ 4)$, we have the Klein group $\mathbb{Z}_{2}\times\mathbb{Z}_{2}$, and for $N = 2r\equiv 2\ (\text{mod}\ 4)$, the spinor generates a $\mathbb{Z}_{4}$ \cite{Bockenhauer:1995hx}. Using the standard simple current orbifold formulas and the higher-genus consistency constraints for multiple generators \cite{Schellekens:1990xy}, it is straightforward to check for even $r$ that the $(\mathbb{Z}_{2}\times\mathbb{Z}_{2})$-orbifold of $D_{r,1}$ agree with the parent theory only in the triality case $r = 4$; we leave these checks to the reader. 

\subsection{$G_{r,1}$ orbifolds}
For the $G$-series WZW models, we have $G_{r,k}\cong \widehat{\mathfrak{g}}_{2k}$, since there is no infinite $G_{r}$ series in the Cartan classification.  Consider $G_{2,1}$, which is given by the Fibonacci MTC \cite{Rowell:2007dge}, with simples $\{\mathbb{1}, \tau\}$ and fusion
\begin{align}\label{fusion_Fib}
    \tau\otimes\tau = \mathbb{1}\oplus \tau.
\end{align}
The quantum dimensions are $(d_{\mathbb{1}}, d_{\tau}) = (1, \varphi)$, where $\varphi = \tfrac{1 + \sqrt{5}}{2}$ is the golden ratio. The $S$- and the $F$-matrix read
\begin{align}\label{F_matrix_Fib}
    S = \frac{1}{\sqrt{2 + \varphi}}\begin{pmatrix}
        1 & \varphi\\
        \varphi & \matminus1
    \end{pmatrix},\qquad F^{\tau}_{\tau\tau\tau} = \begin{pmatrix}
        \varphi^{-1} & \varphi^{-\frac{1}{2}}\\
        \varphi^{-\frac{1}{2}}   & \matminus\varphi^{-1}
    \end{pmatrix}.
\end{align}
For the multiplicity-free fusion \ref{fusion_Fib}, the natural basis of partial traces is 
\begin{align}
    Z^{L_{3}}_{L_{1}, L_{2}}(\tau,\overline{\tau}),\qquad L_{1}, L_{2}\in\{\mathbb{1}, \tau\},\qquad L_{3}\in L_{1}\otimes L_{2}.
\end{align}
Here, we the five partial traces, $\{ Z^{\mathbb{1}}_{\mathbb{1}, \mathbb{1}}, Z^{\tau}_{\mathbb{1}, \tau}, Z^{\tau}_{\tau, \mathbb{1}}, Z^{\mathbb{1}}_{\tau, \tau}, Z^{\tau}_{\tau, \tau}\}$, with $S$ and $T$ transforms acting via \ref{Zl123_SandT}. The only non-trivial mixing that happens here is in the $(\tau, \tau)$-sector, and the relevant $\widetilde{K}$ is precisely the $F$-matrix \ref{F_matrix_Fib} while the rest are trivial since lines involving $\mathbb{1}$ do not create non-trivial recouplings. Since there are only two fusion channels, we have
\begin{align}
    \widetilde{K}^{\tau, \tau}_{\tau, \tau}(L_{3}, L_{4}) = \left(F^{\tau}_{\tau\tau\tau}\right)^{L_{4}}_{L_{3}},\qquad L_{3}, L_{4}\in\{\mathbb{1}, \tau\}.
\end{align}
For $G_{2,1}$, the chiral data are
\begin{align}
    c = \frac{14}{5},\qquad h_{\mathbb{1}} = h_{0} = 0,\qquad h_{\tau} = \frac{2}{5},
\end{align}
with the following $T$-matrix
\begin{align}
    T_{\mathbb{1}} =e^{2\pi i\left(h_{0} -\tfrac{c}{24}\right)} = e^{-\frac{7\pi i}{30}},\qquad  T_{\tau} =e^{2\pi i\left(h_{\tau} -\tfrac{c}{24}\right)} = e^{\frac{17\pi i}{30}},
\end{align}
and we find the following $q$-series expansions for the characters
\begin{align}
    \begin{split}
        \chi_{0}(\tau) =& q^{-\frac{7}{60}}\left(1 + 14q + 42q^{2} + 140q^{3} + 350q^{4} + \ldots\right),\\
        \chi_{\frac{2}{5}}(\tau) =& q^{\frac{17}{60}}\left(7 + 34q + 119q^{2} + 322q^{3} + 819q^{4} + \ldots\right).
    \end{split}
\end{align}
The untwisted torus partition function takes the form
\begin{align}
    Z(\tau, \overline{\tau}) = Z^{\mathbb{1}}_{\mathbb{1}, \mathbb{1}}(\tau, \overline{\tau}) = \vert\chi_{0}\vert^{2} + \vert\chi_{\frac{2}{5}}\vert^{2}.
\end{align}
Inserting a topological line $\tau$ along the spatial cycle acts diagonally on the primaries by the eigenvalue $\lambda_{i}(\tau) = \tfrac{S_{\tau i}}{S_{0i}}$, and with the $S$-matrix \ref{F_matrix_Fib}, we find
\begin{align}
    \lambda_{0}(\tau) = \varphi,\qquad \lambda_{\tau}(\tau) = -\varphi^{-1},
\end{align}
and hence, we find the twining partition function to take the form
\begin{align}
    Z^{\tau}_{\mathbb{1}, \tau} = \varphi\vert\chi_{0}\vert^{2} - \frac{1}{\varphi}\vert\chi_{\frac{2}{5}}\vert^{2}.
\end{align}
Applying a $S$-transform, we find
\begin{align}
    Z^{\tau}_{\tau, \mathbb{1}} = S\cdot Z^{\tau}_{\mathbb{1}, \tau} = \sum\limits_{i,j\in\{\mathbb{1}, \tau\}}M_{ij}^{(\tau, \mathbb{1})}\chi_{i}\overline{\chi_{j}},
\end{align}
with matrix
\begin{align}
    M^{(\tau, \mathbb{1})} = S\begin{pmatrix}
        \varphi & 0\\
        0 & \matminus\varphi^{-1}
    \end{pmatrix}S = \begin{pmatrix}
        0 & 1\\ 1 & 1
    \end{pmatrix}.
\end{align}
This yields
\begin{align}
    Z^{\tau}_{\tau, \mathbb{1}} = 
    \chi_{0}\overline{\chi_{\frac{2}{5}}} + \chi_{\frac{2}{5}}\overline{\chi_{0}} + \vert\chi_{\frac{2}{5}}\vert^{2}.
\end{align}
Next, we perform a $T$ transform to obtain
\begin{align}
    Z^{\mathbb{1}}_{\tau, \tau}(\tau) = Z^{\tau}_{\tau, \mathbb{1}}(\tau + 1).
\end{align}
Fixing this yields
\begin{align}
     Z^{\mathbb{1}}_{\tau, \tau} = e^{-\tfrac{4\pi i}{5}}\chi_{0}\overline{\chi_{\frac{2}{5}}} + e^{\tfrac{4\pi i}{5}}\chi_{\frac{2}{5}}\overline{\chi_{0}} + \vert\chi_{\frac{2}{5}}\vert^{2},
\end{align}
\begin{comment}
Note that per the fusion \ref{fusion_Fib}, the $(\tau, \tau)$-sector has two channels-resolved partial traces, and they must sum to the unresolved $(\tau, \tau)$ trace, and mix under modular moves by the $F$-matrix. Solving these constraints uniquely fixes the two $2\times 2$ coefficient matrices in the $\{\chi_{0}, \chi_{\tau}\}$ basis, and we find
\begin{align}
 \begin{split}
     Z^{\mathbb{1}}_{\tau, \tau}:&\  M^{(\mathbb{1})}_{\tau, \tau} = \begin{pmatrix}
         \varphi + 1 - \frac{1}{2}\sqrt{\varphi} & \frac{1}{2}\left(1 - \sqrt{\varphi}\right)\\
         \frac{1}{2}\left(1 - \sqrt{\varphi}\right) & \frac{1}{2}\left(\varphi + 1\right) - \sqrt{\varphi}
         \end{pmatrix},\\
         Z^{\tau}_{\tau, \tau}:&\  M^{(\tau)}_{\tau, \tau} = \begin{pmatrix}
             -\varphi + \frac{1}{2}\sqrt{\varphi} & \frac{1}{2}\left(1 + \sqrt{\varphi}\right)\\
             \frac{1}{2}\left(1 + \sqrt{\varphi}\right) & \frac{3}{2} - \frac{1}{2}\varphi + \sqrt{\varphi}
         \end{pmatrix}.
    \end{split}   
\end{align}
From these, we indeed obtain 
\begin{align}
    M^{(\mathbb{1}}_{\tau, \tau} + M^{(\tau)}_{\tau, \tau} = N^{2}_{\tau} = \begin{pmatrix}
        1 & 1\\
        1 & 2
    \end{pmatrix},
\end{align}
where $N_{\tau} = \begin{psmallmatrix}
    0 & 1\\ 1 & 1
\end{psmallmatrix}$ is the Fibonacci fusion matrix.
\end{comment}
\noindent Finally, we have
\begin{align}
    Z^{\tau}_{\tau, \tau} = \varphi^{\frac{1}{2}}e^{\tfrac{3\pi i}{5}}\chi_{0}\overline{\chi_{\tfrac{2}{5}}} + \varphi^{\frac{1}{2}}e^{-\frac{3\pi i}{5}}\chi_{\frac{2}{5}}\overline{\chi_{0}} + \varphi^{-\frac{3}{2}}\vert\chi_{\frac{2}{5}}\vert^{2}.
\end{align}
Let us now consider the gauging algebra to be 
\begin{align}
    \mathfrak{A}\equiv \mathbb{1}\oplus \tau,
\end{align}
and crucially
\begin{align}\label{|A|}
    \vert\mathfrak{A}\vert \equiv \sum\limits_{x\subset\mathfrak{A}}d_{x} = d_{\mathbb{1}} + d_{\tau} = 1 + \varphi = \varphi^{2},
\end{align}
is not an integer. In the language of \cite{Perez-Lona:2023djo}, we distinguish strong and weak gauging as follows:
\begin{itemize}
    \item Strong gauging demands integrality/positivity conditions consistent with a genuine unitary orbifold CFT, and in practice, forces serve arithmetic constraints such as integer quantum dimensions.
    \item Weak gauging relaxes these, allowing modular covariant completions that may fail positivity/integrality in some sectors but still exist as consistent modular objects. 
\end{itemize}
Now, from \ref{|A|}, we immediately see that this obstructs strong gaugeability, but it does not obstruct weak gaugeability, where irrational structure constants are allowed at the level of topological data. This is exactly why the Fibonacci MTC is a prototypical example that is weakly but not strongly gaugeable. In a setting where the symmetry action is trivial, each partial trace reduces to the parent partition function multiplied by a purely topological coefficient $\Lambda^{L_{3}}_{L_{1}, L_{2}}$, and the gauged or the orbifold partition function is a normalized sum over all such sectors given by \cite{Perez-Lona:2023djo}
\begin{align}
    Z_{\text{gauged}} = Z_{\text{orb}} = \sum\limits_{L_{1}, L_{2}, L_{3}}\Lambda^{L_{3}}_{L_{1}, L_{2}}Z^{L_{3}}_{L_{1}, L_{2}},
\end{align}
We compute these coefficients to be (see appendix \ref{sec:appendix_Lambda_computations})
\begin{align}\label{Lambdas}
    \left\{\Lambda^{\mathbb{1}}_{\mathbb{1}, \mathbb{1}}, \Lambda^{\mathbb{1}}_{\tau, \tau}, \Lambda^{\tau}_{\mathbb{1}, \tau}, \Lambda^{\tau}_{\tau, \mathbb{1}}, \Lambda^{\tau}_{\tau, \tau}\right\} = \left\{\frac{1}{\varphi}, \frac{1}{\varphi^{2}}, \frac{1}{3}, \frac{1}{3}, \frac{1}{3}\right\}.
\end{align}
Using this, we find the orbifold partition function 
\begin{align}
    Z_{\text{orb}} = \frac{\varphi + 4}{3\varphi}\vert\chi_{0}\vert^{2} + \left(\frac{4}{3} - \frac{1}{3\varphi} + \frac{1}{3\varphi^{\frac{3}{2}}}\right)\vert\chi_{\frac{2}{5}}\vert^{2} + \left(\frac{e^{-\frac{4\pi i}{5}}}{\varphi^{2}} + \frac{1 + \varphi^{\frac{1}{2}}e^{\frac{3\pi i}{5}}}{3}\right)\chi_{0}\overline{\chi_{\frac{2}{5}}} + \left(\frac{e^{\frac{4\pi i}{5}}}{\varphi^{2}} + \frac{1 + \varphi^{\frac{1}{2}}e^{-\frac{3\pi i}{5}}}{3}\right)\chi_{\frac{2}{5}}\overline{\chi_{0}}
\end{align}
From this expression we see that a $q$-series expansion would involve coefficients

\section{Three-character pairings into $E_{8,1}$ and defect insertions}
In this section, we present explicit three-character examples that realize the commutant pair philosophy in a form that is suitable for topological defect computations and for systematic extensions of the replacement rule approach. The guiding precedent is the construction of defect partition functions in the Monster CFT presented in \cite{Lin:2019hks}, where non-trivial defect data of a $c = 24$ meromorphic theory can be accessed indirectly by passing to a suitable commutant pair, after which the resulting defect insertion is implemented at the level of characters. This logic was sharpened into an effective ``replacement rule" in the two-character setting and used to construct defect partition functions for the $c = 8$ meromorphic theory with Kac-Moody algebra $E_{8,1}$ \cite{Hegde:2021sdm}. Our aim is to extend this computational pipeline to three-character commutant pairs whose bilinear pairings reproduce $E_{8,1}$ that arise from the coset framework \cite{Das:2022uoe}.\\

\noindent For $E_{8,1}$, a variety of defect (twining/twisted) partition functions can also be accessed  directly using lattice-VOA and automorphism techniques, i.e., as modular trace functions associated to finite-order automorphisms, bypassing the two-character commutant pairs approach \cite{Dong:1997ea, Volpato:2024goy}. However, these techniques do not immediately provide a uniform treatment across the full list of coset realizations emphasized in the meromorphic coset classification. In particular, the tensor-product and IVOA-type commutant pairs highlighted in the meromorphic coset classification motivate a dedicated analysis from the standpoint of the replacement rule \cite{Burbano:2021loy, Das:2022uoe}.  Our strategy here is concrete. For each commutant pair $(\mathcal{C}, \widetilde{\mathcal{C}})$, we begin from the bilinear pairing relation established by the branching rules in \S \ref{sec:branching_rules}, which provides a direct handle on how defect insertions act on equatorial projections. Following this, we extract the relevant defect data from the lower-central charge theory by computing the eigenvalues of the defect action from the modular $S$-matrix for the relevant simple objects  obtained in the MMS framework \cite{Mathur:1988gt}, which determines the modular representation and hence, the Verlinde operators, and fixes the corresponding defect partition functions via standard RCFT constructions of topological line insertions \cite{Hegde:2021sdm}. Finally, we analyze the resulting defect partition functions as modular functions and determine their modular invariance groups.

\subsection{$\left(B_{1,1}, B_{6,1}\right)$}\label{sec: B_1,1, B_6,1}
We consider a commutant pair with central charges $(c,\widetilde{c}) = \left(\tfrac{3}{2}, \tfrac{13}{2}\right)$, and hence, out of this pair we build
\begin{align}
    \chi^{E_{8}} = \chi^{B_{1}}_{0}\chi^{B_{6}}_{0} + \chi^{B_{1}}_{\tfrac{1}{2}}\chi^{B_{6}}_{\tfrac{1}{2}} + \chi^{B_{1}}_{\tfrac{3}{16}}\chi^{B_{6}}_{\tfrac{13}{16}}.
\end{align}
The defect that we consider is for the theory with the lower central charge, i.e., $B_{1,1}$. 
The $B_{1,1}$ is nothing but the $(A_{1}, 2)$ MTC in \cite{Rowell:2007dge} (see \S$5.3.5$) with non-trivial primaries $(\phi_{0}, \phi_{1},\phi_{2}) = (\mathbb{1},\sigma, \psi)$ with $\left(h_{\mathbb{1}}, h_{\sigma}, h_{\psi}\right) = \left(0,\tfrac{3}{16}, \tfrac{1}{2}\right)$. The corresponding $S$-matrix reads
\begin{align}
    S = \frac{1}{2}\begin{pmatrix}
        1 & \sqrt{2} & 1\\
        \sqrt{2} & 0 & \matminus\sqrt{2}\\
        1 & \matminus\sqrt{2} & 1
    \end{pmatrix}.
\end{align}
Now, we construct the Verlinde lines and look at their action on the primaries. We have
\begin{align}
    \begin{split}
        \widehat{\mathcal{L}}_{0}\vert\phi_{i}\rangle =& \vert\phi_{i}\rangle,\\
        \widehat{\mathcal{L}}_{1}\vert\phi_{0}\rangle =& \sqrt{2}\vert\phi_{0}\rangle,\ \widehat{\mathcal{L}}_{1}\vert\phi_{1}\rangle = 0,\ \ \ \ \ \ \ \ \widehat{\mathcal{L}}_{1}\vert\phi_{2}\rangle = -\sqrt{2}\vert\phi_{2}\rangle,\\
        \widehat{\mathcal{L}}_{2}\vert\phi_{0}\rangle =& \vert\phi_{0}\rangle,\ \ \ \ \ \widehat{\mathcal{L}}_{2}\vert\phi_{1}\rangle = -\sqrt{2}\vert\phi_{1}\rangle,\ \ \widehat{\mathcal{L}}_{2}\vert\phi_{2}\rangle = \vert\phi_{2}\rangle.
    \end{split}
\end{align}
From this, we observe that $\widehat{\mathcal{L}}_{0}$ is the trivial or the identity line, $\widehat{\mathcal{L}}_{2}$ is the invertible $\psi$-line corresponding to the $\mathbb{Z}_{2}$ global symmetry of the 2D RCFT, and $\widehat{\mathcal{L}}_{1}$ is the non-invertible $N$-line that annihilates $\vert\phi_{1}\rangle$. Now, given a Verlinde line, we can obtain the partition function
with a defect insertion (the spatial cycle) as follows
\begin{align}\label{Z_L}
    \mathcal{Z}^{\widehat{\mathcal{L}}_{k}} = \sum\limits_{i}\frac{S_{ki}}{S_{0i}}\chi_{i}(\tau)\widetilde{\chi}_{i}(\overline{\tau}).
\end{align}
Using this, the spatial cycle defect partition functions read
\begin{align}
    \begin{split}
        \mathcal{Z}^{\mathbb{1}} =& \vert\chi_{0}\vert^{2} + \vert\chi_{1}\vert^{2} + \vert\chi_{2}\vert^{2},\\
        \mathcal{Z}^{\sigma} =& \sqrt{2}\left(\vert\chi_{0}\vert^{2} - \vert\chi_{2}\vert^{2}\right),\\
        \mathcal{Z}^{\psi} =& \vert\chi_{0}\vert^{2} - \vert\chi_{1}\vert^{2} + \vert\chi_{2}\vert^{2}.
    \end{split}
\end{align}
We are interested in the $\mathcal{Z}^{\psi}$ defect partition function. Performing the replacement, we have
\begin{align}\label{B_1,1_defect}
    \mathcal{Z}^{E_{8}, (B_{1,1};\psi)} = \chi^{B_{1}}_{0}\chi^{B_{6}}_{0} + \chi^{B_{1}}_{\tfrac{1}{2}}\chi^{B_{6}}_{\tfrac{1}{2}} - \chi^{B_{1}}_{\tfrac{3}{16}}\chi^{B_{6}}_{\tfrac{13}{16}}.
\end{align}
\noindent With the modular form expressions for the characters in \ref{B_r_characters} substituted in \ref{B_1,1_defect}, the defect partition function now reads
\begin{align}\label{q_defect}
    \begin{split}
        \mathcal{Z}^{E_{8}, (B_{1,1} ;\psi)}(\tau) =& \left(\frac{\theta_{3}^{8} + \theta_{4}^{8} - \theta_{2}^{8}}{2\eta^{8}}\right)(\tau) = \left(\frac{\eta(\tau)}{\eta(2\tau)}\right)^{8}\\
    =& q^{-\frac{1}{3}}\left(1 - 8q + 28q^{2} - 64q^{3} + \ldots\right).
    \end{split}
\end{align}
This is an eta-quotient of level $2$, and hence a weight $0$ modular function for $\Gamma_{0}(2)$, but only covariantly, i.e., up to a root of unity or multiplier system. Using the $T$-transform of the Dedekind eta, which reads
\begin{align}
    \eta(\tau + 1) = e^{\frac{\pi i}{12}}\eta(\tau),
\end{align}
we find the following behaviour of the defect partition function
\begin{align}\label{T-behaviour}
    \mathcal{Z}^{E_{8}, (B_{1,1} ;\psi)}(\tau + 1) = e^{-\frac{2\pi i}{3}}\mathcal{Z}^{E_{8}, (B_{1,1} ;\psi)}(\tau).
\end{align}
This immediately tells us that the defect partition function is not invariant under the $T$-transform, and the $T$-multiplier has order $3$. Accordingly, the conductor extracted from the $T$-eigenvalue is 
\begin{align}\label{conductor}
    N_{T} = \text{ord}(\rho(T)) = 3,
\end{align}
which is also reflected in the leading exponent of $q^{-\tfrac{1}{3}}$ in \ref{q_defect}. We can recast the defect partition function \ref{q_defect} in terms of Weber functions, and doing so not only makes the order-$3$ multiplier under $T$ completely transparent but also reveals a surprisingly clean Fricke action $\tau\mapsto-\tfrac{1}{2\tau}$ as an inversion of the defect observable. Using the standard normalization, we define the Weber function as follows \cite{Weber1908}
\begin{align}
    \mathfrak{f}_{2}(\tau) \equiv \sqrt{2}\frac{\eta(2\tau)}{\eta(\tau)}.
\end{align}
In this normalization, $\mathfrak{f}_{2}$ has a simple multiplier under $T$ of order $24$, namely
\begin{align}
    \mathfrak{f}_{2}(\tau + 1) = \zeta_{24}\mathfrak{f}_{2}(\tau),
\end{align}
where $\zeta_{24} = e^{\tfrac{2\pi i}{24}}$, and consequently, we have
\begin{align}
    \mathfrak{f}_{2}(\tau+1)^{8} = e^{\frac{2\pi i}{3}}\mathfrak{f}_{2}(\tau)^{8}.
\end{align}
We immediately see the following relation
\begin{align}
    \mathcal{Z}^{E_{8}, (B_{1,1}; \psi)}(\tau) = \frac{16}{\mathfrak{f}_{2}(\tau)^{8}}.
\end{align}
This identity reproduces the observed $T$-behaviour noted in \ref{T-behaviour}. Hence, from the Weber function point of view, the order-$3$ multiplier under $T$ is nothing but the statement that the defect partition function is proportional to $\mathfrak{f}_{2}^{-8}$, and $\mathfrak{f}_{2}$ itself has a $24^{\text{th}}$ root multiplier under $T$. It is also interesting to note that $\mathfrak{f}_{2}$ is a modular function for $\Gamma(24)$. Now consider the Fricke involution at level $2$, which reads
\begin{align}
    W_{2}: \tau\mapsto -\frac{1}{2\tau}.
\end{align}
Using $\eta(-\tfrac{1}{\tau}) = \sqrt{-i\tau}\eta(\tau)$, we find
\begin{align}
    \mathfrak{f}_{2}(W_{2}\tau) = \frac{\sqrt{2}}{\mathfrak{f}_{2}(\tau)},
\end{align}
and consequently, the defect partition function transforms by a simple inversion
\begin{align}
    \mathcal{Z}^{E_{8}, (B_{1,1}; \psi)}(W_{2}\tau) = \frac{16}{\mathcal{Z}^{E_{8}, (B_{1,1}; \psi)}(\tau)}.
\end{align}
Hence, the Fricke involution $W_{2}$ exchanges the two cusps by $\mathcal{Z}\mapsto\tfrac{16}{\mathcal{Z}}$. While \ref{q_defect} itself is not Fricke invariant, this observation immediately produces the following canonical Fricke-completed defect observable
\begin{align}
    \begin{split}
        \mathcal{Z}_{\text{Fr}}(\tau)\equiv&  \mathcal{Z}^{E_{8}, (B_{1,1}; \psi)}(\tau) + \frac{16}{ \mathcal{Z}^{E_{8}, (B_{1,1}; \psi)}(\tau)} = \frac{16}{\mathfrak{f}_{2}(\tau)^{8}} + \mathfrak{f}_{2}(\tau)^{8}\\
    =& q^{-\frac{1}{3}}\left(1 + 16q^{\frac{2}{3}} - 8q + 128q^{\frac{5}{3}} + 28q^{2} + 576q^{\frac{8}{3}} - 64q^{3} + 2048q^{\frac{11}{3}} + \ldots \right),
    \end{split}
\end{align}
with the property
\begin{align}
    \mathcal{Z}_{\text{Fr}}(W_{2}\tau) =  \mathcal{Z}_{\text{Fr}}(\tau).
\end{align}
From the defect perspective, this is a precise modular duality statement. While the Fricke involution does not leave the defect partition function invariant, it instead acts by a controlled inversion that exchanges the asymptotic regimes that are the cusps. The Fricke-completed defect partition function $\mathcal{Z}_{\text{Fr}}$ is therefore a natural object to compare with Fricke-extended congruence groups $\Gamma_{0}^{+}(N)\equiv \Gamma_{0}(N)\cup \Gamma_{0}(N)W_{N}$, and with Fricke-type MLDE character constructions in recent RCFT literature \cite{Das:2022bxm, Umasankar:2022kzs}. In particular, the Fricke-completion that produces a $\Gamma_{0}^{+}(2)$-invariant object that is formed by pairing an eta-quotient with its Fricke image. For instance, the Hauptmodul of $\Gamma_{0}(2)$ is defined as follows
\begin{align}
    j_{2}(\tau)\equiv \left(\frac{\eta(\tau)}{\eta(2\tau)}\right)^{24} = \mathcal{Z}^{E_{8}, (B_{1,1}; \psi)}(\tau)^{3}.
\end{align}
Now, notice that $j_{2}(\tau)$ has a trivial $T$-multiplier since we cubed away the order-$3$ phase. The Fricke involution acts by $j_{2}(\tau)\leftrightarrow\tfrac{2^{12}}{j_{2}(\tau)}$ so the Fricke-invariant combination is simply the sum 
\begin{align}
    j_{2}(\tau) + \frac{2^{12}}{j_{2}(\tau)}.
\end{align}
%\NBU{This is OEIS A106207}
\noindent Lastly, we also notice that we have the relation
\begin{align}
    \mathcal{Z}_{\text{Fr}}(\tau)^{3} - \frac{48}{\mathcal{Z}_{\text{Fr}}(\tau)^{3}}= j_{2^{+}}(\tau) - 128,
\end{align}
where $j_{2^{+}}(\tau)$ is the Hauptmodul of $\Gamma_{0}^{+}(2)$ \cite{Das:2022bxm}.

\subsection{$\left(D_{2,1}, D_{6,1}\right)$}
We consider a commutant pair with central charges $(c,\widetilde{c}) = \left(2, 6\right)$, and hence, out of this pair we build
\begin{align}
    \chi^{E_{8}} = \chi^{D_{2}}_{0}\chi^{D_{6}}_{0} + \chi^{D_{2}}_{\tfrac{1}{2}}\chi^{D_{6}}_{\tfrac{1}{2}} + 2\chi^{D_{2}}_{\tfrac{1}{4}}\chi^{D_{6}}_{\tfrac{3}{4}}.
\end{align}
%\NBU{See Arpit's paper 2212.03136 for why there's a coefficient $2$. See table (1,2) on pg 24}.
\noindent The defect that we consider is for the theory with the lower central charge, i.e., $D_{2,1}$. The $D_{2,1}$ is nothing but two copies of the Semion MTC with anyon types $\{1, s\}$ in \cite{Rowell:2007dge} (see $5.3.1$) with non-trivial primaries $(\phi_{0}, \phi_{1}, \phi_{2}, \phi_{3}) = (\mathbb{1}, \nu, \sigma, \widetilde{\sigma}\}$, where $\mathbb{1} = (1,1), \nu = (s,s), \sigma = (s,1), \widetilde{\sigma} = (1,s)$ and conformal weights $h = \{0,\tfrac{1}{2}, \tfrac{1}{4}, \tfrac{1}{4}\}$. The corresponding $S$-matrix reads
\begin{align}
    S = \frac{1}{2}\begin{pmatrix}
        1 & 1 & 1 & 1\\
        1 & \matminus1 & 1 & \matminus1\\
        1 & 1 & \matminus1 & \matminus1\\
        1 & \matminus1 & \matminus1 & 1
    \end{pmatrix}.
\end{align}
Now, we construct the Verlinde lines and look at their action on primaries. We have
\begin{align}
    \begin{split}
        \widehat{\mathcal{L}}_{0}\vert\phi_{i}\rangle =& \vert\phi_{i}\rangle,\\
        \widehat{\mathcal{L}}_{1}\vert\phi_{0}\rangle =& \vert\phi_{0}\rangle,\ \widehat{\mathcal{L}}_{1}\vert\phi_{1}\rangle = -\vert\phi_{1}\rangle,\  \ \widehat{\mathcal{L}}_{1}\vert\phi_{2}\rangle = \vert\phi_{2}\rangle, \widehat{\mathcal{L}}_{1}\vert\phi_{3}\rangle = -\vert\phi_{3}\rangle,\\
        \widehat{\mathcal{L}}_{2}\vert\phi_{0}\rangle =& \vert\phi_{0}\rangle,\ \widehat{\mathcal{L}}_{2}\vert\phi_{1}\rangle = \vert\phi_{1}\rangle,\ \widehat{\mathcal{L}}_{2}\vert\phi_{2}\rangle =-\vert\phi_{2}\rangle,\ \widehat{\mathcal{L}}_{2}\vert\phi_{3}\rangle = -\vert\phi_{3}\rangle,\\
        \widehat{\mathcal{L}}_{3}\vert\phi_{0}\rangle =& \vert\phi_{0}\rangle, \widehat{\mathcal{L}}_{3}\vert\phi_{1}\rangle = -\vert\phi_{1}\rangle, \widehat{\mathcal{L}}_{3}\vert\phi_{2}\rangle = -\vert\phi_{2}\rangle, \widehat{\mathcal{L}}_{3}\vert\phi_{3}\rangle = \vert\phi_{3}\rangle.
    \end{split}
\end{align}
From this, we observe that $\widehat{\mathcal{L}}_{0}$ is the trivial or the identity line. Here, we notice that all lines are invertible since there is no operator $\hat{\mathcal{L}_{i}}$ that kills a state $\vert\phi_{j}\rangle$. The two primaties $\sigma$ and $\widetilde{\sigma}$ have identical genus-one characters; we denote this common characte by $\chi_{\tfrac{1}{4}}$, so $\vert\chi_{\tfrac{1}{4}}\vert^{@}$ appears with multiplicity $2$ below. The spatial cycle defect partition functions can be found to be
\begin{align}
    \begin{split}
        \mathcal{Z}^{\mathbb{1}} =& \vert\chi_{0}\vert^{2} + \vert\chi_{\frac{1}{2}}\vert^{2} + 2\vert\chi_{\frac{1}{4}}\vert^{2},\\
        \mathcal{Z}^{\nu} =& \vert\chi_{0}\vert^{2} - \vert\chi_{\frac{1}{2}}\vert^{2},\\
        \mathcal{Z}^{\sigma} =& \vert\chi_{0}\vert^{2} -\vert\chi_{\frac{1}{2}}\vert^{2},\\
        \mathcal{Z}^{\widetilde{\sigma}} =& \vert\chi_{0}\vert^{2} + \vert\chi_{\frac{1}{2}}\vert^{2} - 2\vert\chi_{\frac{1}{4}}\vert^{2}.
    \end{split}
\end{align}
From the Verlinde operators, we see that
\begin{align}
    \widehat{\mathcal{L}}_{i}^{2} = \widehat{\mathcal{L}}_{0},\ \ \widehat{\mathcal{L}}_{1}\widehat{\mathcal{L}}_{2} = \widehat{\mathcal{L}}_{3}.
\end{align}
All three lines $\nu, \sigma, \widetilde{\sigma}$ corresponding to operators $\widehat{\mathcal{L}}_{1}, \widehat{\mathcal{L}}_{2}, \widehat{\mathcal{L}}_{3}$ respectively are invertible. 
Performing the replacement, we have
\begin{align}
    \mathcal{Z}^{E_{8}, (D_{2,1};\nu)} = \chi_{0}^{D_{2}}\chi_{0}^{D_{6}} + \chi_{\frac{1}{2}}^{D_{2}}\chi_{\frac{1}{2}}^{D_{6}} - 2\chi_{\frac{1}{4}}^{D_{2}}\chi_{\frac{3}{4}}^{D_{6}}.
\end{align}
Substituting the modular form expressions for the characters, we find
\begin{align}
    \mathcal{Z}^{E_{8}, (D_{2,1};\nu)} = \left(\frac{\eta(\tau)}{\eta(2\tau)}\right)^{8},
\end{align}
which is exactly the same defect partition function we obtained in \ref{q_defect}.
\begin{comment}
, and hence, its modular invariant subgroup is
\begin{align}
    \Gamma_{E_{8}, (D_{2,1};c)}  = \Gamma_{0}(2)\cap\Gamma^{0}(3).
\end{align}    
\end{comment}

\subsection{$\left(\textbf{III}_{2}, G_{2,1}^{\otimes 2}\right)$}
\subsubsection{Simple current extension $\mathcal{E}_{3}[A_{1,8}]$ and the emergence of $\mathbf{III}_{2}$ characters}
The theory $\mathbf{III}_{2}$ is built by the extension 
$\mathcal{E}_{3}[A_{1,8}]$, where the notation indicates that we perform a chiral algebra extension of the $A_{1,8}$ theory such that the resulting RCFT has exactly three irreducible characters. This is unlike the unextended $A_{1,8}$ theory which has nine integrable highest-weight modules labeled by $\ell = 0,1,\ldots,8$, and hence nine independent Kac-Weyl characters $\chi_{\ell}^{\widehat{\mathfrak{su}}(2)_{8}}$. The extension is implemented by an integer-spin simple current in $A_{1,8}$. For $\widehat{\mathfrak{su}}(2)_{k}$, the conformal weight of the integrable representation labeled by $\ell$ is
\begin{align}
    h_{\ell} = \frac{\ell(\ell + 2)}{4(k + 2)},\qquad \ell = 0,1,\ldots, k.
\end{align}
Specializing to $k = 8$, we find the representation $\ell = 8$, $h_{8} = 2\in\mathbb{Z}$, has integer conformal spin. Hence, it is a simple current and fusion with this permutes the integrable representations. The vacuum sector of the extended chiral algebra contains both the original vacuum and the $\ell =8$ modules
\begin{align}
    \mathcal{H}_{\mathcal{A}_{\text{ext}}} \cong \mathcal{H}_{\ell =0}\oplus \mathcal{H}_{\ell = 8}.
\end{align}
At the level of the characters, the additivity of the trace over direct sums immediately implies that the vacuum character of the extended theory is the following orbit sum for the vacuum orbit $I = \{0,8\}$ 
\begin{align}
    \chi_{0}(\tau) = \text{Tr}_{\mathcal{H}_{\mathcal{A}_{\text{ext}}}}q^{L_{0} - \frac{c}{24}} = \sum\limits_{\ell\in I}\text{Tr}_{\mathcal{H}_{\ell}}q^{L_{0} - \frac{c}{24}} = \sum\limits_{\ell\in I}\chi_{\ell}(\tau),
\end{align}
where $\mathcal{H}_{\mathcal{A}_{\text{ext}}} = \bigoplus\limits_{\ell\in I}\mathcal{H}_{\ell}$, or simply
\begin{align}
    \chi_{0}(\tau) = \chi_{\ell=0}^{\widehat{\mathfrak{su}}(2)_{8}}(\tau) + \chi_{\ell=8}^{\widehat{\mathfrak{su}}(2)_{8}}(\tau).
\end{align}
Let $J$ denote the simple current corresponding to $\ell =8$. Now, fusion by $J$ acts on integrable weights as 
\begin{align}
    J:\ell\mapsto 8-\ell,
\end{align}
and hence, the nine primaries of $A_{1,8}$ decompose into the following orbits
\begin{align}\label{A_1,8_orbits}
    \{0,8\},\qquad \{1,7\},\qquad \{2,6\},\qquad \{3,5\},\qquad \{4\}.
\end{align}
By extension via $J$, states related by the action of $J$ become part of a single irreducible module of the extended algebra, and hence, the extended characters are naturally sums over $J$-orbits. Additionally, only those $A_{1,8}$ modules that are local with respect to the extended chiral algebra survive as untwisted modules of the extension. For the extension we are considering to obtain $\mathbf{III}_{2}$, this locality condition selects precisely the even $\ell$ sectors, and hence, among the orbits \ref{A_1,8_orbits}, the allowed ones are
\begin{align}
    \{0,8\},\qquad \{2,6\},\qquad \{4\},
\end{align}
while odd-$\ell$ orbits $\{1,7\}$ and $\{3,5\}$ are projected out of the untwisted spectrum. Equivalently, the untwisted modules are those with vanishing monodromy charge $Q_{J}(\ell) = h_{\ell} + h_{J} - h_{J\ell}\equiv 0\ (\text{od}\ 1)$, which here selects $\ell$. The result is three irreducible modules of the extended algebra with $A_{1,8}\cong \widehat{\mathfrak{su}}(2)_{8}$ characters that read \cite{Ji:2019eqo}
\begin{align}\label{A_1,8_characters}
   \begin{split}
    \widetilde{\chi}_{0}^{\mathbf{III}_{2}} =& \chi_{\ell = 0}^{\widehat{\mathfrak{su}}(2)_{8}} + \chi_{\ell = 8}^{\widehat{\mathfrak{su}}(2)_{8}},\\
    \widetilde{\chi}_{\frac{1}{5}}^{\mathbf{III}_{2}} =& \chi_{\ell = 2}^{\widehat{\mathfrak{su}}(2)_{8}} + \chi_{\ell = 6}^{\widehat{\mathfrak{su}}(2)_{8}},\\
    \widetilde{\chi}_{\frac{3}{5}}^{\mathbf{III}_{2}} =& \chi_{\ell = 4}^{\widehat{\mathfrak{su}}(2)_{8}},
   \end{split}
\end{align}
where
\begin{align}
    \chi_{\ell}^{\widehat{\mathfrak{su}}(2)_{k}}(\tau) = \frac{q^{\frac{(2\ell + 1)^{2}}{4(k + 2)}}}{\eta(\tau)^{3}}\sum\limits_{n\in\mathbb{Z}}\left[2\ell + 1 + 2n(k + 2)\right]q^{n[2\ell + 1 + n(k+2)]}.
\end{align}
Working out the $q$-series expansions for different $\ell$, we have\footnote{We note that in \cite{Ji:2019eqo}, the expressions are given with respect to $j\in\frac{1}{2}\mathbb{Z}$, but in our convention, $\ell = 2j$}
\begin{align}
    \begin{split}
        \chi_{\ell = 0}^{\widehat{\mathfrak{su}}(2)_{8}} =& q^{-\frac{1}{10}}\left(1 + 3q + 9q^{2} + 22q^{3} + 51q^{4} + \ldots\right),\\
        \chi_{\ell = 2}^{\widehat{\mathfrak{su}}(2)_{8}} =& q^{\frac{1}{10}}\left(3 + 9q + 27q^{2} + 66q^{3} + 153q^{4} + \ldots\right),\\
        \chi_{\ell = 4}^{\widehat{\mathfrak{su}}(2)_{8}} =& q^{\frac{1}{2}}\left(5 + 15q + 45q^{2} + 110q^{3} + 255q^{4} + \ldots\right),\\
        \chi_{\ell = 6}^{\widehat{\mathfrak{su}}(2)_{8}} =& q^{\frac{11}{10}}\left(7 + 21q + 63q^{2} + 141q^{3} + 318q^{4} + \ldots\right),\\
        \chi_{\ell = 8}^{\widehat{\mathfrak{su}}(2)_{8}} =& q^{\frac{19}{10}}\left(9 + 16q + 48q^{2} + 99q^{3} + 217q^{4} + \ldots\right).
    \end{split}
\end{align}
Using this in \ref{A_1,8_characters}, we find $\mathbf{III}_{2}$ characters to have the following $q$-series expansions
\begin{align}
    \begin{split}
        \widetilde{\chi}_{0}^{\mathbf{III}_{2}}(\tau) =& q^{-\frac{1}{10}}\left(1 + 3q + 18q^{2} + 38q^{3} + 99q^{4} + \ldots\right),\\
        \widetilde{\chi}_{\frac{1}{5}}^{\mathbf{III}_{2}}(\tau) =& q^{\frac{1}{10}}\left(3 + 16q + 48q^{2} + 129q^{3} + 294q^{4} + \ldots\right),\\
        \widetilde{\chi}_{\frac{3}{5}}^{\mathbf{III}_{2}}(\tau) =& q^{\frac{1}{2}}\left(5 + 15q + 45q^{2} + 110q^{3} + 255q^{4} + \ldots\right).
    \end{split}
\end{align}
From these expansions, it is immediately clear that the vacuum module has $3$ weight-one currents matching $\text{dim}(\mathfrak{su}(2)) = 3$. The other primaries have ground-state multiplicities $3$ and $5$, naturally matching the $\mathfrak{su}(2)$ spin-$1$ and spin-$2$ reps $\mathbf{3}$ and $\mathbf{5}$, respectively as noted in \ref{26_branch} and \ref{52_branch}. Next, for $G_{2,1}$, we use the fact that we can express its character decomposition in terms of $\widehat{\mathfrak{su}}(2)_{3}\times\widehat{\mathfrak{u}}(1)_{2}$ characters as follows \cite{Ji:2019eqo}
\begin{align}\label{G_2,1_characters}
    \begin{split}
        \chi_{0}^{G_{2,1}} =& \chi_{0}^{\widehat{\mathfrak{u}}(1)_{2}}\chi_{\ell = 0}^{\widehat{\mathfrak{su}}(2)_{3}} + \chi_{1}^{\widehat{\mathfrak{u}}(1)_{2}}\chi_{\ell = 3}^{\widehat{\mathfrak{su}}(2)_{3}},\\
        \chi_{\frac{2}{5}}^{G_{2,1}} =& \chi_{1}^{\widehat{\mathfrak{u}}(1)_{2}}\chi_{\ell = 1}^{\widehat{\mathfrak{su}}(2)_{3}} + \chi_{0}^{\widehat{\mathfrak{u}}(1)_{2}}\chi_{\ell = 2}^{\widehat{\mathfrak{su}}(2)_{3}},
    \end{split}
\end{align}
where the $\widehat{\mathfrak{u}}(1)_{2}$ characters read
\begin{align}
    \begin{split}
        \chi_{0}^{\widehat{\mathfrak{u}}(1)_{2}}(\tau) =& \left(\frac{\theta_{3}}{\eta}\right)(\tau) =q^{-\frac{1}{24}}\left(1 + 3q + 4q^{2} + 7q^{3} + 13q^{4} + \ldots\right)\\
        \chi_{1}^{\widehat{\mathfrak{u}}(1)_{2}}(\tau) =& \left(\frac{\theta_{2}}{\eta}\right)(\tau) = q^{\frac{1}{12}}\left(2 + 2q + 6q^{2} + 8q^{3} + 14q^{4} + \ldots\right).
    \end{split}
\end{align}
Substituting this in \ref{G_2,1_characters}, we obtain the following $q$-series expansions
\begin{align}
    \begin{split}
        \chi_{0}^{G_{2,1}}(\tau) =& q^{-\frac{7}{60}}\left(1 + 14q + 42q^{2} + 140q^{3} + 350q^{4} + \ldots\right),\\
        \chi_{\frac{2}{5}}^{G_{2,1}}(\tau) =& q^{\frac{17}{60}}\left(7 + 34q + 119q^{2} + 322q^{3} + 819q^{4} + \ldots\right).
    \end{split}
\end{align}
Here, we find that the vacuum module has $14$ currents at weight-one matching $\text{dim}(G_{2}) = 14$, and for the non-trivial primary we find ground-state multiplicity $7$ that matches the dimension of the fundamental rep. From these characters, we can now define the three orbit-summed tensor characters of weights $\{0, \tfrac{2}{5}, \tfrac{4}{5}\}$ as follows
\begin{align}\label{G_2,1x2_characters}
    \begin{split}
        \chi_{0}^{G_{2,1}^{\otimes2}}(\tau) =& \left(\chi_{0}^{G_{2,1}}\right)^{2}(\tau),\\
        \chi_{\frac{2}{5}}^{G_{2,1}^{\otimes2}}(\tau) =& 2\left(\chi_{0}^{G_{2,1}}\chi_{\frac{2}{5}}^{G_{2,1}}\right)(\tau),\\
        \chi_{\frac{4}{5}}^{G_{2,1}^{\otimes2}}(\tau) =& \left(\chi_{\frac{2}{5}}^{G_{2,1}}\right)^{2}(\tau).
    \end{split}
\end{align}
We consider a commutant pair with central charges $(c,\widetilde{c}) = \left(\tfrac{12}{5}, \tfrac{28}{5}\right)$, and hence, using \ref{G_2,1x2_characters} and \ref{A_1,8_characters}, we build
\begin{align}
    \chi^{E_{8}} = \widetilde{\chi}^{\textbf{III}_{2}}_{0}\chi^{G_{2}^{\otimes 2}}_{0} + \widetilde{\chi}^{\textbf{III}_{2}}_{\tfrac{1}{5}}\chi^{G_{2}^{\otimes 2}}_{\tfrac{4}{5}} + \widetilde{\chi}^{\textbf{III}_{2}}_{\tfrac{3}{5}}\chi^{G_{2}^{\otimes 2}}_{\tfrac{2}{5}}.
\end{align}
The contribution of the vacuum piece reads
\begin{align}
    \widetilde{\chi}^{\textbf{III}_{2}}_{0}\chi^{G_{2}^{\otimes 2}}_{0} = q^{-\frac{1}{3}}\left(1 + (3 + 28)q + \ldots\right),
\end{align}
which contributes $3 + 28 = 31$ weight-one states with the split $14+14 = 28$ comes from the vacuum tensor product character of $G_{2,1}^{\otimes}$. Hence, we have $31 = 14 + 14+ 3$, and the following branching rule for the vacuum module
\begin{align}
    (\mathbf{14}, \mathbf{1}, \mathbf{1})\oplus (\mathbf{1}, \mathbf{14}, \mathbf{1})\oplus (\mathbf{1}, \mathbf{1}, \mathbf{3}).
\end{align}
Next, from $\widetilde{\chi}_{\tfrac{1}{5}}^{\mathbf{III}_{2}}\chi_{\tfrac{4}{5}}^{G_{2,1}^{\otimes2}}$, we find
\begin{align}
    \chi_{\frac{1}{5}}^{\mathbf{III}_{2}}\chi_{\frac{4}{5}}^{G_{2,1}^{\otimes2}} = q^{-\frac{1}{3}}\left(147q + 2212q^{2} + \ldots\right),
\end{align}
where $147 = 3\times 49 = 3\times (7\cdot 7)$, which is exactly the dimension $(\mathbf{7}, \mathbf{7}, \mathbf{3})$. Finally, the contribution of $\widetilde{\chi}_{\tfrac{3}{5}}^{\mathbf{III}_{2}}\chi_{\tfrac{2}{5}}^{G_{2,1}^{\otimes2}}$ reads
\begin{align}
    \widetilde{\chi}_{\frac{3}{5}}^{\mathbf{III}_{2}}\chi_{\frac{2}{5}}^{G_{2,1}^{\otimes2}} = q^{-\frac{1}{3}}\left(70q + 1530q^{2} + \ldots\right),
\end{align}
where $70= 5\times14 = 5\times (7+7)$, which matches $(\mathbf{7}, \mathbf{1}, \mathbf{5})\oplus (\mathbf{1}, \mathbf{7}, \mathbf{5})$. Putting all the weight-one contributions together, we get $31 + 147 + 70 = 248$, and the decomposition is exactly the branching we found in \ref{E8_to_G2G2A1}.

\subsubsection{MTCs for the commutant pair}
We next turn our discussion to the MTC content behind the bilinear pairing $\mathbf{III}_{2}\otimes G_{2,1}^{\otimes2}\to E_{8,1}$. We first identify the MTC of the lower central charge factor $\mathbf{III}_{2}$, and then explain precisely why $G_{2,1}^{\otimes2}$ is $\text{Fib}^{2}$, the squared Fibonacci MTC, while the associated chiral data naturally organizes into three characters but with four primaries.\\

\noindent Let $\mathcal{C}_{G_{2,1}} = \text{Rep}(G_{2,1})$ denote the UMTC attached to the rational VOA $G_{2,1}$. This is also called the ribbon category of integrable highest-weight $\widehat{\mathfrak{g}}_{2}$ modules at level $1$ arising from the $G_{2}$ WZW model \cite{Huang:2005gs}. This theory has exactly two primaries, $I_{\mathcal{C}_{G_{2,1}}} = \{\mathbb{1}, \tau\}$, with conformal weights $\{0, \tfrac{2}{5}\}$ respectively. Moreover, the non-trivial primary obeys the Fibonacci fusion rule
\begin{align}\label{Fib_fusion}
    \tau\otimes\tau = \mathbb{1}\oplus\tau.
\end{align}
The rank-two modular data of $G_{2,1}$ coincides with the Fibonacci UMTC \cite{Davydov:2011pp}. Hence, we simply have 
\begin{align}\label{Fib_identification}
    \mathcal{C}_{G_{2,1}}\simeq \text{Fib},
\end{align}
where Fib denotes the unitary Fibonacci MTC of rank $2$ and fusion \ref{Fib_fusion}. At the level of chiral algebras, we take the tensor product VOA $G_{2,1}^{\otimes2}$. By general results on tensor product VOAs, its category of representations is equivalent to the Deligne tensor product of the representation categories of the individual factors \cite{McRae:2023vyw}
\begin{align}\label{RepG_2,1x2}
    \text{Rep}\left(G_{2,1}^{\otimes2}\right) \simeq \text{Rep}\left(G_{2,1}\right)\boxtimes\text{Rep}\left(G_{2,1}\right).
\end{align}
Hence, combining with \ref{Fib_identification}, we finally have
\begin{align}
    \text{Rep}\left(G_{2,1}^{\otimes2}\right) \simeq\text{Fib}\boxtimes\text{Fib}.
\end{align}
As a chiral RCFT, $G_{2,1}^{\otimes2}$ has four primaries, corresponding to the following pairs
\begin{align}\label{pairs_primaries}
    (\mathbb{1}, \mathbb{1}),\qquad (\mathbb{1}, \tau),\qquad (\tau, \mathbb{1}), \qquad (\tau, \tau).
\end{align}
However, the associated torus characters satisfy an accidental degeneracy since the following two factors are identical
\begin{align}
    \chi_{(\mathbb{1}, \tau)} = \chi_{\mathbb{1}}\chi_{\tau} = \chi_{\tau}\chi_{\mathbb{1}} = \chi_{(\tau, \mathbb{1})}.
\end{align}
Hence, although there are four primaries, there are only three distinct characters. We choose these to be the vacuum-vacuum character $\chi^{(2)}\equiv \chi_{\mathbb{1}}\chi_{\mathbb{1}}$, the mixed-sector character $\chi_{1}^{(2)}\equiv \chi_{\mathbb{1}}\chi_{\tau} = \chi_{\tau}\chi_{\mathbb{1}}$, and the non-trivial pairing character\footnote{This character corresponds to the two inequivalent simples $(\mathbb{1}, \tau)$ and $(\tau, \mathbb{1})$ that share the same characters. In the language of \cite{Das:2022uoe}, this is the same phenomenon tracked by the pair $(d_{1}, d_{2})$ of multiplicities appearing in the modular invariant built from the three characters.} $\chi_{2}^{(2)} \equiv \chi_{\tau}\chi_{\tau}$. Let $U\subseteq A$ be a vertex subalgebra and let $V \equiv \text{Com}_{A}(U)$ denote its commutant. Then, as shown in \cite{McRae:2021urf}, the tensor subcategory of $\text{Rep}(U)$ generated by $U$-modules appearing in $A$, and the corresponding tensor subcategory of $\text{Rep}(V)$ generated by associated $V$-modules, are related by a braid-reversed tensor equivalence, or a mirror equivalence. Applying this to our embedding, we find that $\mathcal{C}_{\mathbf{III}_{2}}$ is the ribbon reverse of $\mathcal{C}_{G_{2,1}}$. Concretely, if  $\overline{\mathcal{C}}$ denotes the modular category obtained from $\mathcal{C}$ by inverting braiding and twist, denoted as rev, we have that 
\begin{align}
    \mathcal{C}_{\mathbf{III}_{2}}\simeq \overline{\mathcal{C}_{G_{2,1}^{\otimes2}}},\qquad \text{Rep}(\mathbf{III}_{2})\simeq\left(\text{Fib}\boxtimes\text{Fib}\right)^{\text{rev}}.
\end{align}
From \ref{Fib_fusion}, we notice that there aren't any invertible simples and hence, the Picard group is trivial
\begin{align}
    \text{Pic}(\mathbf{III}_{2})\cong1.
\end{align}
Also, $\text{Rep}\left(G_{2,1}^{\otimes2}\right)$ described in \ref{RepG_2,1x2} has a trivial Picard group, hence no non-trivial invertible simples; by mirror equivalence the same holds for $\mathbf{III}_{2}$. This is a strong structural constraint since a trivial Picard group implies that there are no $\mathbb{Z}_{n}$ simple current defects or standard group-like replacement rules based on invertible lines, as we will soon see in the next subsection. Lastly, we note that although there aren't any invertible lines, the theory still has topological lines with non-integer quantum dimensions. In $\text{Fib}^{\boxtimes2}$, the four primaries can be thought of as the pairs \ref{pairs_primaries} with quantum dimensions\footnote{Quantum dimensions satisfy $d_{X\otimes Y} = d_{X}d_{Y}$ and $d_{X\oplus Y} = d_{X} + d_{Y}$, so from the fusion rule \ref{Fib_fusion}, we have $d^{2}_{\tau} = d_{\mathbb{1}} + d_{\tau} = 1 + d_{\tau}$, and solving this yields $d_{\tau} = \varphi$. Since dimensions multiply under the Deligne product $\text{Fib}^{\boxtimes2}$, we find dimensions $\{1, \varphi, \varphi, \varphi^{2}\}$.} $1, \varphi, \varphi, \varphi^{2}$, where $\varphi = \tfrac{1 + \sqrt{5}}{2}$ is the Golden ratio. These are genuine topological line operators, but only $(\mathbb{1}, \mathbb{1})$ is invertible.

\subsubsection{Defect partition functions}
The defect that we consider is for the lower central charge, i.e., $\textbf{III}_{2}$. To find the corresponding $S$-matrix, we refer to the procedure outlined in appendix \ref{sec:appendix_S}. 
Plugging in $c = \tfrac{12}{5}$ corresponding to $\textbf{III}_{2}$ with $(h_{1}, h_{2}) = \left(\tfrac{1}{5}, \tfrac{3}{5}\right)$, we have
\begin{align}
    \begin{split}
        \alpha_{0} =& -\frac{1}{10},\ \ \alpha_{1} = \frac{1}{10},\ \ \alpha_{2} = \frac{1}{2},\\
        \mu_{1} =& \frac{59}{225},\ \ \nu_{1} = \frac{49}{25},\\
        a =& -\frac{3}{10},\ \ g = \frac{2}{5}.
    \end{split}
\end{align}
With this, we find the following $S$-matrix
\begin{align}
    \begin{split}
        S = \frac{1}{\sqrt{5}}\begin{pmatrix}
            \frac{\sqrt{5} - 5}{2} & \frac{\sqrt{5} + 5}{2} & 2\\
            \frac{\sqrt{5} + 5}{2} & \frac{\sqrt{5} - 5}{2} & \matminus2\\
            1 & \matminus1 & 1
        \end{pmatrix},
    \end{split}
\end{align}
where we have used $(n_{1}, n_{2}) = (3,5)$ from \cite{Das:2022uoe}. Note that plugging in $\widetilde{c} = \tfrac{28}{5}$ corresponding to $G_{2,1}^{\otimes2}$ with $(\widetilde{h}_{1}, \widetilde{h}_{2}) = \left(\tfrac{4}{5}, \tfrac{2}{5}\right)$, we have
\begin{align}
    \begin{split}
        \alpha_{0} =& -\frac{7}{30},\ \ \alpha_{1} = \frac{17}{30},\ \ \alpha_{2} = \frac{1}{6},\\
        \mu_{1} =& -\frac{119}{225},\ \ \nu_{1} = \frac{127}{75},\\
        a =& \frac{3}{10},\ \ g = -\frac{12}{5},
    \end{split}
\end{align}
and a computation using these coefficients yields the exact same $S$-matrix except now with $(n_{1}, n_{2}) = (49, 7)$. Now, we construct the Verlinde lines and look at their action on the primaries. We have
\begin{align}
    \begin{split}
        \widehat{\mathcal{L}}_{0}\vert\phi_{i}\rangle =& \vert\phi_{i}\rangle,\\
        \widehat{\mathcal{L}}_{1}\vert\phi_{0}\rangle =& \alpha\vert\phi_{0}\rangle,\ \ \ \ \ \ \ \ \widehat{\mathcal{L}}_{1}\vert\phi_{1}\rangle = \frac{1}{\alpha}\vert\phi_{1}\rangle,\ \ \ \ \ \ \ \ \widehat{\mathcal{L}}_{1}\vert\phi_{2}\rangle = -\vert\phi_{2}\rangle,\\
        \widehat{\mathcal{L}}_{2}\vert\phi_{0}\rangle =& \frac{1}{\varphi - 3}\vert\phi_{0}\rangle,\ \ \widehat{\mathcal{L}}_{2}\vert\phi_{1}\rangle = -\frac{1}{\varphi\sqrt{5}}\vert\phi_{1}\rangle,\ \ \widehat{\mathcal{L}}_{2}\vert\phi_{2}\rangle = \frac{1}{2}\vert\phi_{2}\rangle,
    \end{split}
\end{align}
where $\alpha = \tfrac{\sqrt{5} + 5}{\sqrt{5} - 5} = -(1 + \varphi)$. From this, we observe that $\widehat{\mathcal{L}}_{0}$ is the trivial or the identity line. Here, none of the non-trivial lines are invertible. Using \ref{Z_L}, the spatial cycle defect partition functions read
\begin{align}
    \begin{split}
        \mathcal{Z}^{\mathbb{1}} =& \vert\chi_{0}\vert^{2} + \vert\chi_{1}\vert^{2} + \vert\chi_{2}\vert^{2},\\
        \mathcal{Z}^{\widehat{\mathcal{L}}_{1}} =& \alpha\vert\chi_{0}\vert^{2} + \frac{1}{\alpha}\vert\chi_{1}\vert^{2} - \vert\chi_{2}\vert^{2},\\
        \mathcal{Z}^{\widehat{\mathcal{L}}_{2}} =& \frac{2}{\sqrt{5} - 5}\vert\chi_{0}\vert^{2} - \frac{2}{\sqrt{5} + 5}\vert\chi_{1}\vert^{2} + \frac{1}{2}\vert\chi_{2}\vert^{2}.
    \end{split}
\end{align}
This is exactly what we should expect categorically since $\text{Fib}$ has no non-trivial invertible object, and hence, $\text{Fib}^{\boxtimes2}$ also has only the trivial invertible object. By the mirror equivalence, the same holds for $\mathbf{III}_{2}$. The $\mathbb{Z}_{2}$ defect $g$ that is relevant is the permutation defect that exchanges two primaries sitting over the fixed points $\ell = 4$ of the $A_{1,8}$ simple current extension \cite{Schellekens:1990xy}. Let us denote the two primaries that share the $\chi_{2}$ character by $\phi_{\tfrac{3}{5}}^{+}$ and $\phi_{\tfrac{3}{5}}^{-}$, with characters $\chi_{\tfrac{3}{5},\pm}^{\mathbf{III}_{2}}$ satisfying $\chi_{\tfrac{3}{5},\pm}^{\mathbf{III}_{2}} = \chi_{\tfrac{3}{5}}^{\mathbf{III}_{2}}$. The $\mathbb{Z}_{2}$ permutation defect is then given by
\begin{align}
    g: \phi_{\frac{3}{5}}^{+}\leftrightarrow\phi_{\frac{3}{5}}^{-},\qquad \phi_{0}\mapsto \phi_{0},\qquad \phi_{\frac{1}{5}}\mapsto \phi_{\frac{1}{5}}.
\end{align}
We can now express the resulting swapped defect partition function as follows
\begin{align}
    \mathcal{Z}^{g} = \left\vert\left.\chi_{0}^{\mathbf{III}_{2}}\right\vert\right.^{2} +  \left\vert\left.\chi_{\tfrac{1}{5}}^{\mathbf{III}_{2}}\right\vert\right.^{2} +  2\left\vert\left.\chi_{\tfrac{3}{5}}^{\mathbf{III}_{2}}\right\vert\right.^{2}.
\end{align}
We note here that in the bilinear replacement rule, nothing happens since the permutation symmetry of $\mathbf{III}_{2}$ is invisible. However, this permutation partition function is interesting to analyze. In terms of modular forms, this simplifies to 
\begin{align}
    \mathcal{Z}^{g}(\tau, \overline{\tau}) = \frac{1}{\vert\eta(\tau)\vert^{6}}\left(\vert\Theta_{1,10}' + \Theta_{9,10}'\vert^{2} + \vert\Theta_{3,10}' + \Theta_{7,10}'\vert^{2} + 2\vert\Theta_{5,10}'\vert^{2}\right),
\end{align}
where we have expressed the $\widehat{\mathfrak{su}}(2)_{8}$ characters as
\begin{align}
    \chi^{\widehat{\mathfrak{su}}(2)_{8}}_{\ell}(\tau) = \left(\frac{\Theta_{\ell+1, 10}'}{\eta^{3}}\right),(\tau)
\end{align}
where we have defined the theta derivative sums
\begin{align}
    \Theta_{r,10}'(\tau)\equiv \sum\limits_{n\in\mathbb{Z}}(20n + r)q^{\frac{(20n + r)^{2}}{40}},\qquad r\in\mathbb{Z}.
\end{align}
Expressing this partition function in the twisted-twining notation, we have $Z\left[\begin{smallmatrix}
    e\\g
\end{smallmatrix}\right]$, where $g$ is inserted along the spatial cycle while the temporal boundary condition is left untwisted. Under $\gamma\in \left(\begin{smallmatrix}
    a & b\\c & d
\end{smallmatrix}\right)\in\text{SL}(2, \mathbb{Z})$, the boundary condition transforms as follows \cite{Dijkgraaf:1989hb, Dijkgraaf:1989hb}
\begin{align}
    (g,h)\mapsto (h^{c}g^{d}, h^{a}g^{b}).
\end{align}
Hence, starting from $(g,e)$, we end up on $(g^{d},g^{b})$. Now, to preserve the sector $(g,e)$, we require $g^{b} = e$, and since $g$ has order $2$, this is precisely the condition $b\equiv 0$ (mod $2$). Also, since $\text{det}\ \gamma = ad-bc = 1$ and $b$ is even, it follows that $d$ is odd, and we have $g^{d} = g$. So the spatial twist is automatically preserved once $b\equiv 0$ (mod $2$). Hence, the invariance subgroup is 
\begin{align}
    \Gamma^{0}(2)\equiv \left\{\begin{pmatrix}
        a & b\\ c & d
    \end{pmatrix}\in\text{SL}(2, \mathbb{Z})\vert b\equiv 0\ (\text{mod}\ 2)\right\}.
\end{align}
Hence, the permutation defect partition function $\mathcal{Z}^{g}(\tau, \overline{\tau})$ is invariant under the congruence subgroup $\Gamma^{0}(2)$. The $S$-move exchanges the cycles, i.e.,  it sends the spatial-defect amplitude to the corresponding twinned temporal-defect amplitude $(g,e)\mapsto (e,g^{-1}) = (e,g)$. The $T$-move sends $(g,e)\mapsto (g,g)$, thus exactly matching the subgroup argument.

%\NBU{What specific partition function are we interested in here so that we can perform the replacement rule.} \AD{will think about this..}

%\AD{Explain how the MTC of $G_{2,1}^{\otimes 2}$ is $\text{Fib}^2$ with the RCFT now having 3 characters but 4 primaries..}
%\NBU{AD's point has been addressed and answered in detail}.

\subsection{$\left(A_{2,1}^{\otimes2}, A_{2,1}^{\otimes2}\right)$}
We consider a commutant pair with central charges $(c,\widetilde{c}) = \left(4, 4\right)$, and hence, out of this pair we build
\begin{align}
    \chi^{E_{8}} = \chi^{A_{2}^{\otimes2}}_{0}\chi^{A_{2}^{\otimes2}}_{0} + \chi^{A_{2}^{\otimes2}}_{\tfrac{1}{3}}\chi^{A_{2}^{\otimes2}}_{\tfrac{2}{3}} + \chi^{A_{2}^{\otimes2}}_{\tfrac{2}{3}}\chi^{A_{2}^{\otimes2}}_{\tfrac{1}{3}}.
\end{align}
The defect that we consider is for $A_{2,1}^{\otimes2}$. Plugging in $c = 4$, with $(h_{1}, h_{2}) = \left(\tfrac{1}{3}, \tfrac{2}{3}\right)$, we have
\begin{align}
    \begin{split}
        \alpha_{0} =& -\frac{1}{6},\ \ \alpha_{1} = \frac{1}{6},\ \ \alpha_{2} = \frac{1}{2},\\
        \mu_{1} =& \frac{1}{3},\ \ \nu_{1} = \frac{17}{9},\\
        a =& -\frac{1}{6},\ \ g = 0.
    \end{split}
\end{align}
With this, we find the following $S$-matrix
\begin{align}
    \begin{split}
        S = \frac{1}{3}\begin{pmatrix}
            1 & 4 & 4\\
            1 & 1 & \matminus2\\
            1 & \matminus2 & 1
        \end{pmatrix},
    \end{split}
\end{align}
where we have used $(n_{1}, n_{2}) = (3,9)$ from \cite{Das:2022uoe}. Now, we construct Verlinde lines and look at their action on the primaries. We have
\begin{align}
    \begin{split}
        \widehat{\mathcal{L}}_{0}\vert\phi_{i}\rangle =& \vert\phi_{i}\rangle,\\
        \widehat{\mathcal{L}}_{1}\vert\phi_{0}\rangle =& \vert\phi_{0}\rangle,\ \widehat{\mathcal{L}}_{1}\vert\phi_{1}\rangle = \frac{1}{4}\vert\phi_{1}\rangle,\ \ \ \widehat{\mathcal{L}}_{1}\vert\phi_{2}\rangle = -\frac{1}{2}\vert\phi_{2}\rangle,\\
        \widehat{\mathcal{L}}_{2}\vert\phi_{0}\rangle =& \vert\phi_{0}\rangle,\ \widehat{\mathcal{L}}_{2}\vert\phi_{1}\rangle = -\frac{1}{2}\vert\phi_{1}\rangle,\ \widehat{\mathcal{L}}_{2}\vert\phi_{2}\rangle = \frac{1}{4}\vert\phi_{2}\rangle,
    \end{split}
\end{align}
For $A_{2,1}\cong\widehat{\mathfrak{su}}(3)_{1}$, all three primaries have quantum dimension one and satisfy the following fusion rules
\begin{align}
    \begin{split}
        \phi_{1}\otimes\phi_{1} =& \phi_{2},\\
        \phi_{1}\otimes \phi_{2} =& \phi_{0},\\
        \phi_{2}\otimes\phi_{2} =& \phi_{1}.
    \end{split}
\end{align}
Every primary is invertible and the group of simple currents coincides with the full set of primaries. Hence, the Picard group of $A_{2,1}$ is 
\begin{align}
    \text{Pic}\left(A_{2,1}\right) \cong \mathbb{Z}_{3}.
\end{align}
Next, for the tensor product theory, $A_{2,1}^{\otimes2}$, we label the primaries by pairs $(\phi_{a}, \phi_{b})$ with $a,b\in\{0,1,2\}$, and define fusion component-wise as follows
\begin{align}
    \left(\phi_{a}, \phi_{b}\right)\otimes\left(\phi_{a'}, \phi_{b'}\right) = \left(\phi_{a}\otimes\phi_{a'}, \phi_{b}\otimes\phi_{b'}\right).
\end{align}
Since all primaries in each factor are invertible, every primary in the product theory is again invertible. Hence, the Picard group factorizes as follows
\begin{align}\label{Pic_A2,1}
    \text{Pic}\left(A_{2,1}^{\otimes2}\right)\cong\mathbb{Z}_{3}^{2}.
\end{align}
A convenient choice of generators is provided by the simple currents
\begin{align}
    g_{1} = (\phi_{1}, \phi_{0}),\qquad g_{2} = (\phi_{0}, \phi_{1}),
\end{align}
which obey
\begin{align}
    g_{1}^{3} = g_{2}^{3} = \mathbb{1},\qquad g_{1}g_{2} = g_{2}g_{1}.
\end{align}
Every element of $\text{Pic}(A_{2,1}^{\otimes2})$ can be expressed uniquely as $(\phi_{a}, \phi_{b})$, and hence, the Picard group coincides with the full group of simple currents of the theory. Clearly, there is no order-two, i.e., a $\mathbb{Z}_{2}$ invertible line in the tensor product theory since every line has order $3$. To find the explicit form of the defect partition function, we first regard $\widehat{\mathcal{L}}_{1}$ and $\widehat{\mathcal{L}}_{2}$ as two commuting generators whose powers generate the full $\mathbb{Z}_{3}^{2}$ defect family on the equatorial amplitude. Concretely, for $(m, n)\in \mathbb{Z}_{3}^{2}$, we define the defect operator
\begin{align}
    R_{m,n}\equiv \widehat{\mathcal{L}}_{1}^{m}\widehat{\mathcal{L}}_{2}^{n},
\end{align}
which acts diagonally on the three blocks by the eigenvalues
\begin{align}
    \lambda_{0}(m,n) = 1,\qquad \lambda_{1}(m,n) = \left(\frac{1}{4}\right)^{m}\left(-\frac{1}{2}\right)^{n},\qquad \lambda_{2}(m,n) = \left(-\frac{1}{2}\right)^{m}\left(\frac{1}{4}\right)^{n}.
\end{align}
Threading $R_{m,n}$ along the temporal cycle on the left hemisphere while keeping right hemisphere untouched, we obtain the following holomorphic defect amplitude
\begin{align}\label{Z_m,n}
    \mathcal{Z}_{m,n}(\tau) = \chi_{0}(\tau)\chi_{0}(\tau) + \lambda_{1}(m,n)\chi_{\frac{1}{3}}(\tau)\chi_{\frac{2}{3}}(\tau) + \lambda_{2}(m,n)\chi_{\frac{2}{3}}(\tau)\chi_{\frac{1}{3}}(\tau),
\end{align}
where the characters were defined in \ref{characters_A_2,1}. We can now immediately write down all the representative elements, namely the identity line $(0,0)$, the first generator $(1,0)$, the second generator $(0,1)$, and the diagonal element $(1,1)$. The identity line reads
\begin{align}
    \begin{split}
    \mathcal{Z}_{0,0} =&  \chi_{0}\chi_{0} + \chi_{\frac{1}{3}}\chi_{\frac{2}{3}} + \chi_{\frac{2}{3}}\chi_{\frac{1}{3}}\\
    =& \left(\frac{\Theta_{0}^{4} + 12\Theta_{0}^{2}\Theta_{1}^{2} + 8\Theta_{0}\Theta_{1} + 6\Theta_{1}^{4}}{\eta^{8}}\right)(\tau)\\
    =& \chi^{E_{8}}(\tau),
    \end{split}
\end{align}
where the theta series were defined in \ref{theta_A_2}. The remaining defect partition functions read
\begin{align}\label{Z_cases}
    \begin{split}
        \mathcal{Z}_{1,0} =&  \chi_{0}\chi_{0} + \frac{1}{4}\chi_{\frac{1}{3}}\chi_{\frac{2}{3}} -\frac{1}{2}\chi_{\frac{2}{3}}\chi_{\frac{1}{3}},\\
        \mathcal{Z}_{0,1} =&  \chi_{0}\chi_{0} - \frac{1}{2}\chi_{\frac{1}{3}}\chi_{\frac{2}{3}} + \frac{1}{4}\chi_{\frac{2}{3}}\chi_{\frac{1}{3}},\\
        \mathcal{Z}_{1,1}=&  \chi_{0}\chi_{0} - \frac{1}{8}\chi_{\frac{1}{3}}\chi_{\frac{2}{3}} - \frac{1}{8}\chi_{\frac{2}{3}}\chi_{\frac{1}{3}}.
    \end{split}
\end{align}
This makes completely concrete how the Picard group \ref{Pic_A2,1} deforms the equatorial gluing. Plugging in the expressions for the characters, we find that the defect partition functions \ref{Z_cases} are equal to each other, and we obtain the following unique modular form expression
\begin{align}
    \begin{split}
        \mathcal{Z}^{E_{8}, (A_{2,1}^{\otimes2}; (m,n))}(\tau) =& \left(\frac{\Theta_{0}^{4} + 3\Theta_{0}^{2}\Theta_{1}^{2} - \Theta_{0}\Theta_{1}^{3} + \tfrac{15}{4}\Theta_{1}^{4}}{\eta^{8}}\right)(\tau)\\ 
        =& q^{-\frac{1}{3}}\left(1 + 27q^{\frac{2}{3}} + 5q + \frac{1215}{4}q^{\frac{4}{3}} + 594q^{\frac{5}{3}} - 7q^{2} + \ldots\right),
    \end{split}
\end{align}
for $(m,n)\neq (0,0)$. The theta series $\Theta_{0,1}$ are modular forms of $\Gamma(3)$, and hence
\begin{align}
    F(\tau)\equiv \left(\eta^{8}\mathcal{Z}^{E_{8}, (A_{2,1}^{\otimes2}; (m,n))}\right)(\tau),
\end{align}
is a weight-$4$ modular form on $\Gamma(3)$. Hence, we have
\begin{align}
    \Gamma_{E_{8}, \left(A_{2,1}^{\otimes2}; (m,n)\right)} = \Gamma(3),
\end{align}
to be the subgroup under which the defect partition functions, for $(m,n)\neq (0,0)$, are invariant under. The fact that $\mathcal{Z}_{(m,n)}$ for $(m,n)\neq (0,0)$ is invariant only under $\Gamma(3)$ means that the corresponding TDLs are $\mathbb{Z}_{3}$-graded, or equivalently, they define torus amplitudes in a fixed non-trivial flat $\mathbb{Z}_{3}$ background. This can be made explicit by writing the $\text{SL}(2, \mathbb{Z})$ action as $T: (m,n)\mapsto (m,m+n)$ mod $3$, and $S: (m,n)\mapsto (n,-m)$ mod $3$. Hence, the full $\text{SL}(2, \mathbb{Z})$ mapping class group acts non-trivially by permuting the defect-twisted sectors, and $\Gamma(3)$, generated by $T^{3}$ and $ST^{3}S^{-1}$, is precisely the kernel of this action.

\section{Conclusion and Discussions}
In this work, we developed an \textit{equatorial projection} framework for accessing topological defect data in RCFTs through \textit{commutant pairs}. The central idea is to replace direct, often intractable, computations in higher-rank meromorphic theory by a controlled manipulation of characters in lower-central charge factor, where modular data and defect eigenvalues are explicitly computable. The resulting equatorial amplitudes provide a concrete bridge between the topological line data and the holomorphic character technology that is naturally available in RCFTs.\\

\noindent 
The first conceptual outcome is that the equatorial projection principle isolates the precise sense in which defect insertions are transported under modular transformations. Rather than assuming commutativity between defect operators and modular data, the framework tracks the conjugation action of the modular group on insertions, thereby making modular covariance manifest and clarifying how diagonal Verlinde line insertions on one hemisphere are carried to fusion actions in the crossed channels. This perspective is particularly well-suited for commutant pair constructions, where one has a canonical bilinear pairing into a parent theory, and consequently, a natural family of one-sided projected amplitudes that closes under $\text{SL}(2, \mathbb{Z})$.\\

\noindent 
Building on this foundation, we presented explicit orbifold constructions and branching rule computations that realize a range of commutant pairs inside $E_{8,1}$, emphasizing  cases in which at least one factor is a tensor-product CFT in the meromorphic coset classification of \cite{Das:2022slz}. These examples provide a systematic mechanism for producing defect partition functions whose modular properties can be analyzed directly as modular functions, and they exhibit in a concrete way how algebraic data, such as Kac-Moody subalgebras, simple current extensions, and fusion rules interact with geometric constraints such as lattice realizations and automorphism actions. The resulting defect observables illustrate that even with a fixed meromorphic theory, distinct commutant decompositions can illuminate different sectors of defect data and can lead to qualitatively new modular features.
A key theme running through our three-character pairing examples is that the commutant-pair philosophy naturally extends beyond the two-character setting where the replacement rule was originally developed. In the three-character case, the modular representation, the defect eigenvalues extracted from the MMS construction \cite{Mathur:1988gt}, and the possibility of character degeneracies as in tensor product examples introduce additional structure where defect families may be $\mathbb{Z}_{n}$-graded, modular covariance may only close on congruence subgroups, and permutation-type defects can be invisible to naive bilinear replacements yet detectable through refined twisted-twining considerations. The commutant pair viewpoint organizes these phenomena efficiently as it tells us \textit{which} lower central charge theories to analyze, \textit{which} defect sectors are natural from the categorical perspective, and \textit{how} their insertions are implemented at the level of equatorial characters. More broadly, the commutant pair framework appears particularly fruitful for identifying novel defect partition functions in higher-rank meromorphic CFTs. As the central charge grows, the interplay between algebraic structure such as the organization of Kac-Moody currents and their embeddings, and the geometric structure such as lattice realizations and their automorphism groups becomes increasingly intricate. Particularly in this regime, commutant pairs of tensor-product type provide a natural organizing principle as they supply controlled decompositions in which defects can be imported from simpler factors while still probing genuinely higher-rank phenomena in the parent theory. Along these lines, lattice automorphism groups such as those of $E_{8}$ offer a powerful extension mechanism for the replacement rule. While \cite{Hegde:2021sdm} exploited such symmetries to construct defect partition functions in the two-character commutant pairs, the same philosophy can be pushed to further to the three-character (and higher-character) pairings, opening the door to defect observables that have not been explicitly constructed in meromorphic theories at central charges $c = 16, 24, 32$, and beyond.

\section*{Acknowledgments}
We are grateful to David Poland, Mikhail Isachenkov, Bowen Shi, Sunil Mukhi, Suresh Govindrajan, Jishu Das, Timmavajjula Venkata Karthik and Igor Frenkel for multiple detailed and illuminating discussions, as well as for their suggestions throughout the development of this work. NU would also like to thank Colin Coane, Evan Craft, Daniel Quenani, Ryan Everly, Ian Moult, Matthew Mitchell, and Sri Tata for helpful conversations and for providing a valuable sounding board for ideas at various stages. For early discussions and encouragement, NU is grateful to Gordon Rogelberg, Shankar Ganesh, Meng Cheng, Tony Liu, Brett Oertel, and Nayan Myerson-Jain. NU also thanks Xingping Yang for mathematical discussions and insightful guidance on the MTC literature, and Karan Bhatia for discussions and patiently answering questions about MTCs. Thanks to Matthew Mitchell and Mark Gonzalez for help generating the TikZ image. Finally, NU thanks Ana Karina Raygoza Cortez for discussions and steadfast support. NU is supported by DOE grant DE-SC0017660. The work of AD is supported by the STFC Consolidated Grant ST/T000600/1 -- ``Particle Theory at the Higgs Centre''.\\

\appendix
\begin{comment}
\section{To-do...}
\begin{enumerate}
    \item S-matrix for all $D_{r,1}$. Naveen has already done for $D_{4,1}$.
    \item Understand Karan's email and work out S-matrix for tensor-product RCFTs
    \item Figure out the replacement rule
\end{enumerate}
\end{comment}

\section{Branching Rules for commutant pairs in $E_{8,1}$}\label{sec:branching_rules}
A branching rule is the decomposition of an irreducible representation of a Lie algebra upon restriction to a subalgebra. When a symmetry algebra $\mathfrak{g}$ is restricted to a subalgebra $\mathfrak{h}\subset\mathfrak{g}$, each $\mathfrak{g}$-representation decomposes into a direct sum of $\mathfrak{h}$-representations. Concretely, if $\mathfrak{g}\supset\mathfrak{h}\oplus \mathfrak{h}'$ and $R$ is a $\mathfrak{g}$-irrep, then
\begin{align}
    R_{\downarrow\mathfrak{h}\oplus\mathfrak{h}'}\cong\bigoplus\limits_{(\lambda, \lambda')}N_{\lambda,\lambda'}^{R}(\lambda, \lambda'),
\end{align}
with multiplicities $N_{\lambda, \lambda'}^{R}\in\mathbb{Z}_{\geq0}$. In a WZW model, the relevant object is the affine algebra $\widehat{\mathfrak{g}}_{k}$. The vacuum module $\mathcal{H}_{0}^{\mathfrak{g}}$ decomposes under an affine subalgebra $\widehat{\mathfrak{h}}_{k_{h}}\oplus \widehat{\mathfrak{h}}'_{k_{h'}}$ as follows
\begin{align}
    \mathcal{H}_{0}^{\mathfrak{g}}\cong\bigoplus\limits_{(\lambda, \lambda')}\mathcal{H}_{\lambda}^{\mathfrak{h}}\otimes \mathcal{H}_{\lambda'}^{\mathfrak{h}'},
\end{align}
and at the level of characters, this becomes the bilinear identity
\begin{align}
    \chi_{0}^{\mathfrak{g}}(\tau) = \sum\limits_{(\lambda, \lambda')}N_{\lambda, \lambda'}\chi_{\lambda}^{\mathfrak{h}}(\tau)\chi_{\lambda'}^{\mathfrak{h}'}(\tau),
\end{align}
with $N_{\lambda, \lambda'}\in\mathbb{Z}_{\geq0}$ the branching multiplicities. Here, we are restricting the affine algebra $E_{8,1}$ to an affine algebra $\mathcal{A}\otimes A'\subset E_{8,1}$ as a conformal embedding. In the commutant pair situations that underlie the $c = 8$ story, the branching functions are constants, and we find a finite sum of products of RCFT characters. Once the bilinear decomposition of characters is known, then we can interpret the coefficients in the $\chi_{E_{8,1}}$ that read
\begin{align}\label{q-series}
    \chi_{E_{8,1}}(\tau) = q^{-\frac{1}{3}}\left(1 + 248q + 4124q^{2} + 34752q^{3} + \ldots\right),
\end{align}
by tracking what representations appear at each affine grade. This is precisely the logic used in the two-character analysis of $\chi_{E_{8,1}}$ in \cite{Hegde:2021sdm}, where the authors interpret the coefficient $248$ in the $q$-series expansion \ref{q-series} using finite-dimensional branching of the $E_{8}$ adjoint $\mathbf{248}$ under an appropriate maximal subalgebra, and then lift the same pattern to the affine character decomposition. A key conceptual point is that the coefficient $248$ counts the weight-one currents in the holomorphic theory. Hence, when we restrict those currents to $\widehat{\mathfrak{h}}_{k_{h}}\oplus\widehat{\mathfrak{h}}'_{k_{h'}}$, they split into the adjoint currents of each factor plus additional weight-one primaries in non-trivial $(\lambda, \lambda')$ sectors, and this splitting is governed by the branching of the finite-dimensional $\mathbf{248}$. So a finite branching rule for $\mathbf{248}$ is the shadow of a full affine branching of the vacuum module.\\

\noindent 
The vacuum module of the holomorphic $E_{8,1}$ WZW model admits only a rigid set of decompositions when we restrict to affine algebras that occur as mutually commuting factors. Say we have the embedding of chiral algebras, $\mathcal{A}\otimes \widetilde{\mathcal{A}}\subset E_{8,1}$, with $\mathcal{A}$ and $\widetilde{\mathcal{A}}$ forming a commuting pair, the $E_{8,1}$ vacuum representation branches into a finite direct sum of $\mathcal{A}\otimes\widetilde{\mathcal{A}}$-modules. At genus-one this takes the form of a bilinear character identity of the following form
\begin{align}
    \chi_{E_{8,1}}(\tau) = \sum\limits_{i\in \ I_\mathcal{A}}\chi_{i}^{\mathcal{A}}(\tau)\widetilde{\chi}_{\phi(i)}^{\widetilde{\mathcal{A}}}(\tau),
\end{align}
where $\phi: I_{\mathcal{A}}\to I_{\widetilde{\mathcal{A}}}$ is the gluing bijection. With this set up, the restriction of a finite-dimensional $E_{8}$-representation to a maximal subalgebra already predicts how the weight-one states reorganize into currents of $\mathcal{A}$ and $\widetilde{\mathcal{A}}$ together with additional primaries that extend $\mathcal{A}\times\widetilde{\mathcal{A}}$ to the full $E_{8,1}$ algebra. For two-character commutant pairs, this strategy was developed in detail in \cite{Hegde:2021sdm}, where the authors start from \ref{q-series} and explain $248$ by decomposing the $E_{8}$ adjoint under subalgebras $A_{1}\oplus E_{7}$ as follows
\begin{align}
    \mathbf{248}\to (\mathbf{3}, \mathbf{1})\oplus (\mathbf{1}, \mathbf{133})\oplus (\mathbf{2}, \mathbf{56}),
\end{align}
and they repeat this for other commutant pairs such as $A_{2}\oplus E_{6}$, $G_{2}\oplus F_{4}$, and then match those decompositions into bilinear identities for $\chi_{E_{8,1}}$ written in terms of corresponding RCFT characters. Our goal now is to repeat this for three-character commutant pairs. In this setting, the character identity fixes which affine modules appear, while the finite branching explains why the $q$-series coefficients line up the way they do. Hence, the character bilinear identity and the finite branching mutually reinforce each other. In this section, we extend that logic to three-character commutant pairs that appear in the $c = 8$ landscape. These cases are more delicate than the two-character cases for two related reasons. First, several of the relevant chiral theories, such as $B_{r,1}$, $\mathbf{III_{2}}$, and tensor-product constructions such as $A_{2,1}^{\otimes2}$ have non-trivial simple current structure and/or non-trivial braided autoequivalence groups. As a consequence, the naively diagonal pairing is not a priori the only possibility. Second, the three-character setting introduces genuine degeneracy phenomena since distinct primaries can share the same conformal weight and even the same genus-one character as a $q$-series expansion, and the correct branching must track these labels at the level of representations rather than only at the level of weights. We obtain a set of unique, representation-theoretically determined branching rules that simultaneously reproduce the $\chi_{E_{8,1}}$ character as a bilinear sum, and match the expected finite-dimensional decomposition of $\mathbf{248}$. The explicit branchings also play a structural role in the rest of the paper. The present section fixes the commutant pair branchings once and for all, and thereby sets up a concrete computational platform for tracking the action of TDLs on equatorial projections.

\begin{comment}
At the level of the vacuum modules, this takes the following form
\begin{align}
    \mathcal{H}_{E_{8,1}} \cong \bigoplus\limits_{\lambda}\mathcal{H}_{\lambda}^{\mathcal{A}}\otimes \mathcal{H}_{\widetilde{\lambda}}^{\widetilde{\mathcal{A}}},
\end{align}
and at the level of the characters, it becomes a bilinear decomposition of the $E_{8,1}$ vacuum character
\begin{align}
    \chi_{E_{8,1}}(\tau) = \sum\limits_{\lambda}\chi_{\lambda}^{\mathcal{A}}(\tau)\widetilde{\chi}_{\widetilde{\lambda}}^{\widetilde{\mathcal{A}}}(\tau).
\end{align}
Once this bilinear decomposition is known, then we can interpret the coefficients in the $\chi_{E_{8,1}}$ that read
\begin{align}
    \chi_{E_{8,1}}(\tau) = q^{-\frac{1}{3}}\left(1 + 248q + 4124q^{2} + 34752q^{3} + \ldots\right),
\end{align}
by tracking what representations appear at each affine grade. 
\end{comment}

\subsection{$(D_{2,1}, D_{6,1})$}
Consider the $D$-series WZW model at level $1$ described by $D_{r,1}\cong\widehat{\mathfrak{so}}(2r)_{1}$. First, for $D_{2,1}$, the finite Lie algebra $\mathfrak{so}(4)$ splits as $\mathfrak{su}(2)\oplus\mathfrak{su}(2)$ since $\text{Spin}(4)\cong\text{SU}(2)\times\text{SU}(2)$. At level $1$, the corresponding current algebra decomposes accordingly
\begin{align}
    \widehat{\mathfrak{so}}(4)_{1}\cong\widehat{\mathfrak{su}}(2)_{1}\oplus \widehat{\mathfrak{su}}(2)_{1},
\end{align}
with matching levels because the embedding has Dynkin index one in this case. A standard maximal group chain reads\footnote{We use the finite-dimensional chain only to organize  branchings; the conformal embedding $E_{8,1}\supset D_{8,1}$, and then we restrict $D_{8}$ to $D_{2}\oplus D_{6}$ at the level of representations.}
\begin{align}
    E_{8}\supset\text{Spin}(16)\supset\text{Spin}(4)\times\text{Spin}(12),
\end{align}
and at level $1$, we have the following conformal embedding \cite{ChuZheng:2008}
\begin{align}
    E_{8,1}\supset D_{8,1}\cong\widehat{\mathfrak{so}}(16)_{1},
\end{align}
and at the level of finite Lie algebras, the adjoint branches as follows \cite{Slansky1981GroupTF}
\begin{align}
    \mathbf{248}_{\downarrow\mathfrak{so}(16)} = \mathbf{120}\oplus \mathbf{128}.
\end{align}
Next, inside $\mathfrak{so}(16)$, we take $\mathfrak{so}(16)\supset\mathfrak{so}(4)\oplus\mathfrak{so}(12)$ which is equivalent to the statement that $D_{8}\supset D_{2}\oplus D_{6}$. To figure out this branching, we consider the explicit $q$-series expansions of the corresponding WZW characters. For the theory, $D_{2,1}\cong\widehat{\mathfrak{so}}(4)_{1}$, we have primaries $I_{D_{r,1}} = \{\mathbb{1}, \nu,\sigma,\widetilde{\sigma}\}$ with conformal weights $h = \{0,\tfrac{1}{2}, \tfrac{1}{4}, \tfrac{1}{4}\}$. The $q$-series expansion of $D_{2,1}\cong\widehat{\mathfrak{so}}(4)_{1}$ characters reads
\begin{align}
    \begin{split}
        \chi_{\mathbb{1}}^{D_{2}}(\tau) =& q^{-\frac{1}{12}}\left(1 + 6q + 17q^{2} + 38q^{3} + \ldots\right),\\
        \chi_{\nu}^{D_{2}}(\tau) =& q^{-\frac{1}{12}}\left(4q^{\frac{1}{2}} + 8q^{\frac{3}{2}} + 28q^{\frac{5}{2}} + \ldots\right),\\
        \chi_{\sigma}^{D_{2}}(\tau) = \chi_{\widetilde{\sigma}}^{D_{2}}(\tau) =& q^{-\frac{1}{12}}\left(2q^{\frac{1}{4}} + 8q^{\frac{5}{4}} + 20q^{\frac{9}{4}} + \ldots\right).
    \end{split}
\end{align}
Here, the coefficient $6$ in $\chi_{\mathbb{1}}^{D_{2}}$ at $\mathcal{O}(q)$ is $\text{dim}(\mathfrak{so}(4)) = 6$, the adjoint currents at level $1$. The $q$-series expansion of $D_{6,1}\cong\widehat{\mathfrak{so}}(12)_{1}$ characters reads
\begin{align}
    \begin{split}
        \chi_{\mathbb{1}}^{D_{6}}(\tau) =& q^{-\frac{1}{4}}\left(1 + 66q + 639q^{2} + 3774q^{3} + \ldots\right),\\
        \chi_{\nu}^{D_{6}}(\tau) =& q^{-\frac{1}{4}}\left(12q^{\frac{1}{2}} + 232q^{\frac{3}{2}} + 1596q^{\frac{5}{2}} + \ldots\right),\\
        \chi_{\sigma}^{D_{6}}(\tau) = \chi_{\widetilde{\sigma}}^{D_{6}}(\tau) =& q^{-\frac{1}{4}}\left(32q^{\frac{3}{4}} + 384q^{\frac{7}{4}} + 2496q^{\frac{11}{4}} + \ldots\right).
    \end{split}
\end{align}
Again, the $66$ in $\chi_{\mathbb{1}}^{D_{6}}$ at $\mathcal{O}(q)$ is $\text{dim}(\mathfrak{so}(12)) = 66$. Finally, for $D_{8,1}\cong\widehat{\mathfrak{so}}(16)_{1}$ with $c= 8$, we have the following $q$-series expansions
\begin{align}
    \begin{split}
        \chi_{\mathbb{1}}^{D_{2}} =& q^{-\frac{1}{3}}\left(1 + 120q  + 2076q^{2} + 17344q^{3} + \ldots\right),\\
        \chi_{\sigma}^{D_{2}}(\tau) = \chi_{\widetilde{\sigma}}^{D_{8}}(\tau) =& q^{-\frac{1}{3}}\left(128q + 2048q^{2} + 17408q^{3} + \ldots \right).
    \end{split}
\end{align}
From this, we immediately see that at the first excited level, $120$ is the $\text{dim}(\mathfrak{so}(16))$ adjoint, and $128$ is the dimension of the chiral spinor of $\mathfrak{so}(16)$. For the branching, $D_{8,1}\to D_{2,1}\oplus D_{6,1}$, the vacuum and spinor modules decompose as follows
\begin{align}
\begin{split}
    \chi_{\mathbb{1}}^{D_{8}} =& \chi_{\mathbb{1}}^{D_{2}}\widetilde{\chi}_{\mathbb{1}}^{D_{6}} + \chi_{\nu}^{D_{2}}\widetilde{\chi}_{\nu}^{D_{6}},\\
    \chi_{\sigma}^{D_{8}} =& \chi_{\sigma}^{D_{2}}\widetilde{\chi}_{\sigma}^{D_{6}} + \chi_{\widetilde{\sigma}}^{D_{2}}\widetilde{\chi}_{\widetilde{\sigma}}^{D_{6}},
\end{split}
\end{align}
where we denote the second factor characters by $\widetilde{\chi}^{D_{6}}$ for clarity. We now expand the products and look at the coefficient $q^{1}$ after factoring out $q^{-\tfrac{1}{3}}$. First, the contribution of $\chi_{\mathbb{1}}^{D_{2}}\widetilde{\chi}_{\mathbb{1}}^{D_{6}}$ to the vacuum piece reads
\begin{align}
    \chi_{\mathbb{1}}^{D_{2}}\widetilde{\chi}_{\mathbb{1}}^{D_{6}} =& q^{-\frac{1}{3}}\left[1 + \left(6 + 66\right) q + \ldots\right].
\end{align}
Since these currents come from $D_{2,1}$ and are neutral under $D_{6,1}$, their representation under $D_{2,1}\oplus D_{6,1}$ is $(\mathbf{6},\mathbf{1})$, and similarly, we also have currents that come from $D_{6,1}$ that are neutral under $D_{2,1}$, and hence, they transform as $(\mathbf{1}, \mathbf{66})$. Next, we have the contribution
\begin{align}
    \chi_{\nu}^{D_{2}}\widetilde{\chi}_{\nu}^{D_{6}} = q^{-\frac{1}{3}}\left[(4\cdot 12)q + \ldots\right]
\end{align}
The remaining $4\cdot 12 = 48$ currents mix the two factors and transform as $(\mathbf{4}, \mathbf{12})$. Hence, we see that $120 = 6 + 66 + 48$, and the branching rule
\begin{align}\label{120}
    \mathbf{120} \to (\mathbf{6}, \mathbf{1})\oplus (\mathbf{1},\mathbf{66})\oplus (\mathbf{4}, \mathbf{12}).
\end{align}
Next, for the chiral spinor $\mathbf{128}$ of $\text{Spin}(16)$, we have
\begin{align}
    \begin{split}
        \chi_{\sigma}^{D_{2}}\widetilde{\chi}_{\sigma}^{D_{6}} =& q^{-\frac{1}{3}}\left(2\cdot 32  + \ldots\right),\\
        \chi_{\widetilde{\sigma}}^{D_{2}}\widetilde{\chi}_{\widetilde{\sigma}}^{D_{6}} =& q^{-\frac{1}{3}}\left(2\cdot 32  + \ldots\right),
    \end{split}
\end{align}
from which the leading multiplicities can be read off to be $128 = 64 + 64 = (2\cdot 32) + (2'\cdot 32')$. Here, $2,2'$ denote the two chiral spinors of $\text{Spin}(4)$, while $32, 32'$ denote the two chiral spinors of $\text{Spin}(12)$. We now have the following branching for the spinor module
\begin{align}\label{128}
    \mathbf{128}\to (\mathbf{2}, \mathbf{32}) \oplus (\mathbf{2'}, \mathbf{32'}).
\end{align}
The branching rules of $\mathbf{120}$ and $\mathbf{128}$ can alternatively be worked out from the adjoint representation and Clifford algebra factorization, respectively. We present this in appendix \ref{sec:appendix_branching}. From \ref{120} and \ref{128}, we find that we have the branching rule
\begin{align}
    \mathbf{248}\to (\mathbf{6}, \mathbf{1})\oplus (\mathbf{1},\mathbf{66})\oplus (\mathbf{4}, \mathbf{12}) \oplus  (\mathbf{2}, \mathbf{32}) \oplus (\mathbf{2'}, \mathbf{32'}).
\end{align}
For this commutant pair, we have two distinct spinor primaries $\sigma$ and $\widetilde{\sigma}$ that both have the same conformal weight $h_{\sigma} = h_{\widetilde{\sigma}} = \tfrac{r}{8}$ for $D_{r,1}$, and their graded dimensions coincide, i.e., $\chi^{D{_r}}_{\sigma} = \chi^{D_{r}}_{c}$. Now, when we rewrite the characters using conformal weight labels, we set
\begin{align}
    \begin{split}
        \chi_{\frac{1}{4}}^{D_{2}}\equiv& \chi_{\sigma}^{D_{2}} = \chi_{\widetilde{\sigma}}^{D_{2}},\\
        \chi_{\frac{3}{4}}^{D_{6}}\equiv& \chi_{\sigma}^{D_{6}} = \chi_{\widetilde{\sigma}}^{D_{6}},
    \end{split}
\end{align}
and we have
\begin{align}
    \chi^{E_{8}} = \chi_{0}^{D_{2}}\chi_{0}^{D_{6}} + \chi_{\frac{1}{2}}^{D_{2}}\chi_{\frac{1}{2}}^{D_{6}} + 2\chi_{\frac{1}{4}}^{D_{2}}\chi_{\frac{3}{4}}^{D_{6}}.
\end{align}

\subsection{$(B_{1,1}, B_{6,1})$}
We now have the conformal embedding
\begin{align}
    B_{1,1}\oplus B_{6,1}  \cong \widehat{\mathfrak{so}}(3)_{1}\oplus \widehat{\mathfrak{so}}(13)_{1}\subset\widehat{\mathfrak{so}}(16)_{1},
\end{align}
since $B_{r,1}\cong\widehat{\mathfrak{so}}(2r+1)_{1}$. We closely follow the methodology presented in appendix \ref{sec:appendix_branching} to find the branching rules. The block embedding $\mathfrak{so}(16)\supset\mathfrak{so}(3)\oplus \mathfrak{so}(13)$ is realized as 
\begin{align}
    \mathbf{16}\to (\mathbf{3},\mathbf{1})\oplus (\mathbf{1}, \mathbf{13}).
\end{align}
Using dimensions
\begin{align}
    \text{dim}(\Lambda^{2}\mathbf{3}) = \text{dim}(\mathfrak{so}(3)) = 3,\qquad \text{dim}(\Lambda^{2}\mathbf{13}) = \text{dim}(\mathfrak{so}(13)) = 78,\qquad \text{dim}(\mathbf{3}\otimes \mathbf{13}) = 39,
\end{align}
and applying the identity \ref{adjoint_identity}, we find the following branching
\begin{align}
    \mathbf{120}\to (\mathbf{3}, \mathbf{1})\oplus (\mathbf{3}, \mathbf{13})\oplus (\mathbf{1}, \mathbf{78}).
\end{align}
Next, for the branching of the spinor $\mathbf{128}$, we use Clifford factorization. With  an odd $\oplus$ odd split $16 = 3\oplus 13$, we obtain the chiral spinor of $\text{Spin}(16)$ to be 
\begin{align}
    \begin{split}
        S^{(16)}_{\pm} =& S^{(3)}\otimes S^{(13)},
    \end{split}
\end{align}
i.e. each chiral spinor restricts to the same irreducible $(\mathbf{2}, \mathbf{64})$. Here, $S^{(3)}$ is the unique spinor of $\text{Spin}(3)\cong \text{SU}(2)$, the accidental isomorphism, and $S^{(13)}$ is the unique spinor of $\text{Spin}(13)$. Apart from the standard low-rank accidental isomorphism, e.g., $\text{Spin}(3)\cong \text{SU}(2)$, $\text{Spin}(4)\cong \text{SU}(2)\times\text{SU}(2)$, $\text{Spin}(6)\cong \text{SU}(4)$, no further identifications with $\text{SU}(n)$ occur for $\text{Spin}(N)$  at higher rank, and hence, $\text{Spin}(13)$ has no such simplification. So we use the general spinor-dimension formula which reads \cite{Okubo:1994ss}
\begin{align}
    \text{dim}(S^{(N)}) = 2^{\lfloor\frac{N}{2}\rfloor},
\end{align}
and for $N = 3$ and $N =13$, we have $\text{dim}(S^{(3)}) = 2$ and $\text{dim}(S^{(13)}) = 64$, respectively. Hence, we have the branching
\begin{align}
    \mathbf{248}\to (\mathbf{3},\mathbf{1})\oplus (\mathbf{1}, \mathbf{78})\oplus (\mathbf{3}, \mathbf{13})\oplus (\mathbf{2}, \mathbf{64}),
\end{align}
where $(\mathbf{2}, \mathbf{64})$ denotes the tensor product of unique spinors of $\text{Spin}(3)$ and $\text{Spin}(13)$.

\subsection{$(\mathbf{III}_{2}, G_{2,1}^{\otimes2})$}
The commutant pair $(\mathrm{III}_{2},\,G_{2,1}^{\otimes 2})$ inside $E_{8,1}$
is reflected at weight one by the decomposition of the $E_8$ adjoint $\mathbf{248}$ under the finite subalgebra
\begin{align}
  \mathfrak{e}_8\ \supset\ \mathfrak{g}_2^{(L)}\oplus\mathfrak{g}_2^{(R)}\oplus \mathfrak{a}_1\,.
\end{align}
The VOA $\mathbf{III}_{2}$ is a $\mathbb{Z}_{2}$ simple current extension of $A_{1,8}$ \cite{Das:2022uoe}, and together with the conformal embedding
\begin{align}
    G_{2,1}\otimes G_{2,1}\otimes A_{1,8}\subset E_{8,1},
\end{align}
this yields the commutant pair $(\mathbf{III}_{2}, G_{2,1}^{\otimes2})$. One effective way to obtain this branching is to use the exceptional maximal chain
\begin{align}
  \mathfrak{e}_8\ \supset\ \mathfrak{g}_2\oplus\mathfrak{f}_4,
  \qquad
  \mathfrak{f}_4\ \supset\ \mathfrak{a}_1\oplus\mathfrak{g}_2,
\end{align}
and branch $\mathbf{248}$ in two steps. First, we consider $\mathbf{248}$ under $\mathfrak{g}_2\oplus\mathfrak{f}_4$. The adjoint of $E_8$ branches as \cite{Slansky1981GroupTF, Hegde:2021sdm}
\begin{align}\label{E8_to_G2F4}
  \mathbf{248}\ \to
  (\mathbf{14},\mathbf{1})\ \oplus\ (\mathbf{7},\mathbf{26})\ \oplus\ (\mathbf{1},\mathbf{52}),
\end{align}
where $\mathbf{14}$ is the $G_2$ adjoint, $\mathbf{7}$ its fundamental, and
$\mathbf{26},\mathbf{52}$ are the fundamental and adjoint of $F_4$. The second step is to branch the $F_4$ representations under $\mathfrak{a}_1\oplus\mathfrak{g}_2$.  Under $F_{4}\supset A_{1}\times G_{2}^{(R)}$, we use \ref{26_branch} and \ref{52_branch} and then reorder factors to $(G_{2}^{(L)}, G_{2}^{(R)}, A_{1})$. We now have the following standard decompositions \cite{Slansky1981GroupTF}
\begin{align}
  \mathbf{26}=&
  (\mathbf{5},\mathbf{1})\ \oplus\ (\mathbf{3},\mathbf{7}),
  \label{26_branch}
  \\
  \mathbf{52}=&
  (\mathbf{3},\mathbf{1})\ \oplus\ (\mathbf{1},\mathbf{14})\ \oplus\ (\mathbf{5},\mathbf{7}),
  \label{52_branch}
\end{align}
where $\mathbf{3}$ and $\mathbf{5}$ are the spin-$1$ and spin-$2$ irreps of
$A_1\cong \mathfrak{su}(2)$. We can now assemble to obtain $\mathfrak{g}_2^{(L)}\oplus\mathfrak{g}_2^{(R)}\oplus \mathfrak{a}_1$. To do this, we insert \ref{26_branch}--\ref{52_branch} into \ref{E8_to_G2F4}, and writing the factors as $(G_2^{(L)},G_2^{(R)},A_1)$, we obtain
\begin{align}
  \mathbf{248}\ \to\ 
  (\mathbf{14},\mathbf{1},\mathbf{1})
  \oplus
  (\mathbf{1},\mathbf{14},\mathbf{1})
  \oplus
  (\mathbf{1},\mathbf{1},\mathbf{3})
  \oplus
  (\mathbf{7},\mathbf{7},\mathbf{3})
  \oplus
  (\mathbf{7},\mathbf{1},\mathbf{5})
  \oplus
  (\mathbf{1},\mathbf{7},\mathbf{5}).
  \label{E8_to_G2G2A1}
\end{align}
A dimension check yields
\begin{align}
  14+14+3 + (7\cdot 7\cdot 3) + (7\cdot 5) + (7\cdot 5)
  = 14+14+3+147+35+35
  = 248.
\end{align}
In the affine setting, the corresponding conformal embedding
$G_{2,1}\otimes G_{2,1}\otimes A_{1,8}\subset E_{8,1}$, together with the $\mathbb{Z}_2$
simple current extension $A_{1,8}\leadsto \mathrm{III}_2$, is the origin of the
commutant pair $(\mathrm{III}_2,\,G_{2,1}^{\otimes 2})$.

\subsection{$(A_{4,1}, A_{4,1})$}
The commutant pair $(A_{4,1},A_{4,1})$ corresponds, at weight one, to the finite-dimensional maximal-rank subalgebra
\begin{align}
  \mathfrak{e}_8 \ \supset\  \mathfrak{su}(5)\oplus \mathfrak{su}(5).
\end{align}
So the relevant branching rule is the decomposition of the $E_8$ adjoint $\mathbf{248}$ into irreducibles of $\mathfrak{su}(5)\oplus \mathfrak{su}(5)$.   This branching appears e.g. in  equation $(3.7)$ in \cite{Grimm:2010ez} and is tabulated in \cite{Slansky1981GroupTF}
\begin{align}
  \mathbf{248}\ \to\
  (\mathbf{24},\mathbf{1})\ \oplus\ (\mathbf{1},\mathbf{24})\
  \oplus\ (\mathbf{10},\mathbf{5})\ \oplus\ (\overline{\mathbf{10}},\overline{\mathbf{5}})\
  \oplus\ (\mathbf{5},\mathbf{10})\ \oplus\ (\overline{\mathbf{5}},\overline{\mathbf{10}}),
  \label{E8_to_SU5xSU5}
\end{align}
Recall that $\mathbf{10}\cong \Lambda^{2}\mathbf{5}$ and $\overline{\mathbf{10}}\cong \Lambda^{2}\overline{\mathbf{5}}$, so the  mixed summands in \ref{E8_to_SU5xSU5} can be viewed as
\begin{align}
  (\mathbf{5},\mathbf{10}) \cong (\mathbf{5},\Lambda^{2}\mathbf{5}),\qquad
  (\mathbf{10},\mathbf{5}) \cong (\Lambda^{2}\mathbf{5},\mathbf{5}),
\end{align}
and similarly for the conjugates.  The dimension check then becomes immediate
\begin{align}
  24 + 24 + 4\ \text{dim}(\mathbf{5})\text{dim}(\mathbf{10}) = 24 + 24 + 4\cdot 50 = 248.
\end{align}

\subsection{$(A_{2,1}^{\otimes 2}, A_{2,1}^{\otimes 2})$}
Let's begin at the level of finite-dimensional Lie algebras, where we use the following standard maximal-rank chain
\begin{align}
  \mathfrak{e}_8 \supset \mathfrak{su}(3)_{(4)} \oplus \mathfrak{e}_6,
  \qquad
  \mathfrak{e}_6 \supset \mathfrak{su}(3)_{(1)} \oplus \mathfrak{su}(3)_{(2)} \oplus \mathfrak{su}(3)_{(3)}.
\end{align}
Here, the subscripts $(i)$ label $\mathfrak{su}(3)$ factors and are not affine levels. This is a familiar maximal-rank construction where the inclusion $E_{8}\supset \text{SU}(3)\times E_{6}$ can be seen from the extended Dynkin diagram, or equivalently from the $E_{8}$ root lattice admitting an $A_{2}\oplus E_{6}$ root subsystem. The second inclusion, $E_{6}\supset\text{SU}(3)^{3}$ is the so-called trinification maximal-rank subgroup \cite{Dinh:2019jdg, Babu:2023zsm}. At the affine level, the same maximal-rank structure lifts to the following conformal embedding \cite{Cordova:2018qvg}
\begin{align}
    A_{2,1}\times E_{6,1}\subset E_{8,1}.
\end{align}
Using the fact that $E_{6}$ contains $\text{SU}(3)$ at maximal-rank, we can think of $E_{6,1}$ as containing three commuting affine $A_{2,1}$ factors, and hence, we obtain the following conformal embedding
\begin{align}
    A_{2,1}^{\otimes4}\subset E_{8,1}.
\end{align}
This uses that each $\mathfrak{su}(3)\hookrightarrow \mathfrak{e}_{6}$ in the trinification embedding has Dynkin index $1$, so at level $1$ the induced affine subalgebras are $A_{2,1}$ and $c(A_{2,1}^{\otimes3}) = 6 = c(E_{6,1})$ \cite{Mukhopadhyay:2016jog} (see \S$2.2$). Partitioning the four $A_{2,1}$ factors into groups of two yields the commutant pair $(A_{2,1}^{\otimes 2}, A_{2,1}^{\otimes2})\subset E_{8,1}$. Let's first consider $E_{8}\supset\text{SU}(3)\times E_{6}$, whose standard branching rule reads \cite{Slansky1981GroupTF}
\begin{align}\label{248_E8E6SU(3)}
    \mathbf{248}\to (\mathbf{8}, \mathbf{1})\oplus (\mathbf{1}, \mathbf{78})\oplus (\mathbf{3}, \mathbf{27})\oplus (\overline{\mathbf{3}}, \overline{\mathbf{27}}),
\end{align}
where $\mathbf{8}$ is the adjoint representation of $\text{SU}(3)$ called the octet, $\mathbf{78}$ is the adjoint of $E_{6}$, and $\mathbf{27}$ is the fundamental of $E_{6}$. Here, the $(\mathbf{8}, \mathbf{1})$ and $(\mathbf{1}, \mathbf{78})$ summands are literally the current subalgebras $\mathfrak{su}(3)$ and $\mathfrak{e}_{6}$ sitting inside $\mathfrak{e}_{8}$. We also have $(\mathbf{3}, \mathbf{27})$ and $(\overline{\mathbf{3}}, \overline{\mathbf{27}})$ that comprise the off-diagonal complement transforming under both factors. Let's be more precise. The Lie algebra decomposes as follows
\begin{align}
    \mathfrak{e}_{8}\cong\mathfrak{su}(3)\oplus\mathfrak{e}_{6}\oplus \mathfrak{m},
\end{align}
where $\mathfrak{m}$ is the complement of the subalgebra $\mathfrak{h} \equiv \mathfrak{su}(3)\oplus\mathfrak{e}_{6}$ inside $\mathfrak{e}_{8}$. The adjoint action of $\mathfrak{h}$ on $\mathfrak{e}_{8}$ preserves this splitting, and hence, $\mathfrak{m}$ is a representation of $\text{SU}(3)\times E_{6}$. In the decomposition \ref{248_E8E6SU(3)}, $(\mathbf{8}, \mathbf{1})$ is the adjoint of $\text{SU}(3)$, $(\mathbf{1}, \mathbf{78})$ is the adjoint of $E_{6}$. The remaining subspace $\mathfrak{m}$ is the off-diagonal complement, $\mathfrak{m}\cong (\mathbf{3}, \mathbf{27})\oplus (\overline{\mathbf{3}}, \overline{\mathbf{27}})$. A dimension check yields the desired result, $248 = 8 + 78 + 3\cdot 27 + 3\cdot 27$. Next, we branch the $E_{6}$ representations under the subgroup $\text{SU}(3)_{(1)}\times\text{SU}(3)_{(2)}\times\text{SU}(3)_{(3)}$, and the fundamental adjoint decompose as follows \cite{Babu:2023zsm}
\begin{align}  \label{E6_27,27o,78_to_SU3^3}
\begin{split}
  \mathbf{27}
  &\to
  (\mathbf{3},\mathbf{3},\mathbf{1})\oplus(\mathbf{1},\overline{\mathbf{3}},\mathbf{3})
  \oplus(\overline{\mathbf{3}},\mathbf{1}, \overline{\mathbf{3}}),\\
  \overline{\mathbf{27}}
  &\to
  (\overline{\mathbf{3}},\overline{\mathbf{3}},\mathbf{1})\oplus(\mathbf{1},\mathbf{3},\overline{\mathbf{3}})
  \oplus(\mathbf{3},\mathbf{1}, \mathbf{3}),
  \\
    \mathbf{78}
  &\to
  (\mathbf{8},\mathbf{1},\mathbf{1})\oplus(\mathbf{1},\mathbf{8},\mathbf{1})\oplus(\mathbf{1},\mathbf{1},\mathbf{8})
  \oplus(\mathbf{3},\overline{\mathbf{3}},\mathbf{3})\oplus(\overline{\mathbf{3}},\mathbf{3},\overline{\mathbf{3}}),
  \end{split}
\end{align}
Substituting this into the $E_{8}$ branching yields an explicit decomposition of $\mathbf{248}$ under
\begin{align}
    \left(\text{SU}(3)\times\text{SU}(3)_{(1)}\right)\times\left(\text{SU}(3)_{(2)}\times\text{SU}(3)_{(3)}\right),    
\end{align}
or equivalently, under $A_{2,1}^{\otimes4}$. Writing representations as $(R_{1}, R_{2}; R_{3}, R_{4})$, where $R_{i}$ is an irrep of the $i^{\text{th}}$ $\text{SU}(3)$ factor, we obtain the following branching rule (see appendix \ref{sec:appendix_A2,1_branching} for an explicit proof)
\begin{align}\label{248_A2,1}
\begin{split}
    \mathbf{248} \to (\mathbf{8}, \mathbf{1}; \mathbf{1}, \mathbf{1})\oplus& (\mathbf{1}, \mathbf{8}; \mathbf{1}, \mathbf{1})\oplus (\mathbf{1}, \mathbf{1}; \mathbf{8}, \mathbf{1})\oplus (\mathbf{1}, \mathbf{1}; \mathbf{1}, \mathbf{8})\\
    &\oplus(\mathbf{3},\overline{\mathbf{3}};\overline{\mathbf{3}},\mathbf{1})\oplus(\overline{\mathbf{3}},\mathbf{3};\mathbf{3},\mathbf{1})\\
    &\oplus(\mathbf{3},\mathbf{3};\mathbf{1},\mathbf{3})\oplus(\mathbf{1},\overline{\mathbf{3}};\mathbf{3},\mathbf{3})\oplus(\overline{\mathbf{3}},\mathbf{1};\overline{\mathbf{3}},\mathbf{3})\\
    &\oplus(\overline{\mathbf{3}},\overline{\mathbf{3}};\mathbf{1},\overline{\mathbf{3}})\oplus(\mathbf{1},\mathbf{3};\overline{\mathbf{3}},\overline{\mathbf{3}})\oplus(\mathbf{3},\mathbf{1};\mathbf{3},\overline{\mathbf{3}}).
\end{split}
\end{align}
The dimension of a product irrep is simply
\begin{align}
    \text{dim}(R_{1}, R_{2}; R_{3}, R_{4}) = \text{dim}\ R_{1}\cdot \text{dim}\ R_{2}\cdot \text{dim}\ R_{3}\cdot \text{dim}\ R_{4}.
\end{align}
Using $\text{dim}\ \mathbf{1}= 1$, $\text{dim}\ \mathbf{3} = \text{dim}\ \overline{\mathbf{3}} = 3$, and $\text{dim}\ \mathbf{8} = 8$, we obtain the correct dimension count
\begin{align}
    4\cdot (8\cdot 1^{3}) + 8\cdot (1\cdot 3^{3}) = 32 + 216 = 248.
\end{align}
Let's now verify the $\mathfrak{su}(3)^{\oplus 4}$ branching of the $E_{8}$ adjoint directly from characters. This pairing will soon be helpful when we want to explicitly compute the defect partition functions due to TDLs. The $E_{8}$ vacuum module decomposes into a finite sum of tensor product modules of the two $A_{2,1}^{\otimes}$ blocks. Each $A_{2,1}\cong \widehat{\mathfrak{su}}(3)_{1}$ has central charge $c = 2$, and hence we consider the blocks
\begin{align}
    \mathcal{V} \equiv A_{2,1}^{\otimes2},\qquad\widetilde{\mathcal{V}}\equiv A_{2,1}^{\otimes2},
\end{align}
with central charges $c(\mathcal{V}) = c(\widetilde{\mathcal{V}}) = 4$. As a shorthand, we will denote $c(\widetilde{\mathcal{V}}) = \widetilde{c}$. Together, these blocks fill the $c = 8$, matching $E_{8}$. To write this pairing explicitly, we can organize $\mathcal{V}$ modules by their $\mathbb{Z}_{3}$ simple current charge. Concretely, $\widehat{\mathfrak{su}}(3)_{1}$ has three integrable highest-weight modules which are labeled by charge $0,1,2$ (mod $3$), with fusion given by addition mod $3$ \cite{DiFrancesco:1997nk}. In the trinification convention, we can think of these as $\mathbf{1}, \mathbf{3}, \overline{\mathbf{3}}$, with $\overline{\mathbf{3}}$ being the charge $2$ object. Let $\chi_{0}, \chi_{1}, \chi_{2}$ be the unflavoured characters of charges $0,1,2$ respectively. These have the following $q$-series expansions\footnote{At level $1$, $\mathbf{3}$ and $\overline{\mathbf{3}}$ have the same unflavoured character as a $q$-series, though we keep the charge labels to track fusion and branching.}
\begin{align}\label{characters_A_2,1}
    \begin{split}
        \chi_{0}(\tau) =& \left(\frac{\Theta_{0}}{\eta^{2}}\right)(\tau) = q^{-\frac{1}{12}}\left(1 + 8q + 17q^{2} + 46q^{3} + \ldots\right),\\
        \chi_{1}(\tau) = \chi_{2}(\tau)  =& \left(\frac{\Theta_{1}}{\eta^{2}}\right)(\tau) = q^{\frac{1}{4}}\left(3 + 9q + 27q^{2} + 57q^{3} + \ldots\right),
    \end{split}
\end{align}
where \cite{Hirschhorn1993CubicAO}
\begin{align}\label{theta_A_2}
    \begin{split}
        \Theta_{0}(\tau) =& \sum\limits_{m,n\in\mathbb{Z}}q^{m^{2} + mn + n^{2}}\equiv a(q)\\
        =& 1 + 6q + 6q^{3} + 6q^{4} + \ldots,\\
        \Theta_{1}(\tau) =& q^{\frac{1}{3}}\sum\limits_{m,n\in\mathbb{Z}}q^{m^{2} + mn + n^{2} + m + n}\equiv q^{\frac{1}{3}}c(q)\\
        =& 3q^{\frac{1}{3}} + 3q^{\frac{4}{3}} + 6q^{\frac{7}{3}} + 6q^{\frac{13}{3}} + \ldots,
    \end{split}
\end{align}
are the theta series, which are weight-$1$ modular forms (with a character) for $\Gamma(3)$. Here, the overall exponents reflect $\tfrac{c}{24} = \tfrac{1}{12}$ and the lowest conformal weight $h = \tfrac{1}{3}$ since $-\tfrac{1}{12} + \tfrac{1}{3} = \tfrac{1}{4}$. For the tensor product module $\mathcal{V}$, the total charge is simply $a + b$ (mod $3$) since its simple currents are additive. Hence, we can package together all modules of a given charge and write
\begin{align}\label{characters_A_2,1_x2}
    \begin{split}
        \chi^{A_{2,1}^{\otimes2}}_{0}(\tau) \equiv& \left(\chi_{0}\chi_{0} + \chi_{1}\chi_{2} + \chi_{2}\chi_{1}\right)(\tau)\\
        =& q^{-\frac{1}{6}}\left(1 + 18q^{\frac{2}{3}} + 16q + 108 q^{\frac{5}{3}} + 98q^{2} + 486q^{\frac{8}{3}} + 364q^{3} + \ldots\right),\\
        \chi^{A_{2,1}^{\otimes2}}_{\frac{1}{3}}(\tau) \equiv& \left(\chi_{0}\chi_{1} + \chi_{1}\chi_{0} + \chi_{2}\chi_{2}\right)(\tau)\\
        =& q^{\frac{1}{6}}\left(6 + 9q^{\frac{1}{3}} + 66q + 54q^{\frac{4}{3}} + 300q^{2} + 243q^{\frac{7}{3}} + 1128q^{3} + \ldots\right)\\
        =& \chi^{A_{2,1}^{\otimes2}}_{\tfrac{2}{3}}(\tau).
    \end{split}
\end{align}
Now, for the commutant pair $(\mathcal{V}, \widetilde{\mathcal{V}})\subset E_{8,1}$, the $E_{8,1}$ vacuum character reads
\begin{align}
    \chi_{E_{8,1}}(\tau) = \chi_{0}^{\mathcal{V}}(\tau)\chi_{0}^{\widetilde{\mathcal{V}}}(\tau) + \chi_{\frac{1}{3}}^{\mathcal{V}}(\tau)\chi_{\frac{2}{3}}^{\widetilde{\mathcal{V}}}(\tau) + \chi_{\frac{2}{3}}^{\mathcal{V}}(\tau)\chi_{\frac{1}{3}}^{\widetilde{\mathcal{V}}}(\tau).
\end{align}
Here, the vacuum pairing has the following $q$-series expansion
\begin{align}
    \chi_{0}^{\mathcal{V}}(\tau)\chi_{0}^{\widetilde{\mathcal{V}}}(\tau) = q^{-\frac{1}{3}}\left(1 + 36q^{\frac{2}{3}} + \mathbf{32}q + \ldots\right),
\end{align}
where we have a contribution of exactly $32$ weight-one states. These are precisely the currents of the algebra $\mathfrak{su}(3)^{\oplus 4}$, so they transform as the direct sum of four adjoints
\begin{align}
    (\mathbf{8}, \mathbf{1}; \mathbf{1}, \mathbf{1})\oplus (\mathbf{1}, \mathbf{8}; \mathbf{1}, \mathbf{1})\oplus(\mathbf{1}, \mathbf{1}; \mathbf{8}, \mathbf{1})\oplus (\mathbf{1}, \mathbf{1}; \mathbf{1}, \mathbf{8}).
\end{align}
Next, we consider the $q$-series expansion of the first cross term
\begin{align}
    \chi_{\frac{1}{3}}^{\mathcal{V}}(\tau)\chi_{\frac{2}{3}}^{\widetilde{\mathcal{V}}}(\tau) = q^{\frac{1}{3}}(36q^{\frac{2}{3}} +  \mathbf{108}q + \ldots),
\end{align}
where the contribution is $108$ weight-one states. The other cross term, $\chi_{\tfrac{2}{3}}^{\mathcal{V}}\chi_{\tfrac{1}{3}}^{\widetilde{\mathcal{V}}}$ also contributes $108$ weight-one states, and hence we have a total of $248 - 32 = 216$ decomposed as 
\begin{align}
    216 = 8\times 27 = 8\times (3\cdot 3\cdot 3).
\end{align}
This is exactly what we must expect from the branching \ref{248_A2,1}. The complement of $\mathfrak{su}(3)^{\oplus 4}$ inside $\mathfrak{e}_{8}$ transforms as a sum of tri-fundamental representations, three factors of $\mathbf{3}$ or $\overline{\mathbf{3}}$ and one trivial factor $\mathbf{1}$, each of dimension $27$. These eight tri-fundamentals are shown in \ref{248_A2,1} and comprise a $216$-dimensional complement.

\section{$G_{2,1}$ orbifold topological coefficients}\label{sec:appendix_Lambda_computations}
We start with 
\begin{align}
    \Lambda^{L_{3}}_{L_{1}, L_{2}}\equiv \mu^{L_{3}}_{L_{1}, L_{2}}\Delta_{L_{3}}^{L_{1}, L_{2}}.
\end{align}
Given $\mu_{*}: \mathfrak{A}\otimes\mathfrak{A}\to \mathfrak{A}$, and $\Delta_{F}: \mathfrak{A}\otimes\mathfrak{A}\to \mathfrak{A}$, where the subscript $F$ stands for a Frobenius algebra, the components are defined by the maps
\begin{align}
    \begin{split}
        \mu^{L_{3}}_{L_{1}, L_{2}}:&\  \text{Hom}_{\text{Rep}(\mathcal{H})}(L_{1}, \mathfrak{A})\otimes\text{Hom}_{\text{Rep}(\mathcal{H})}(L_{2}, \mathfrak{A})\longrightarrow \text{Hom}_{\text{Rep}(\mathcal{H})}(L_{3}, \mathfrak{A}),\\
        \Delta^{L_{3}}_{L_{1}, L_{2}}:&\ \text{Hom}_{\text{Rep}(\mathcal{H})}(L_{1}, \mathfrak{A})\longrightarrow \text{Hom}_{\text{Rep}(\mathcal{H})}(L_{2}, \mathfrak{A})\otimes\text{Hom}_{\text{Rep}(\mathcal{H})}(L_{3}, \mathfrak{A}),
    \end{split}
\end{align}
where $\mathcal{H}$ is a Hopf algebra. For the multiplicity-free case, we write the multiplication by scalars $\mu^{L_{3}}_{L_{1}, L_{2}}\in \mathbb{C}$ multiplying the basis of intertwiners $L_{1}\otimes L_{2}\to L_{3}$. Unitarity forces
\begin{align}
    \mu^{\mathbb{1}}_{\mathbb{1}, \mathbb{1}} = 1,\qquad \mu^{\tau}_{\mathbb{1},\tau} = \mu^{\tau}_{\tau, \mathbb{1}} = 1,\qquad \mu^{\mathbb{1}}_{\mathbb{1}, \tau} = \mu^{\mathbb{1}}_{\tau, \mathbb{1}} = 0,
\end{align}
and the only non-trivial multiplication constants are
\begin{align}
    \mu^{\mathbb{1}}_{\tau, \tau} \equiv x,\qquad \mu^{\tau}_{\tau, \tau} \equiv y.
\end{align}
Associativity $\mu\circ (\mu\otimes \text{id}) = \mu\circ(\text{id}\otimes\mu)$ on the Fusion rule \ref{fusion_Fib} becomes a $2$-component  constraint because $\text{dim}\ \text{Hom}(\tau^{\otimes3}, \tau) = 2$. Concretely, in the left-associated basis, the coefficient vector is $(x, y^{2})$, while in the right-associated basis, it is the same vector but related by the basis change $F^{\tau}_{\tau\tau\tau}$. Hence, we require
\begin{align}
    \begin{pmatrix}
    x\\
    y^{2}
    \end{pmatrix} = F^{\tau}_{\tau\tau\tau}\begin{pmatrix}
    x\\
    y^{2}
    \end{pmatrix} .
\end{align}
Using \ref{F_matrix_Fib}, we find the $(F - 1)$-kernel condition yields
\begin{align}
    x= \frac{y^{2}}{\sqrt{\varphi}}.
\end{align}
So associativity forces an irrational relation between the two channels of $\tau\tau\to (\mathbb{1}\oplus \tau)$. We make the choice 
\begin{align}\label{gauge_x,y}
    (x,y) = \left(\varphi^{-\frac{1}{2}}, 1\right),
\end{align}
Next, we write $\Delta$ in components $\Delta_{L_{3}}^{L_{1}, L_{2}}$ multiplying basis intertwiners $L_{3}\to L_{21}\otimes L_{2}$. In the language of \cite{Perez-Lona:2023djo}, the \textit{bubble move} is now the following axiom 
\begin{align}\label{muDeltaid}
    \mu\circ \Delta = \text{id}_{\mathfrak{A}}.
\end{align}
In the multiplicity-free case, a very explicit way to solve this is to choose $\Delta$ as a bimodule section of $\mu$ channel-by-channel\footnote{This is the categorical analogue of choosing a separability idempotent.}. Concretely, in the $1$ output, the only contributing pairs are $(\mathbb{1}, \mathbb{1})$ and $(\tau, \tau)\to \mathbb{1}$. So, we impose
\begin{align}
    \mu^{\mathbb{1}}_{\mathbb{1}, \mathbb{1}}\Delta^{\mathbb{1}, \mathbb{1}}_{\mathbb{1}} + \mu^{\mathbb{1}}_{\tau, \tau}\Delta^{\tau, \tau}_{\mathbb{1}} = 1.
\end{align}
A canonical symmetric choice is to take $\Delta$ proportional to the transpose of $\mu$ in that two-dimensional space, i.e., 
\begin{align}
    \Delta^{\mathbb{1}, \mathbb{1}}_{\mathbb{1}} = \frac{1}{1 + x^{2}},\qquad \Delta^{\tau, \tau}_{\mathbb{1}} = \frac{x}{1 + x^{2}}.
\end{align}
Next, in the $\tau$ output, the contributing pairs are $(\mathbb{1}, \tau)$, $(\tau, \mathbb{1})$, and $(\tau \tau)\to \tau$. so we impose
\begin{align}
    \Delta^{\mathbb{1}, \tau}_{\tau} + \Delta^{\tau, \mathbb{1}}_{\tau} + y\Delta^{\tau, \tau}_{\tau} = 1,
\end{align}
and again take the symmetric transpose choice
\begin{align}
    \Delta^{\mathbb{1}, \tau}_{\tau} = \Delta^{\tau, \mathbb{1}}_{\tau} = \frac{1}{2 + y^{2}},\qquad \Delta^{\tau, \tau}_{\tau} = \frac{y}{2 + y^{2}}.
\end{align}
This is exactly the move that makes \ref{muDeltaid} manifest in components. This has the same structural role as the $\tfrac{1}{\vert G\vert}$ normalization in ordinary group orbifolds. We can now compute the topological coefficients $\Lambda^{L_{3}}_{L_{1}, L_{2}}$ explictly. 
\begin{enumerate}
    \item Vacuum output $L_{3} = 1$:\\
    Only $(\mathbb{1}, \mathbb{1})$ and $(\tau, \tau)$ contribute:
    \begin{align}
        \Lambda^{\mathbb{1}}_{\mathbb{1}, \mathbb{1}} = \mu^{\mathbb{1}}_{\mathbb{1},\mathbb{1}}\Delta^{\mathbb{1},\mathbb{1}}_{\mathbb{1}} = \frac{1}{1 + x^{2}},\qquad \Lambda^{\mathbb{1}}_{\tau, \tau} = \mu^{\mathbb{1}}_{\tau, \tau}\Delta^{\tau, \tau}_{\mathbb{1}} = \frac{x^{2}}{1 + x^{2}}.
    \end{align}

    \item $\tau$ output $L_{3} = \tau$:\\
    Contributions from $(\mathbb{1}, \tau)$, $(\tau, \mathbb{1})$, $(\tau, \tau)\to \tau$:
    \begin{align}
        \begin{split}
            \Lambda^{\tau}_{\mathbb{1},\tau} =& \mu^{\tau}_{\mathbb{1},\tau}\Delta^{\tau, \mathbb{1}}_{\tau} = \frac{1}{2 + y^{2}},\\
            \Lambda^{\tau}_{\tau, \mathbb{1}} =& \mu^{\tau}_{\tau, \mathbb{1}}\Delta^{\tau, \mathbb{1}}_{\tau} = \frac{1}{2 + y^{2}},\\
            \Lambda^{\tau}_{\tau\tau} =& \mu^{\tau}_{\tau, \tau}\Delta^{\tau, \tau}_{\tau} = \frac{y^{2}}{2 + y^{2}}.
        \end{split}
    \end{align}
\end{enumerate}
With our choice \ref{gauge_x,y}, and using the fact that $1 + \varphi = \varphi^{2}$, we find the list \ref{Lambdas}.

\section{More computations on branching rules}\label{sec:appendix_branching}
\subsection*{The $\mathbf{120}$ from $\Lambda^{2}(\mathbf{16})$}

The branching rule for the $\mathbf{120}$ of $\text{Spin}(16)$ under
$\text{Spin}(16)\supset \text{Spin}(4)\times \text{Spin}(12)$ is most easily obtained from the
identification of the adjoint representation with two-forms. Indeed,
$\mathfrak{so}(N)$ consists of antisymmetric endomorphisms of $\mathbb{R}^{N}$, and therefore admits a canonical identification as an $\mathfrak{so}(N)$-module with $\Lambda^{2}(\mathbb{R}^{N})$ \cite{Fulton1991RepresentationTA}
\begin{align}
    \text{Ad}_{\mathfrak{so}(N)}\ \cong\ \Lambda^{2}(\mathbf{N}).
\end{align}
Under $\mathfrak{so}(16)\supset\mathfrak{so}(4)\oplus\mathfrak{so}(12)$, the
vector representation decomposes as
\begin{align}
    \mathbf{16}\ \to\ (\mathbf{4},\mathbf{1})\oplus (\mathbf{1},\mathbf{12}).
\end{align}
Applying $\Lambda^2$ and using the standard identity
\begin{align}\label{adjoint_identity}
    \Lambda^{2}(A\oplus B)\ \cong\ \Lambda^{2}A\ \oplus\ (A\otimes B)\ \oplus\ \Lambda^{2}B,
\end{align}
we obtain
\begin{align}
    \Lambda^{2}(\mathbf{16})
    &\to \Lambda^{2}(\mathbf{4},\mathbf{1})
        \oplus \big((\mathbf{4},\mathbf{1})\otimes(\mathbf{1},\mathbf{12})\big)
        \oplus \Lambda^{2}(\mathbf{1},\mathbf{12}) \nonumber\\
    &\cong (\Lambda^{2}\mathbf{4},\mathbf{1})\ \oplus\ (\mathbf{4},\mathbf{12})\ \oplus\ (\mathbf{1},\Lambda^{2}\mathbf{12}).
\end{align}
Taking dimensions, we find
\begin{align}
    \dim(\Lambda^{2}\mathbf{4})=\dim(\mathfrak{so}(4))=6,\qquad
    \dim(\Lambda^{2}\mathbf{12})=\dim(\mathfrak{so}(12))=66,\qquad
    \dim(\mathbf{4}\otimes\mathbf{12})=48,
\end{align}
and therefore
\begin{align}
    \mathbf{120}\ \to\ (\mathbf{6},\mathbf{1})\ \oplus\ (\mathbf{1},\mathbf{66})\ \oplus\ (\mathbf{4},\mathbf{12}),
\end{align}
as claimed.

\subsection*{The $\mathbf{128}$ from Clifford factorization and chirality}
To describe the branching of the chiral spinor $\mathbf{128}$, it is useful
to recall that spinor representations are naturally constructed as modules of
the Clifford algebra \cite{Fulton1991RepresentationTA}. Let $V$ be a real even-dimensional quadratic space and
$\text{Cl}(V)$ its Clifford algebra. If $V$ splits orthogonally as
\begin{align}
    V\ \cong\ V_1\ \oplus\ V_2,
\end{align}
then there is a canonical graded algebra isomorphism
\begin{align}
    \text{Cl}(V)\ \cong\ \text{Cl}(V_1)\,\widehat{\otimes}\,\text{Cl}(V_2)\,,
\end{align}
where $\widehat{\otimes}$ denotes the $\mathbb{Z}_{2}$-graded tensor product. In
particular, for $V=\mathbb{R}^{2n}$ with the standard inner product and a split
$\mathbb{R}^{2n}\cong \mathbb{R}^{2r}\oplus \mathbb{R}^{2n-2r}$, we have
\begin{align}
    \text{Cl}(\mathbb{R}^{2n})\ \cong\ \text{Cl}(\mathbb{R}^{2r})\,\widehat{\otimes}\,\text{Cl}(\mathbb{R}^{2n-2r}).
\end{align}
Let $S^{(2m)}$ denote a complex spinor module for $\text{Cl}(\mathbb{R}^{2m})$. The above
factorization implies that, as a module for $\text{Cl}(\mathbb{R}^{2r})\widehat{\otimes}\text{Cl}(\mathbb{R}^{2n-2r})$, we have
\begin{align}
    S^{(2n)}\ \cong\ S^{(2r)}\otimes S^{(2n-2r)}.
\end{align}
When $2m$ is even, the chirality operator $\Gamma_{(2m)}$, which is the product of the $2m$
gamma matrices, yields the decomposition into Weyl spinors
$S^{(2m)}=S^{(2m)}_{+}\oplus S^{(2m)}_{-}$. Under the orthogonal split, chirality factorizes (up to a convention-dependent overall sign) as follows
\begin{align}
    \Gamma_{(2n)}\ =\ \Gamma_{(2r)}\otimes \Gamma_{(2n-2r)},
\end{align}
so that the $\pm$ eigenspaces obey the equal-chirality selection rule
\begin{align}
    S^{(2n)}_{+}\ \cong\ \big(S^{(2r)}_{+}\otimes S^{(2n-2r)}_{+}\big)\ \oplus\
    \big(S^{(2r)}_{-}\otimes S^{(2n-2r)}_{-}\big),\label{chirality_rule}\\
    S^{(2n)}_{-}\ \cong\ \big(S^{(2r)}_{+}\otimes S^{(2n-2r)}_{-}\big)\ \oplus\
    \big(S^{(2r)}_{-}\otimes S^{(2n-2r)}_{+}\big).\nonumber
\end{align}
We now specialize to $2n=16$ and the split $\mathbb{R}^{16}\cong \mathbb{R}^{4}\oplus \mathbb{R}^{12}$,
corresponding to $\text{Spin}(16)\supset \text{Spin}(4)\times \text{Spin}(12)$. The chiral
spinors of $\text{Spin}(4)\cong \text{SU}(2)_L\times \text{SU}(2)_R$ are $(\mathbf{2},\mathbf{1})$
and $(\mathbf{1},\mathbf{2})$, which we denote by $\mathbf{2}$ and $\mathbf{2}'$,
respectively, while the chiral spinors of $\text{Spin}(12)$ are the inequivalent
Weyl spinors $\mathbf{32}$ and $\mathbf{32}'$. Taking $S^{(16)}_{+}\cong \mathbf{128}$,
\ref{chirality_rule} gives
\begin{align}
    \mathbf{128}\ \to\ (\mathbf{2},\mathbf{32})\ \oplus\ (\mathbf{2}',\mathbf{32}'),
\end{align}
which is the desired branching.

\subsection*{Branching rule computation for $A_{2,1}^{\otimes4}$}\label{sec:appendix_A2,1_branching}
In this appendix, we present an explicit computation that helps us understand the $E_{8}$ adjoint $\mathbf{248}$ as a representation of four commuting $\text{SU}(3)$ factors, in a way that makes the following commutant pair split manifest
\begin{align}\label{commutant pair_split}
    \text{SU}(3)^{4} = \left(\text{SU}(3)_{(1)}\times \text{SU}(3)_{(2)}\right)\times \left(\text{SU}(3)_{(3)}\times \text{SU}(3)_{(4)}\right).
\end{align}
We first define
\begin{align}
    H\equiv \text{SU}(3)_{(1)}\times \text{SU}(3)_{(2)}\times\text{SU}(3)_{(3)}\times \text{SU}(3)_{(4)},
\end{align}
and consider the embedding
\begin{align}\label{iota_embed}
    \iota: H\hookrightarrow \text{SU}(3)_{(4)}\times E_{6}.
\end{align}
We know the following trinification embedding defined by
\begin{align}
    \jmath: \text{SU}(3)_{(1)}\times \text{SU}(3)_{(2)}\times\text{SU}(3)_{(3)}\hookrightarrow E_{6},
\end{align}
and hence, the action of \ref{iota_embed} reads
\begin{align}
    \iota(g_{1}, g_{2}, g_{3}, g_{4}) \equiv (g_{4}, \jmath(g_{1}, g_{2}, g_{3}))\in \text{SU}(3)_{(4)}\times E_{6}.
\end{align}
This is the map using which we will restrict the representations. Now, we fix a $E_{8}$-representation $V$ so that restricting $V$ to $H$ is to simply view $V$ as a rep of the subgroup $\text{SU}(3)_{(4)}\times E_{6}\subset E_{8}$, and then restrict further along the embedding \ref{iota_embed}. Hence, whenever an element $(g_{1}, g_{2}, g_{3}, g_{4})\in H$ acts on $V$, it acts through an element $(g_{4}, \jmath(g_{1}, g_{2}, g_{3}))\in\text{SU}(3)_{(4)}\times E_{6}\subset E_{8}$. Let us denote an irrep of $H$ to be $(R_{1}, R_{2}, R_{3}, R_{4})$, where $R_{i}$ is the $i^{\text{th}}$ irrep of $\text{SU}(3)_{(i)}$. To keep the commutant pair split \ref{commutant pair_split} visible, we rewrite this irrep as $(R_{1}, R_{2}; R_{3}, R_{4})$.\\

\noindent 
For groups $G, K$, let 
\begin{align}
    A: G\to \text{GL}(V),\qquad B: K\to \text{GL}(W),
\end{align}
be their representations on finite-dimensional vector spaces $V, W$. Their external tensor product, defined as $A\boxtimes B$, is the representation of $G\times K$ on $V\otimes W$ defined as follows
\begin{align}
    \left(A\boxtimes B\right)(g,k)(v\otimes w)\equiv A(g)v\otimes B(k)w.
\end{align}
Hence, if $R$ is a rep of $\text{SU}(3)_{(4)}$ and $S$ is a rep of $E_{6}$, the representation $(R, S)$ of $\text{SU}(3)_{(4)}\times E_{6}$ is precisely $R\boxtimes S$. We can now compute the restriction $\text{Res}^{\text{SU}(3)_{(4)}\times E_{6}}_{H}$ on each term in the $E_{8}$ branching. The restriction commutes with direct sums, and the external tensor products distribute over direct sums as follows
\begin{align}
    \begin{split}
        R\boxtimes\left(\bigoplus\limits_{i}S_{i}\right)\cong& \bigoplus\limits_{i}(R\boxtimes S_{i}),\\
    \left(\bigoplus\limits_{i}R_{i}\right)\boxtimes S_{i}\cong& \bigoplus\limits_{i}\left(R_{i}\boxtimes S\right).
    \end{split}
\end{align}
Now, for a representation $R\boxtimes S$, its restriction to $H$ is given by
\begin{align}
    \text{Res}_{H}^{\text{SU}(3)_{(4)}\times E_{6}}\left(R\boxtimes S\right) = R_{(4)}\boxtimes \text{Res}_{H}^{\text{SU}(3)_{(4)}\times E_{6}}(s),
\end{align}
where $\text{SU}(3)^{3}$ is embedded in $E_{6}$ by $\jmath$. We now present the explicit restrictions. The term-by-term restriction is presented below:
\begin{enumerate}
    \item $(\mathbf{8}, \mathbf{1})$:\\
    \begin{align}
        \begin{split}
            (\mathbf{8}, \mathbf{1})\vert_{H} =& \mathbf{8}_{(4)}\boxtimes(\mathbf{1}\circ\jmath)\\
            =& \mathbf{8}_{(4)}\boxtimes \text{Res}_{\text{SU}(3)^{3}}(\mathbf{1})\\
            =& (\mathbf{1}, \mathbf{1}; \mathbf{1}, \mathbf{8}).
        \end{split}
    \end{align}

    \item $(\mathbf{1}, \mathbf{78})$:\\
    \begin{align}
        \begin{split}
            (\mathbf{1}, \mathbf{78})\vert_{H} =& \mathbf{1}_{(4)}\boxtimes (\mathbf{78}\circ \jmath)\\
            =& \mathbf{1}_{(4)}\boxtimes \text{Res}_{\text{SU}{(3)^{3}}}(\mathbf{78})\\
            =& \mathbf{1}_{(4)}\boxtimes\left[(\mathbf{8},\mathbf{1},\mathbf{1})\oplus(\mathbf{1},\mathbf{8},\mathbf{1})\oplus(\mathbf{1},\mathbf{1},\mathbf{8})\oplus(\mathbf{3},\overline{\mathbf{3}},\mathbf{3})\oplus(\overline{\mathbf{3}},\mathbf{3},\overline{\mathbf{3}})\right]\\
            =& (\mathbf{8},\mathbf{1};\mathbf{1},\mathbf{1})\oplus(\mathbf{1},\mathbf{8};\mathbf{1},\mathbf{1})\oplus(\mathbf{1},\mathbf{1};\mathbf{8},\mathbf{1})\oplus(\mathbf{3},\overline{\mathbf{3}};\mathbf{3},\mathbf{1})\oplus(\overline{\mathbf{3}},\mathbf{3};\overline{\mathbf{3}},\mathbf{1}).
        \end{split}
    \end{align}

    \item $(\mathbf{3}, \mathbf{27})$:\\
    \begin{align}\label{3,27}
        \begin{split}
            (\mathbf{3}, \mathbf{27})\vert_{H} =& \mathbf{3}_{(4)}\boxtimes \text{Res}_{\text{SU}(3)^{3}}(\mathbf{27})\\
            =& \mathbf{3}_{(4)}\boxtimes\left[(\mathbf{3},\mathbf{3},\mathbf{1})\oplus(\mathbf{1},\overline{\mathbf{3}},\mathbf{3})
  \oplus(\overline{\mathbf{3}},\mathbf{1}, \overline{\mathbf{3}})\right]\\
            =& (\mathbf{3},\mathbf{3};\mathbf{1}, \mathbf{3})\oplus(\mathbf{1}, \overline{\mathbf{3}}; \mathbf{3}, \mathbf{3})\oplus(\overline{\mathbf{3}}, \mathbf{1}; \overline{\mathbf{3}}, \mathbf{3}).
        \end{split}
    \end{align}
\end{enumerate}
Putting this together precisely yields the branching \ref{248_A2,1}. The $(\overline{\mathbf{3}}, \overline{\mathbf{27}})$ restriction is the complex conjugate of \ref{3,27}, and supplies the remaining summands.

\section{Computing the $S$-matrix}\label{sec:appendix_S}
The $S$-matrix construction for three-character theories in \cite{Mathur:1988gt} reads
\begin{align}
    S = \begin{pmatrix}
        \alpha_{00} & \frac{1}{n_{1}}\frac{N_{1}}{N_{0}}\alpha_{01} & \frac{1}{n_{2}}\frac{N_{2}}{N_{0}}\alpha_{02}\\
        n_{1}\frac{N_{0}}{N_{1}}\alpha_{10} & \alpha_{11} & \frac{n_{1}}{n_{2}}\frac{N_{2}}{N_{1}}\alpha_{12}\\
        n_{2}\frac{N_{0}}{N_{2}}\alpha_{20} & \frac{n_{2}}{n_{1}}\frac{N_{1}}{N_{2}}\alpha_{21} & \alpha_{22}
    \end{pmatrix},
\end{align}
where
\begin{align}
    \begin{split}
        \alpha_{00} =& \frac{s(a)s\left(a + \frac{1}{2}g\right)}{s(2a)s\left(2a + \frac{1}{2}g\right)},\qquad
        \alpha_{01} = -\frac{s(a)^{2}}{s(2a)s(2a + g)},\qquad \alpha_{02} = \frac{s(a)s\left(a + \frac{1}{2}g\right)}{s\left(2a + \frac{1}{2}g\right)s(2a + g)},\\
        \alpha_{10} =& -2\frac{s\left(3a + \frac{1}{2}g\right)s\left(a + \frac{1}{2}g\right)c\left(\frac{1}{2}g\right)}{s(2a)s\left(2a + \frac{1}{2}g\right)},\\
        \alpha_{11} =& \frac{s\left(3a + \frac{1}{2}g\right)^{2}s(a)}{s(2a)s\left(2a + \frac{1}{2}g\right)} - \frac{s\left(a + \frac{1}{2}g\right)(a)}{s\left(2a + \frac{1}{2}g\right)s(2a + g)},\\
        \alpha_{12} =& 2\frac{s\left(a + \frac{1}{2}g\right)^{2}c\left(\frac{1}{2}g\right)}{s\left(2a + \frac{1}{2}g\right)s(2a + g)},\\
        \alpha_{20} =& \frac{s\left(3a + \frac{1}{2}g\right)s(3a + g)}{s(2a)s\left(2a + \frac{1}{2}g\right)},\qquad   \alpha_{21} = \frac{s(3a + g)s(a)}{s(2a)s(2a + g)},\qquad  \alpha_{22} = \frac{s(a)s\left(a + \frac{1}{2}g\right)}{s\left(2a + \frac{1}{2}g\right)s(2a + g)}.
    \end{split}
\end{align}
Here, $s(x) = \sin(\pi x)$, $c(x) = \cos(\pi x)$, and the constants $a$ and $g$ read
\begin{align}
    \begin{split}
        a =& -\frac{1}{4}(6\alpha_{0} + 1)\pm \frac{1}{4}\sqrt{-12\alpha_{0}^{2} + 4\alpha_{0} + 9 - 4\nu_{1}},\\
        g =& -6\alpha_{0} - 2 - 6a,
    \end{split}
\end{align}
where the sign in front of the square root is decided on the cases, $(+, -) = (h_{1}>h_{2}, h_{1}<h_{2})$. The normalization constants
$N_{0,1,2}$ read
\begin{align}
    \begin{split}
        N_{0} =& \frac{\Gamma(g)}{\Gamma\left(\frac{1}{2}g\right)}\frac{\Gamma(-1 - 3a - g)\Gamma\left(-1-3a - \frac{1}{2}g\right)\Gamma(1 + a)\Gamma\left(1 + a + \frac{1}{2}g\right)}{\Gamma(-2a)\Gamma\left(-2a - \frac{1}{2}g\right)},\\
        N_{1} =& (16)^{1 + 2a}\frac{\Gamma(-1-3a-g)\Gamma(1 + a)^{3}}{\Gamma(-2a-g)\Gamma(2 + 2a)},\\
        N_{2} =& (16)^{2 + 4a + g}\frac{\Gamma(g)}{\Gamma\left(\frac{1}{2}g\right)}\frac{\Gamma(1 + a)^{2}\Gamma\left(1 + a + \frac{1}{2}g\right)^{2}}{\Gamma\left(2 + 2a + \frac{1}{2}g\right)\Gamma(2 + 2a + g)}.
    \end{split}
\end{align}
We also have
\begin{align}
    \alpha_{0} = -\frac{c}{24},\qquad \nu_{1} = \frac{20}{9} - \mu_{1},
\end{align}
and finally, $\mu_{1}$ can be explicitly computed as follows
\begin{align}
    \mu_{1} = \frac{2}{9} - 4\left(\alpha_{0}\alpha_{1} + \alpha_{1}\alpha_{2} + \alpha_{2}\alpha_{0}\right),
\end{align}
with 
\begin{align}
    \alpha_{i} = -\frac{c}{24} + h_{i},\ h_{0} = 0.
\end{align}

%\section{}\labelf{sec: appendix_Hauptmodul}

\bibliographystyle{JHEP}

\bibliography{Tdls}

\providecommand{\href}[2]{#2}\begingroup\raggedright\begin{thebibliography}{10}

\bibitem{Hegde:2021sdm}
S.~Hegde and D.~P. Jatkar, \emph{{Defect partition function from TDLs in
  commutant pairs}},
  \href{https://doi.org/10.1142/S0217732322501930}{\emph{Mod. Phys. Lett. A}
  {\bfseries 37} (2022) 2250193}
  [\href{https://arxiv.org/abs/2101.12189}{{\ttfamily 2101.12189}}].

\bibitem{Lin:2019hks}
Y.-H. Lin and S.-H. Shao, \emph{{Duality Defect of the Monster CFT}},
  \href{https://doi.org/10.1088/1751-8121/abd69e}{\emph{J. Phys. A} {\bfseries
  54} (2021) 065201} [\href{https://arxiv.org/abs/1911.00042}{{\ttfamily
  1911.00042}}].

\bibitem{Schellekens:1992db}
A.~N. Schellekens, \emph{{Meromorphic c = 24 conformal field theories}},
  \href{https://doi.org/10.1007/BF02099044}{\emph{Commun. Math. Phys.}
  {\bfseries 153} (1993) 159}
  [\href{https://arxiv.org/abs/hep-th/9205072}{{\ttfamily hep-th/9205072}}].

\bibitem{Polyakov:1981rd}
A.~M. Polyakov, \emph{{Quantum Geometry of Bosonic Strings}},
  \href{https://doi.org/10.1016/0370-2693(81)90743-7}{\emph{Phys. Lett. B}
  {\bfseries 103} (1981) 207}.

\bibitem{Moore:1988qv}
G.~W. Moore and N.~Seiberg, \emph{{Classical and Quantum Conformal Field
  Theory}}, \href{https://doi.org/10.1007/BF01238857}{\emph{Commun. Math.
  Phys.} {\bfseries 123} (1989) 177}.

\bibitem{Mathur:1988na}
S.~D. Mathur, S.~Mukhi and A.~Sen, \emph{{On the Classification of Rational
  Conformal Field Theories}},
  \href{https://doi.org/10.1016/0370-2693(88)91765-0}{\emph{Phys. Lett.}
  {\bfseries B213} (1988) 303}.

\bibitem{Bantay:2005vk}
P.~Bantay and T.~Gannon, \emph{{Conformal characters and the modular
  representation}},
  \href{https://doi.org/10.1088/1126-6708/2006/02/005}{\emph{JHEP} {\bfseries
  02} (2006) 005} [\href{https://arxiv.org/abs/hep-th/0512011}{{\ttfamily
  hep-th/0512011}}].

\bibitem{Chandra:2018ezv}
A.~R. Chandra and S.~Mukhi, \emph{{Curiosities above c = 24}},
  \href{https://doi.org/10.21468/SciPostPhys.6.5.053}{\emph{SciPost Phys.}
  {\bfseries 6} (2019) 053} [\href{https://arxiv.org/abs/1812.05109}{{\ttfamily
  1812.05109}}].

\bibitem{Witten:2007kt}
E.~Witten, \emph{{Three-Dimensional Gravity Revisited}},
  \href{https://arxiv.org/abs/0706.3359}{{\ttfamily 0706.3359}}.

\bibitem{Maloney:2007ud}
A.~Maloney and E.~Witten, \emph{{Quantum Gravity Partition Functions in Three
  Dimensions}}, \href{https://doi.org/10.1007/JHEP02(2010)029}{\emph{JHEP}
  {\bfseries 02} (2010) 029} [\href{https://arxiv.org/abs/0712.0155}{{\ttfamily
  0712.0155}}].

\bibitem{Gaberdiel:2007ve}
M.~R. Gaberdiel, \emph{{Constraints on extremal self-dual CFTs}},
  \href{https://doi.org/10.1088/1126-6708/2007/11/087}{\emph{JHEP} {\bfseries
  11} (2007) 087} [\href{https://arxiv.org/abs/0707.4073}{{\ttfamily
  0707.4073}}].

\bibitem{Burbano:2021loy}
I.~M. Burbano, J.~Kulp and J.~Neuser, \emph{{Duality defects in E$_{8}$}},
  \href{https://doi.org/10.1007/JHEP10(2022)187}{\emph{JHEP} {\bfseries 10}
  (2022) 186} [\href{https://arxiv.org/abs/2112.14323}{{\ttfamily
  2112.14323}}].

\bibitem{Stigner:2010nz}
C.~Stigner, \emph{{Factorization constraints and boundary conditions in
  rational CFT}},  \href{https://arxiv.org/abs/1006.5923}{{\ttfamily
  1006.5923}}.

\bibitem{Bantay:2001ni}
P.~Bantay, \emph{{The Kernel of the modular representation and the Galois
  action in RCFT}},
  \href{https://doi.org/10.1007/s00220-002-0760-x}{\emph{Commun. Math. Phys.}
  {\bfseries 233} (2003) 423}
  [\href{https://arxiv.org/abs/math/0102149}{{\ttfamily math/0102149}}].

\bibitem{Das:2022bxm}
A.~Das and N.~B. Umasankar, \emph{{Two- \& Three-character solutions to MLDEs
  and Ramanujan-Eisenstein Identities for Fricke Groups}},
  \href{https://arxiv.org/abs/2211.15369}{{\ttfamily 2211.15369}}.

\bibitem{Umasankar:2022kzs}
N.~B. Umasankar, \emph{{Modular Linear Differential Equations for Hecke and
  Fricke Groups}},  \href{https://arxiv.org/abs/2210.07186}{{\ttfamily
  2210.07186}}.

\bibitem{Barkeshli:2014cna}
M.~Barkeshli, P.~Bonderson, M.~Cheng and Z.~Wang, \emph{{Symmetry
  Fractionalization, Defects, and Gauging of Topological Phases}},
  \href{https://doi.org/10.1103/PhysRevB.100.115147}{\emph{Phys. Rev. B}
  {\bfseries 100} (2019) 115147}
  [\href{https://arxiv.org/abs/1410.4540}{{\ttfamily 1410.4540}}].

\bibitem{Gaiotto:2014kfa}
D.~Gaiotto, A.~Kapustin, N.~Seiberg and B.~Willett, \emph{{Generalized Global
  Symmetries}}, \href{https://doi.org/10.1007/JHEP02(2015)172}{\emph{JHEP}
  {\bfseries 02} (2015) 172} [\href{https://arxiv.org/abs/1412.5148}{{\ttfamily
  1412.5148}}].

\bibitem{Benedetti:2024dku}
V.~Benedetti, H.~Casini, Y.~Kawahigashi, R.~Longo and J.~M. Magan,
  \emph{{Modular invariance as completeness}},
  \href{https://doi.org/10.1103/PhysRevD.110.125004}{\emph{Phys. Rev. D}
  {\bfseries 110} (2024) 125004}
  [\href{https://arxiv.org/abs/2408.04011}{{\ttfamily 2408.04011}}].

\bibitem{Shao:2023gho}
S.-H. Shao, \emph{{What's Done Cannot Be Undone: TASI Lectures on
  Non-Invertible Symmetries}},  in \emph{{Theoretical Advanced Study Institute
  in Elementary Particle Physics 2023}: {Aspects of Symmetry}}, 8, 2023,
  \href{https://arxiv.org/abs/2308.00747}{{\ttfamily 2308.00747}}.

\bibitem{Iqbal:2024pee}
N.~Iqbal, \emph{{Jena lectures on generalized global symmetries: principles and
  applications}},  7, 2024, \href{https://arxiv.org/abs/2407.20815}{{\ttfamily
  2407.20815}}.

\bibitem{Das:2022slz}
A.~Das, C.~N. Gowdigere and S.~Mukhi, \emph{{New meromorphic CFTs from
  cosets}}, \href{https://doi.org/10.1007/JHEP07(2022)152}{\emph{JHEP}
  {\bfseries 07} (2022) 152}
  [\href{https://arxiv.org/abs/2207.04061}{{\ttfamily 2207.04061}}].

\bibitem{Rastelli:2025nyn}
L.~Rastelli, B.~C. Rayhaun, M.~Sacchi and G.~Zafrir, \emph{{$2+2=4$}},
  \href{https://arxiv.org/abs/2601.00058}{{\ttfamily 2601.00058}}.

\bibitem{Chandra:2025qpv}
A.~R. Chandra, S.~Mukhi and P.~Singh, \emph{{Generalised 4d Partition Functions
  and Modular Differential Equations}},
  \href{https://arxiv.org/abs/2512.02107}{{\ttfamily 2512.02107}}.

\bibitem{Deb:2025ddc}
A.~Deb, \emph{{Generalized Schur limit, modular differential equations and
  quantum monodromy traces}},
  \href{https://arxiv.org/abs/2512.02102}{{\ttfamily 2512.02102}}.

\bibitem{Verlinde:1988sn}
E.~P. Verlinde, \emph{{Fusion Rules and Modular Transformations in 2D Conformal
  Field Theory}},
  \href{https://doi.org/10.1016/0550-3213(88)90603-7}{\emph{Nucl. Phys.}
  {\bfseries B300} (1988) 360}.

\bibitem{Drukker:2010jp}
N.~Drukker, D.~Gaiotto and J.~Gomis, \emph{{The Virtue of Defects in 4D Gauge
  Theories and 2D CFTs}},
  \href{https://doi.org/10.1007/JHEP06(2011)025}{\emph{JHEP} {\bfseries 06}
  (2011) 025} [\href{https://arxiv.org/abs/1003.1112}{{\ttfamily 1003.1112}}].

\bibitem{Gaiotto:2014lma}
D.~Gaiotto, \emph{{Open Verlinde line operators}},
  \href{https://arxiv.org/abs/1404.0332}{{\ttfamily 1404.0332}}.

\bibitem{Frohlich:2009gb}
J.~Frohlich, J.~Fuchs, I.~Runkel and C.~Schweigert, \emph{{Defect Lines,
  Dualities and Generalised Orbifolds}},  in \emph{{16th International Congress
  on Mathematical Physics}}, pp.~608--613, 2010,
  \href{https://arxiv.org/abs/0909.5013}{{\ttfamily 0909.5013}},
  \href{https://doi.org/10.1142/9789814304634_0056}{DOI}.

\bibitem{Chang:2018iay}
C.-M. Chang, Y.-H. Lin, S.-H. Shao, Y.~Wang and X.~Yin, \emph{{Topological
  Defect Lines and Renormalization Group Flows in Two Dimensions}},
  \href{https://doi.org/10.1007/JHEP01(2019)026}{\emph{JHEP} {\bfseries 01}
  (2019) 026} [\href{https://arxiv.org/abs/1802.04445}{{\ttfamily
  1802.04445}}].

\bibitem{Chen:2023jht}
J.~Chen, B.~Haghighat and Q.-R. Wang, \emph{{Para-fusion Category and
  Topological Defect Lines in $\mathbb Z_N$-parafermionic CFTs}},
  \href{https://arxiv.org/abs/2309.01914}{{\ttfamily 2309.01914}}.

\bibitem{Frenkel2000VertexAA}
E.~Frenkel, \emph{Vertex algebras and algebraic curves},  2000.

\bibitem{Frenkel:1988xz}
I.~Frenkel, J.~Lepowsky and A.~Meurman, \emph{Vertex Operator Algebras and the
  Monster}. Academic Press, Boston, USA, 1988.

\bibitem{Kac1997VertexAF}
V.~G. Kac, \emph{Vertex algebras for beginners},  1997,
  \href{https://api.semanticscholar.org/CorpusID:118198735}{https://api.semanticscholar.org/CorpusID:118198735}.

\bibitem{Huang1993ATO}
Y.-Z. Huang and J.~Lepowsky, \emph{A theory of tensor products for module
  categories for a vertex operator algebra, i}, {\emph{Selecta Mathematica}
  {\bfseries 1} (1993) 699}.

\bibitem{Huang2014BraidedTC}
Y.-Z. Huang, A.~A. Kirillov and J.~Lepowsky, \emph{Braided tensor categories
  and extensions of vertex operator algebras}, {\emph{Communications in
  Mathematical Physics} {\bfseries 337} (2014) 1143}.

\bibitem{Zhu:1996gaq}
Y.~Zhu, \emph{{Modular invariance of characters of vertex operator algebras}},
  \href{https://doi.org/10.1090/s0894-0347-96-00182-8}{\emph{J. Am. Math. Soc.}
  {\bfseries 9} (1996) 237}.

\bibitem{Mason:2007}
G.~Mason, \emph{{Vector-valued Modular Forms and Linear Differential
  Operators}},
  \href{https://doi.org/10.1142/S1793042107000973}{\emph{International Journal
  of Number Theory} {\bfseries 03} (2007) 377}.

\bibitem{Bantay:2007zz}
P.~Bantay and T.~Gannon, \emph{{Vector-valued modular functions for the modular
  group and the hypergeometric equation}},
  \href{https://doi.org/10.4310/CNTP.2007.v1.n4.a2}{\emph{Commun. Num. Theor.
  Phys.} {\bfseries 1} (2007) 651}.

\bibitem{Mathur:1988rx}
S.~D. Mathur, S.~Mukhi and A.~Sen, \emph{{Differential Equations for
  Correlators and Characters in Arbitrary Rational Conformal Field Theories}},
  \href{https://doi.org/10.1016/0550-3213(89)90022-9}{\emph{Nucl. Phys.}
  {\bfseries B312} (1989) 15}.

\bibitem{Mathur:1988gt}
S.~D. Mathur, S.~Mukhi and A.~Sen, \emph{{Reconstruction of Conformal Field
  Theories From Modular Geometry on the Torus}},
  \href{https://doi.org/10.1016/0550-3213(89)90615-9}{\emph{Nucl. Phys.}
  {\bfseries B318} (1989) 483}.

\bibitem{Hampapura:2015cea}
H.~R. Hampapura and S.~Mukhi, \emph{{On 2d Conformal Field Theories with Two
  Characters}}, \href{https://doi.org/10.1007/JHEP01(2016)005}{\emph{JHEP}
  {\bfseries 01} (2016) 005}
  [\href{https://arxiv.org/abs/1510.04478}{{\ttfamily 1510.04478}}].

\bibitem{Das:2022uoe}
A.~Das, C.~N. Gowdigere and S.~Mukhi, \emph{{Meromorphic cosets and the
  classification of three-character CFT}},
  \href{https://doi.org/10.1007/JHEP03(2023)023}{\emph{JHEP} {\bfseries 03}
  (2023) 023} [\href{https://arxiv.org/abs/2212.03136}{{\ttfamily
  2212.03136}}].

\bibitem{Gannon:2013jua}
T.~Gannon, \emph{{The theory of vector-modular forms for the modular group}},
  \href{https://doi.org/10.1007/978-3-662-43831-2_9}{\emph{Contrib. Math.
  Comput. Sci.} {\bfseries 8} (2014) 247}
  [\href{https://arxiv.org/abs/1310.4458}{{\ttfamily 1310.4458}}].

\bibitem{Das:2020wsi}
A.~Das, C.~N. Gowdigere and J.~Santara, \emph{{Wronskian Indices and Rational
  Conformal Field Theories}},
  \href{https://doi.org/10.1007/JHEP04(2021)294}{\emph{JHEP} {\bfseries 04}
  (2021) 294} [\href{https://arxiv.org/abs/2012.14939}{{\ttfamily
  2012.14939}}].

\bibitem{Naculich:1988xv}
S.~G. Naculich, \emph{{Differential Equations for Rational Conformal
  Characters}}, \href{https://doi.org/10.1016/0550-3213(89)90150-8}{\emph{Nucl.
  Phys.} {\bfseries B323} (1989) 423}.

\bibitem{Pan:2023jjw}
Y.~Pan and Y.~Wang, \emph{{Flavored modular differential equations}},
  \href{https://arxiv.org/abs/2306.10569}{{\ttfamily 2306.10569}}.

\bibitem{Das:2023qns}
A.~Das, C.~N. Gowdigere, S.~Mukhi and J.~Santara, \emph{{Modular differential
  equations with movable poles and admissible RCFT characters}},
  \href{https://doi.org/10.1007/JHEP12(2023)143}{\emph{JHEP} {\bfseries 12}
  (2023) 143} [\href{https://arxiv.org/abs/2308.00069}{{\ttfamily
  2308.00069}}].

\bibitem{Govindarajan:2025jlq}
S.~Govindarajan, A.~Sadanandan and J.~Santara, \emph{{Quasi-Characters for
  three-character Rational Conformal Field Theories}},
  \href{https://arxiv.org/abs/2510.24248}{{\ttfamily 2510.24248}}.

\bibitem{Turaev:1994xb}
V.~G. Turaev, \emph{{Quantum invariants of knots and three manifolds}},
  vol.~18. 1994.

\bibitem{Bakalov2000LecturesOT}
B.~N. Bakalov and A.~A. Kirillov, \emph{Lectures on tensor categories and
  modular functors},  2000,
  \href{https://api.semanticscholar.org/CorpusID:52201867}{https://api.semanticscholar.org/CorpusID:52201867}.

\bibitem{Kitaev:2005hzj}
A.~Kitaev, \emph{{Anyons in an exactly solved model and beyond}},
  \href{https://doi.org/10.1016/j.aop.2005.10.005}{\emph{Annals Phys.}
  {\bfseries 321} (2006) 2}
  [\href{https://arxiv.org/abs/cond-mat/0506438}{{\ttfamily
  cond-mat/0506438}}].

\bibitem{Fuchs:2002cm}
J.~Fuchs, I.~Runkel and C.~Schweigert, \emph{{TFT construction of RCFT
  correlators 1. Partition functions}},
  \href{https://doi.org/10.1016/S0550-3213(02)00744-7}{\emph{Nucl. Phys. B}
  {\bfseries 646} (2002) 353}
  [\href{https://arxiv.org/abs/hep-th/0204148}{{\ttfamily hep-th/0204148}}].

\bibitem{Fuchs:2003id}
J.~Fuchs, I.~Runkel and C.~Schweigert, \emph{{TFT construction of RCFT
  correlators. 2. Unoriented world sheets}},
  \href{https://doi.org/10.1016/j.nuclphysb.2003.11.026}{\emph{Nucl. Phys. B}
  {\bfseries 678} (2004) 511}
  [\href{https://arxiv.org/abs/hep-th/0306164}{{\ttfamily hep-th/0306164}}].

\bibitem{Felder:1999mq}
G.~Felder, J.~Frohlich, J.~Fuchs and C.~Schweigert, \emph{{Correlation
  functions and boundary conditions in RCFT and three-dimensional topology}},
  \href{https://doi.org/10.1023/A:1014903315415}{\emph{Compos. Math.}
  {\bfseries 131} (2002) 189}
  [\href{https://arxiv.org/abs/hep-th/9912239}{{\ttfamily hep-th/9912239}}].

\bibitem{Gannon:1999cp}
T.~Gannon, \emph{{The Cappelli-Itzykson-Zuber A-D-E classification}},
  \href{https://doi.org/10.1142/S0129055X00000265}{\emph{Rev. Math. Phys.}
  {\bfseries 12} (2000) 739}
  [\href{https://arxiv.org/abs/math/9902064}{{\ttfamily math/9902064}}].

\bibitem{Frohlich2004PicardGI}
J.~Frohlich, J.~Fuchs, I.~Runkel and C.~Schweigert, \emph{Picard groups in
  rational conformal field theory},  2004,
  \href{https://api.semanticscholar.org/CorpusID:13921194}{https://api.semanticscholar.org/CorpusID:13921194}.

\bibitem{Cheng:2020rpl}
M.~Cheng and D.~J. Williamson, \emph{{Relative anomaly in ( 1+1 )d rational
  conformal field theory}},
  \href{https://doi.org/10.1103/PhysRevResearch.2.043044}{\emph{Phys. Rev.
  Res.} {\bfseries 2} (2020) 043044}
  [\href{https://arxiv.org/abs/2002.02984}{{\ttfamily 2002.02984}}].

\bibitem{Edie-Michell:2022abq}
C.~Edie-Michell, \emph{{Auto-equivalences of the modular tensor categories of
  type A, B, C and G}},
  \href{https://doi.org/10.1016/j.aim.2022.108364}{\emph{Adv. Math.} {\bfseries
  402} (2022) 108364}.

\bibitem{Albert:2025umy}
J.~Albert, Y.~Honda, J.~Kaidi and Y.~Zheng, \emph{{Haagerup Symmetry in
  $(E_8)_1$?}},  \href{https://arxiv.org/abs/2512.08225}{{\ttfamily
  2512.08225}}.

\bibitem{Sonoda:1988fq}
H.~Sonoda, \emph{{SEWING CONFORMAL FIELD THEORIES. 2.}},
  \href{https://doi.org/10.1016/0550-3213(88)90067-3}{\emph{Nucl. Phys. B}
  {\bfseries 311} (1988) 417}.

\bibitem{fuchs2002tft}
J.~Fuchs, I.~Runkel and C.~Schweigert, \emph{{TFT} construction of {RCFT}
  correlators {I}: Partition functions}, {\emph{Nuclear Physics B} {\bfseries
  646} (2002) 353}.

\bibitem{Moore:1988uz}
G.~W. Moore and N.~Seiberg, \emph{{Polynomial Equations for Rational Conformal
  Field Theories}},
  \href{https://doi.org/10.1016/0370-2693(88)91796-0}{\emph{Phys. Lett.}
  {\bfseries B212} (1988) 451}.

\bibitem{Belavin:1984vu}
A.~A. Belavin, A.~M. Polyakov and A.~B. Zamolodchikov, \emph{{Infinite
  Conformal Symmetry in Two-Dimensional Quantum Field Theory}},
  \href{https://doi.org/10.1016/0550-3213(84)90052-X}{\emph{Nucl. Phys.}
  {\bfseries B241} (1984) 333}.

\bibitem{Green:1987sp}
M.~B. Green, J.~H. Schwarz and E.~Witten, \emph{{SUPERSTRING THEORY. VOL. 1:
  INTRODUCTION}}, Cambridge Monographs on Mathematical Physics. 7, 1988.

\bibitem{Knus2009TrialitarianAO}
M.~A. Knus, \emph{Trialitarian automorphisms of lie algebras},
  {\emph{Transformation Groups} {\bfseries 14} (2009) 361}.

\bibitem{Chang:2022hud}
C.-M. Chang, J.~Chen and F.~Xu, \emph{{Topological defect lines in two
  dimensional fermionic CFTs}},
  \href{https://doi.org/10.21468/SciPostPhys.15.5.216}{\emph{SciPost Phys.}
  {\bfseries 15} (2023) 216}
  [\href{https://arxiv.org/abs/2208.02757}{{\ttfamily 2208.02757}}].

\bibitem{Dijkgraaf:1989hb}
R.~Dijkgraaf, C.~Vafa, E.~P. Verlinde and H.~L. Verlinde, \emph{{The Operator
  Algebra of Orbifold Models}},
  \href{https://doi.org/10.1007/BF01238812}{\emph{Commun. Math. Phys.}
  {\bfseries 123} (1989) 485}.

\bibitem{Robbins:2019zdb}
D.~Robbins and T.~Vandermeulen, \emph{{Orbifolds from Modular Orbits}},
  \href{https://doi.org/10.1103/PhysRevD.101.106021}{\emph{Phys. Rev. D}
  {\bfseries 101} (2020) 106021}
  [\href{https://arxiv.org/abs/1911.05172}{{\ttfamily 1911.05172}}].

\bibitem{DiFrancesco:1997nk}
P.~Di~Francesco, P.~Mathieu and D.~Senechal, \emph{{Conformal Field Theory}},
  Graduate Texts in Contemporary Physics. Springer-Verlag, New York, 1997,
  \href{https://doi.org/10.1007/978-1-4612-2256-9}{10.1007/978-1-4612-2256-9}.

\bibitem{Perez-Lona:2023djo}
A.~Perez-Lona, D.~Robbins, E.~Sharpe, T.~Vandermeulen and X.~Yu, \emph{{Notes
  on gauging noninvertible symmetries. Part I. Multiplicity-free cases}},
  \href{https://doi.org/10.1007/JHEP02(2024)154}{\emph{JHEP} {\bfseries 02}
  (2024) 154} [\href{https://arxiv.org/abs/2311.16230}{{\ttfamily
  2311.16230}}].

\bibitem{Kirillov2008}
A.~A. Kirillov, \emph{An Introduction to Lie Groups and Lie Algebras},
  Cambridge Studies in Advanced Mathematics. Cambridge University Press,
  Cambridge, 2008.

\bibitem{Davydov2010TheWG}
A.~Davydov, M.~Mueger, D.~Nikshych and V.~Ostrik, \emph{The witt group of
  non-degenerate braided fusion categories},  2010,
  \href{https://api.semanticscholar.org/CorpusID:119165523}{https://api.semanticscholar.org/CorpusID:119165523}.

\bibitem{Rowell:2007dge}
E.~Rowell, R.~Stong and Z.~Wang, \emph{{On Classification of Modular Tensor
  Categories}}, \href{https://doi.org/10.1007/s00220-009-0908-z}{\emph{Commun.
  Math. Phys.} {\bfseries 292} (2009) 343}
  [\href{https://arxiv.org/abs/0712.1377}{{\ttfamily 0712.1377}}].

\bibitem{Schellekens:1990xy}
A.~N. Schellekens and S.~Yankielowicz, \emph{{Simple Currents, Modular
  Invariants and Fixed Points}},
  \href{https://doi.org/10.1142/S0217751X90001367}{\emph{Int. J. Mod. Phys. A}
  {\bfseries 5} (1990) 2903}.

\bibitem{Gato-Rivera:1991bqv}
B.~Gato-Rivera and A.~N. Schellekens, \emph{{Complete classification of simple
  current modular invariants for (Z(p))**k}},
  \href{https://doi.org/10.1007/BF02099282}{\emph{Commun. Math. Phys.}
  {\bfseries 145} (1992) 85}.

\bibitem{Bartsch:2023pzl}
T.~Bartsch, M.~Bullimore and A.~Grigoletto, \emph{{Higher representations for
  extended operators}},  \href{https://arxiv.org/abs/2304.03789}{{\ttfamily
  2304.03789}}.

\bibitem{Henningson:2007qr}
M.~Henningson and N.~Wyllard, \emph{{Bound states in N = 4 SYM on T**3:
  Spin(2n) and the exceptional groups}},
  \href{https://doi.org/10.1088/1126-6708/2007/07/084}{\emph{JHEP} {\bfseries
  07} (2007) 084} [\href{https://arxiv.org/abs/0706.2803}{{\ttfamily
  0706.2803}}].

\bibitem{Bockenhauer:1995hx}
J.~Bockenhauer, \emph{{An Algebraic formulation of level one Wess-Zumino-Witten
  models}}, \href{https://doi.org/10.1142/S0129055X96000330}{\emph{Rev. Math.
  Phys.} {\bfseries 8} (1996) 925}
  [\href{https://arxiv.org/abs/hep-th/9507047}{{\ttfamily hep-th/9507047}}].

\bibitem{Dong:1997ea}
C.-y. Dong, H.-s. Li and G.~Mason, \emph{{Modular invariance of trace functions
  in orbifold theory}},
  \href{https://doi.org/10.1007/s002200000242}{\emph{Commun. Math. Phys.}
  {\bfseries 214} (2000) 1}
  [\href{https://arxiv.org/abs/q-alg/9703016}{{\ttfamily q-alg/9703016}}].

\bibitem{Volpato:2024goy}
R.~Volpato, \emph{{Vertex algebras, topological defects, and Moonshine}},
  \href{https://arxiv.org/abs/2412.21141}{{\ttfamily 2412.21141}}.

\bibitem{Weber1908}
H.~Weber, \emph{Lehrbuch der Algebra}, vol.~III. Friedrich Vieweg und Sohn,
  Braunschweig, 1908.

\bibitem{Ji:2019eqo}
W.~Ji and X.-G. Wen, \emph{{Non-invertible anomalies and mapping-class-group
  transformation of anomalous partition functions}},
  \href{https://doi.org/10.1103/PhysRevResearch.1.033054}{\emph{Phys. Rev.
  Research.} {\bfseries 1} (2019) 033054}
  [\href{https://arxiv.org/abs/1905.13279}{{\ttfamily 1905.13279}}].

\bibitem{Huang:2005gs}
Y.-Z. Huang, \emph{{Rigidity and modularity of vertex tensor categories}},
  \href{https://doi.org/10.1142/S0219199708003083}{\emph{Commun. Contemp.
  Math.} {\bfseries 10} (2008) 871}
  [\href{https://arxiv.org/abs/math/0502533}{{\ttfamily math/0502533}}].

\bibitem{Davydov:2011pp}
A.~Davydov and T.~Booker, \emph{{Commutative Algebras in Fibonacci
  Categories}},  \href{https://arxiv.org/abs/1103.3537}{{\ttfamily 1103.3537}}.

\bibitem{McRae:2023vyw}
R.~McRae, \emph{{Deligne tensor products of categories of modules for vertex
  operator algebras}},  \href{https://arxiv.org/abs/2304.14023}{{\ttfamily
  2304.14023}}.

\bibitem{McRae:2021urf}
R.~McRae, \emph{{A general mirror equivalence theorem for coset vertex operator
  algebras}}, \href{https://doi.org/10.1007/s11425-022-2181-0}{\emph{Sci. China
  Math.} {\bfseries 67} (2024) 2237}
  [\href{https://arxiv.org/abs/2107.06577}{{\ttfamily 2107.06577}}].

\bibitem{ChuZheng:2008}
Y.-J. Chu and Z.-J. Zheng, \emph{Vertex operator algebra analogue of embedding
  $d_8$ into $e_8$},  \href{https://arxiv.org/abs/0808.1458}{{\ttfamily
  0808.1458}}.

\bibitem{Slansky1981GroupTF}
R.~C. Slansky, \emph{Group theory for unified model building}, {\emph{Physics
  Reports} {\bfseries 79} (1981) 1}.

\bibitem{Okubo:1994ss}
S.~Okubo, \emph{{Representation of Clifford algebras and its applications}},
  {\emph{Math. Jap.} {\bfseries 41} (1995) 59}
  [\href{https://arxiv.org/abs/hep-th/9408165}{{\ttfamily hep-th/9408165}}].

\bibitem{Grimm:2010ez}
T.~W. Grimm and T.~Weigand, \emph{{On Abelian Gauge Symmetries and Proton Decay
  in Global F-theory GUTs}},
  \href{https://doi.org/10.1103/PhysRevD.82.086009}{\emph{Phys. Rev. D}
  {\bfseries 82} (2010) 086009}
  [\href{https://arxiv.org/abs/1006.0226}{{\ttfamily 1006.0226}}].

\bibitem{Dinh:2019jdg}
D.~N. Dinh, D.~T. Huong, N.~T. Duy, N.~T. Nhuan, L.~D. Thien and P.~Van~Dong,
  \emph{{Flavor changing in the flipped trinification}},
  \href{https://doi.org/10.1103/PhysRevD.99.055005}{\emph{Phys. Rev. D}
  {\bfseries 99} (2019) 055005}
  [\href{https://arxiv.org/abs/1901.07969}{{\ttfamily 1901.07969}}].

\bibitem{Babu:2023zsm}
K.~S. Babu, B.~Bajc and V.~Susi{\v{c}}, \emph{{Trinification from E$_{6}$
  symmetry breaking}},
  \href{https://doi.org/10.1007/JHEP07(2023)011}{\emph{JHEP} {\bfseries 07}
  (2023) 011} [\href{https://arxiv.org/abs/2305.16398}{{\ttfamily
  2305.16398}}].

\bibitem{Cordova:2018qvg}
C.~C{\'o}rdova, P.-S. Hsin and K.~Ohmori, \emph{{Exceptional
  Chern-Simons-Matter Dualities}},
  \href{https://doi.org/10.21468/SciPostPhys.7.4.056}{\emph{SciPost Phys.}
  {\bfseries 7} (2019) 056} [\href{https://arxiv.org/abs/1812.11705}{{\ttfamily
  1812.11705}}].

\bibitem{Mukhopadhyay:2016jog}
S.~Mukhopadhyay and R.~Wentworth, \emph{{Generalized theta functions, strange
  duality, and odd orthogonal bundles on curves}},
  \href{https://arxiv.org/abs/1608.04990}{{\ttfamily 1608.04990}}.

\bibitem{Hirschhorn1993CubicAO}
M.~D. Hirschhorn, F.~G. Garvan and J.~M. Borwein, \emph{Cubic analogues of the
  jacobian theta function $\theta$(z, q)}, {\emph{Canadian Journal of
  Mathematics} {\bfseries 45} (1993) 673 }.

\bibitem{Fulton1991RepresentationTA}
W.~Fulton and J.~W. Harris, \emph{Representation theory: A first course},
  1991,
  \href{https://api.semanticscholar.org/CorpusID:118744956}{https://api.semanticscholar.org/CorpusID:118744956}.

\end{thebibliography}\endgroup

\end{document}